\documentclass[aps,twocolumn,prd,showpacs,
nofootinbib,superscriptaddress,amsmath,amssymb]{revtex4}

\usepackage[letterpaper]{hyperref}
\usepackage{graphicx}

\usepackage[paper=letterpaper,text={7.2in,9.34in}]{geometry}
\renewcommand{\tabcolsep}{8pt}
\usepackage{stkernel}

\newcommand{\onecolwidth}{0.485\textwidth}
\newcommand{\onecolwidthsmaller}{0.445\textwidth}
\newcommand{\twocolwidth}{0.97\textwidth}

\newcommand{\lowe}{\textsc{le}}
\newcommand{\highe}{\textsc{he}}
\newcommand{\geant}{\textsc{Geant4}}
\newcommand{\twopanel}[2]{%
  \includegraphics[width=\onecolwidth]{#1}\hfill%
  \includegraphics[width=\onecolwidth]{#2}%
}
\newcommand{\twopanelsmaller}[2]{%
  \includegraphics[width=\onecolwidthsmaller]{#1}\hfill%
  \includegraphics[width=\onecolwidthsmaller]{#2}%
}
\newcommand{\fourpanel}[4]{%
  \twopanel{#1}{#2}\\
  \twopanel{#3}{#4}%
}

\begin{document}


\title{Fermi LAT observations of cosmic-ray electrons from 7 GeV to 1 TeV}

\author{M.~Ackermann}
\affiliation{W. W. Hansen Experimental Physics Laboratory, Kavli Institute for Particle Astrophysics and Cosmology, Department of Physics and SLAC National Accelerator Laboratory, Stanford University, Stanford, California 94305, USA}
\author{M.~Ajello}
\affiliation{W. W. Hansen Experimental Physics Laboratory, Kavli Institute for Particle Astrophysics and Cosmology, Department of Physics and SLAC National Accelerator Laboratory, Stanford University, Stanford, California 94305, USA}
\author{W.~B.~Atwood}
\affiliation{Santa Cruz Institute for Particle Physics, Department of Physics and Department of Astronomy and Astrophysics, University of California at Santa Cruz, Santa Cruz, California 95064, USA}
\author{L.~Baldini}
\affiliation{Istituto Nazionale di Fisica Nucleare, Sezione di Pisa, I-56127 Pisa, Italy}
\author{J.~Ballet}
\affiliation{Laboratoire AIM, CEA-IRFU/CNRS/Universit\'e Paris Diderot, Service d'Astrophysique, CEA Saclay, 91191 Gif sur Yvette, France}
\author{G.~Barbiellini}
\affiliation{Istituto Nazionale di Fisica Nucleare, Sezione di Trieste, I-34127 Trieste, Italy}
\affiliation{Dipartimento di Fisica, Universit\`a di Trieste, I-34127 Trieste, Italy}
\author{D.~Bastieri}
\affiliation{Istituto Nazionale di Fisica Nucleare, Sezione di Padova, I-35131 Padova, Italy}
\affiliation{Dipartimento di Fisica ``G. Galilei", Universit\`a di Padova, I-35131 Padova, Italy}
\author{B.~M.~Baughman}
\affiliation{Department of Physics, Center for Cosmology and Astro-Particle Physics, The Ohio State University, Columbus, Ohio 43210, USA}
\author{K.~Bechtol}
\affiliation{W. W. Hansen Experimental Physics Laboratory, Kavli Institute for Particle Astrophysics and Cosmology, Department of Physics and SLAC National Accelerator Laboratory, Stanford University, Stanford, California 94305, USA}
\author{F.~Bellardi}
\affiliation{Istituto Nazionale di Fisica Nucleare, Sezione di Pisa, I-56127 Pisa, Italy}
% ...

\collaboration{Fermi LAT Collaboration}
\noaffiliation


\begin{abstract}
We present the results of our analysis of cosmic-ray electrons using 
about $8\times 10^6$ electron candidates detected in the first 12 months 
on-orbit by the Fermi Large Area Telescope. This work extends our 
previously published cosmic-ray electron spectrum down to 7 GeV, giving a 
spectral range of approximately 2.5 decades up to 1 TeV. 
We describe in detail the analysis and its validation using beam-test
and on-orbit data. In addition, we describe the spectrum measured via 
a subset of events selected for the best energy resolution 
as a cross-check on the measurement using the full event sample.
Our electron spectrum can be described with
a power law $\propto {\rm E}^{-3.08 \pm 0.05}$ 
with no prominent spectral features within systematic uncertainties.
Within the limits of our uncertainties,
we can accommodate a slight spectral hardening at around 100~GeV and a
slight softening above 500~GeV.
\end{abstract}




\maketitle


\section{Introduction}\label{sec:introduction}

We report here a new analysis of our cosmic-ray electron 
(CRE, includes positrons) data sample, at energies between 7~GeV and 1~TeV
based on measurements made using data 
from the first full year of on-orbit operations of the 
\emph{Fermi} Gamma-ray Space Telescope's 
Large Area Telescope (LAT)~\cite{LAT}. 
Fermi was launched on June 11, 2008, into a circular orbit at 565 km 
altitude and $25.6^\circ$ inclination.  
This paper extends the energy range of our previous
measurement~\cite{PRL} down to 7 GeV, and provides more detailed information 
about our previous analysis based on the 
first six months of operations. 
In our earlier work, we reported that the CRE spectrum 
between 20 GeV and 1 TeV has a 
harder spectral index (best fit 3.04 in the case of a single 
power law) than previously indicated 
(in the range 3.1 to 3.4)~\cite{nishimura,barwick,mueller}, 
showing an excess of CREs at energies above 100~GeV
with respect to most  pre-Fermi experiments. 
The extension down to 7 GeV takes us close to the lowest geomagnetic 
cutoff energy accessible to the Fermi satellite. 
This part of the spectrum is important for understanding the 
heliospheric transport of CREs.

High-energy ($\gtrsim 100$~GeV) CREs lose their energy rapidly 
($-dE/dt \propto E^2$\/) by synchrotron radiation on Galactic magnetic fields 
and by inverse Compton scattering on the interstellar radiation field.
The typical distance over which a 1~TeV CRE loses half its total energy is 
estimated to be 300--400 pc (see e.g.~\cite{ahar}) when it propagates 
within about 1 kpc of the Sun.
This makes them a unique tool for probing nearby Galactic space. 
Lower-energy CREs are affected more readily by energy-dependent diffusive 
losses, convective processes in the interstellar medium, and perhaps 
reacceleration by second-order Fermi processes during transport 
from their sources to us. Since all these processes can affect 
the CRE spectrum after its injection by the sources, the observed spectrum 
is sensitive to the environment, i.e., to where and how electrons 
(and positrons) originate and propagate through the Galaxy.   

Recent results from the ATIC~\cite{ATIC}, 
PPB-BETS~\cite{PPB}, HESS~\cite{HESS,HESS2}, PAMELA~\cite{Pamela}, 
and Fermi LAT~\cite{PRL} collaborations have shed new light on the origin of CREs. 
The ATIC and PPB-BETS teams reported evidence for an excess of electrons 
in the range 300--700 GeV compared to the background expected from 
a conventional homogeneous distribution of cosmic-ray (CR) sources.
The HESS team reported a spectrum that steepens above $\sim 900$~GeV, 
a result which is consistent with an absence of sources of electrons above 
$\sim 1$~TeV within 300--400 pc.
The PAMELA Collaboration reports that the ratio of the positron flux 
to the total flux of electrons and positrons increases 
with energy~\cite{Pamela}, a result which has significant implications.
The majority of CR positrons (and some electrons) are thought to be 
produced via inelastic collisions between CR nuclei and interstellar 
gas (e.g.~\cite{MS1998}).
For this case of secondary production, the source spectrum for the CR positrons
mirrors that of the CR nuclei and is steeper than the injection spectrum of
primary CREs. After propagation, the secondary CR positron spectrum
remains steeper, and this should give a $e^+/(e^+ + e^-)$ ratio 
that falls with energy.
Therefore, some additional component of CR positrons appears to be required.
The Fermi result either requires a reconsideration of the source spectrum 
and/or the propagation model or indicates the presence of a nearby source.
However, the excess of events reported by ATIC and PPB-BETS 
was not detected by the LAT.
 
The measurements described above disagree 
in their details with most previous models 
(e.g.~\cite{MS1998,ahar,kob}) in which CREs 
were assumed, for the sake of simplicity, to be produced in sources 
homogeneously distributed throughout the Galaxy. 
Many recent papers have revisited the CR source modeling, 
exploring the possibility of nearby sources whose nature
could be astrophysical (e.g. pulsars) or ``exotic'' 
(see~\cite{Dario} and references therein).

In this paper, we describe the procedures for event energy reconstruction, 
electron candidate selection, and our assessment of the instrument response 
functions.  An important cross-check of our analysis is provided by 
a subset of events having longer path lengths through the calorimeter 
and therefore better energy resolution than the full data set. 
The consistency of the spectrum derived using this subset and that derived 
using the full data set indicates that the energy resolution assumed in our 
previously published work~\cite{PRL} for events $\geq 50$~GeV is 
indeed adequate.
Finally we discuss the inferred spectrum of CR electrons and its possible 
interpretation.  

Section~\ref{sec:approach} describes various aspects of our analysis method. 
Section~\ref{sec:spec_analysis} contains a thorough discussion of our efforts 
to minimize and characterize the systematic uncertainties in the analysis.  
The results are presented and discussed in Section~\ref{sec:result}.

\section{Analysis approach}\label{sec:approach}
\subsection{Overview}\label{sec:overview}

The LAT is a pair-conversion gamma-ray telescope designed to 
measure gamma rays 
in the energy range from 20~MeV to greater than 300~GeV. 
Although the LAT was designed to detect photons, 
it was recognized very early that it would be a capable detector 
of high-energy electrons~\cite{Durban,Alex}.
The LAT is composed of a $4 \times 4$ array of identical towers that 
measure the arrival direction and energy of each photon. 
Each tower is comprised of a tracker and a calorimeter module. 
A tracker module has 18 $x$-$y$ planes 
of silicon-strip detectors, interleaved with tungsten converter foils, 
with a total of 1.5 radiation lengths (X$_0$) of material 
for normally-incident particles.
In order to limit the power consumption and reduce the data volume, 
the tracker information at the single-strip level is digital 
(i.e., the  pulse height is not recorded). 
However, some information about the charge deposition in the silicon detectors 
is provided by the measurement of the time over threshold (TOT) 
of the trigger signal from each of the tracker planes; 
see~\cite{2006ITNS...53..466B} for further details on the architecture of the 
tracker electronics system.
A calorimeter module with 8.6 X$_0$ for normal incidence, has 96 CsI(Tl) 
crystals, hodoscopically arranged in 8 layers, aligned alternately 
along the $x$ and $y$ axes of the instrument.
A segmented anticoincidence detector (ACD), which tags $>99.97$\% of
the charged particles, covers the tracker module array. 
The electronic subsystem includes a robust programmable hardware trigger and 
software filters. The description of the detector calibrations 
that are not covered in this paper can be found in~\cite{Eduardo}.

The CRE analysis is based on the
gamma-ray analysis, as described in~\cite{LAT}.
The main challenge of the analysis is to identify and separate 
7--1000~GeV electrons from all other species, mainly CR protons. 
The analysis involves a trade-off between the efficiency for detecting 
electrons and that for rejecting interacting hadrons. 
The high flux of CR protons and helium~\cite{ams,bess} compared to that 
of CREs dictates that the hadron rejection must be $10^3$--$10^4$, 
increasing with energy.

The development of the LAT included careful and accurate Monte Carlo (MC) 
modeling. The details of the MC simulations are described in 
Sec.~\ref{sec:background}.
To validate the responses of the instrument, we built and modeled a beam test 
unit using spare flight towers. This unit was subjected to comprehensive 
calibration data taking using beams of photons, electrons, protons and nuclei. 
Beam-test data were compared to the results of the MC modeling, 
and the detector response was modified in the MC code, 
as discussed in Sec.~\ref{sec:beam}.
In Sec.~\ref{sec:selections} we discuss the cuts 
that select the final data sample.

\subsection{Monte Carlo simulations}\label{sec:background}

Monte Carlo simulations played an essential role in the design of the LAT 
and optimization of the data analysis.
These simulations have been used to develop the electron selection algorithms 
to remove interacting hadron background, and to determine the instrument 
response functions including efficiency, effective area, 
and solid angle for spectral reconstruction. 

We generate input distributions of gamma rays and charged particles with 
fully configurable spatial, temporal, and spectral properties, 
which allow us to simulate CR particles, beam-test data, ground calibration 
data, and even a complete gamma-ray sky. 
The simulated events are fed into a detailed model of the instrument 
with all its materials down to individual screws as well as a simplified model 
of the Fermi spacecraft and material below the LAT. 
The model of the instrument and the physical interaction processes are based 
on the \geant\ package~\cite{Geant_1}, widely used in high-energy physics. 
The details of the MC simulations for the electron analysis are 
given in~\cite{carmelo}.  
The MC simulations produce both raw and processed data,  include 
the effects of statistical processes such as Landau fluctuations in
energy loss, and simulate the on-board 
processing and trigger algorithms.
Output from the simulation is fed into the same reconstruction chain as data, 
thus producing the output quantities that can be compared with data 
from flight, calibration runs, beam tests, etc.

In the present analysis we have used three types of simulations: 
electrons only, full CR and Earth albedo particle populations, 
and protons only.
For simulating CREs, other CR particles, and Earth albedo particles, a model 
of the energetic particle populations in the Fermi orbit has been 
developed~\cite{LAT}.   
The modeled fluxes of the particles were constructed using the results from 
CR experiments, when available.
Where data were missing (e.g., the angular distribution of 
albedo protons below the geomagnetic cutoff from Galactic cosmic rays 
interacting with the atmosphere), published simulations were used 
(e.g. see~\cite{zuccon}). The model includes all the components of charged 
Galactic cosmic rays (protons, antiprotons, electrons, positrons, and nuclei 
up through iron) from the lowest geomagnetic cutoff rigidity seen by 
the spacecraft up to 10~TeV, together with reentrant and splash Earth 
albedo particles (neutrons, gamma rays, positrons, electrons, and protons) 
within the energy range 10~MeV to 20~GeV (their rates become negligible 
at higher energy). The fluxes are taken to be the same as those observed 
near solar minimum (i.e., maximum Galactic cosmic-ray intensities), 
the condition that applied for the data-taking period 
covered in this paper.

To study the effective acceptance for electrons and also
to characterize the residual background from hadrons, we needed 
a large sample of simulated events. 
Our total Monte Carlo simulations for the present analysis used 
approximately 400 CPUs for 80 days, corresponding to $\sim 90$ CPU years 
computing time, and was the most resource-intensive part of the analysis.
To enhance the number of simulated events at high energies, 
we often use input power-law spectra with equal numbers of counts per decade 
($dN/dE \propto E^{-1}$). The results then easily can be weighted to be valid 
for the spectral index of interest.

\subsection{Beam test validation}\label{sec:beam}

\begin{figure*}[hbtp]
\fourpanel%
{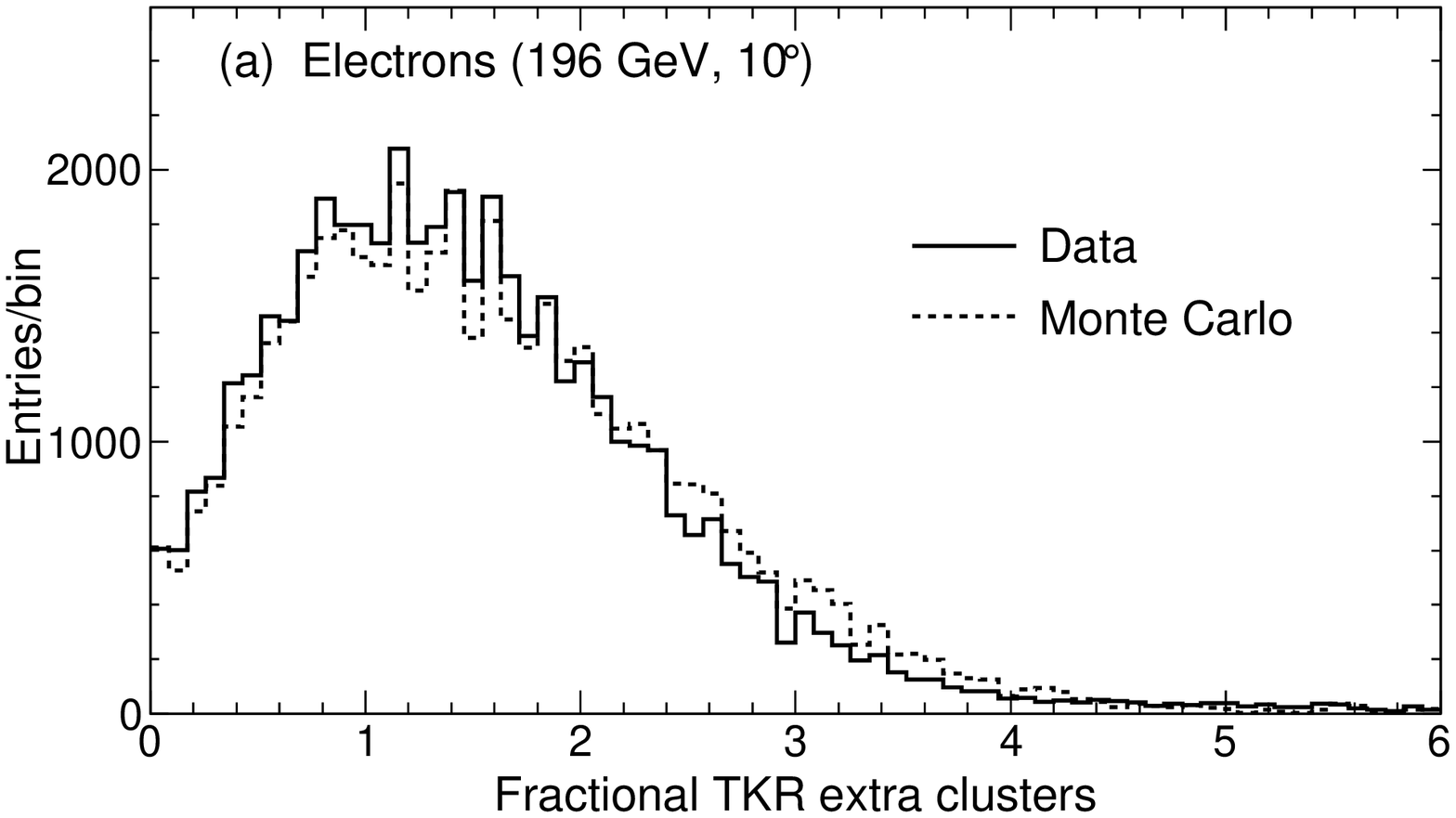}%
{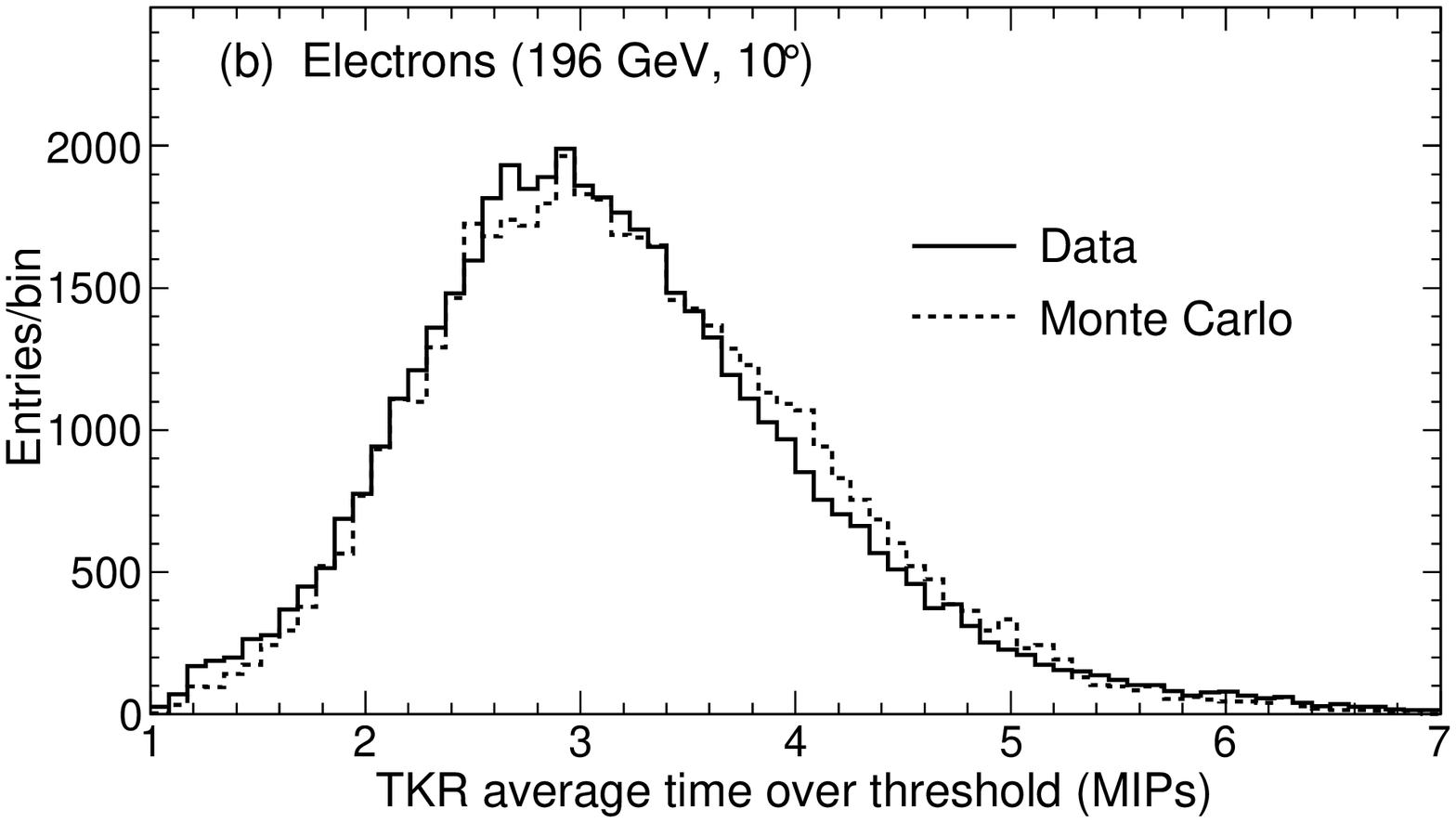}%
{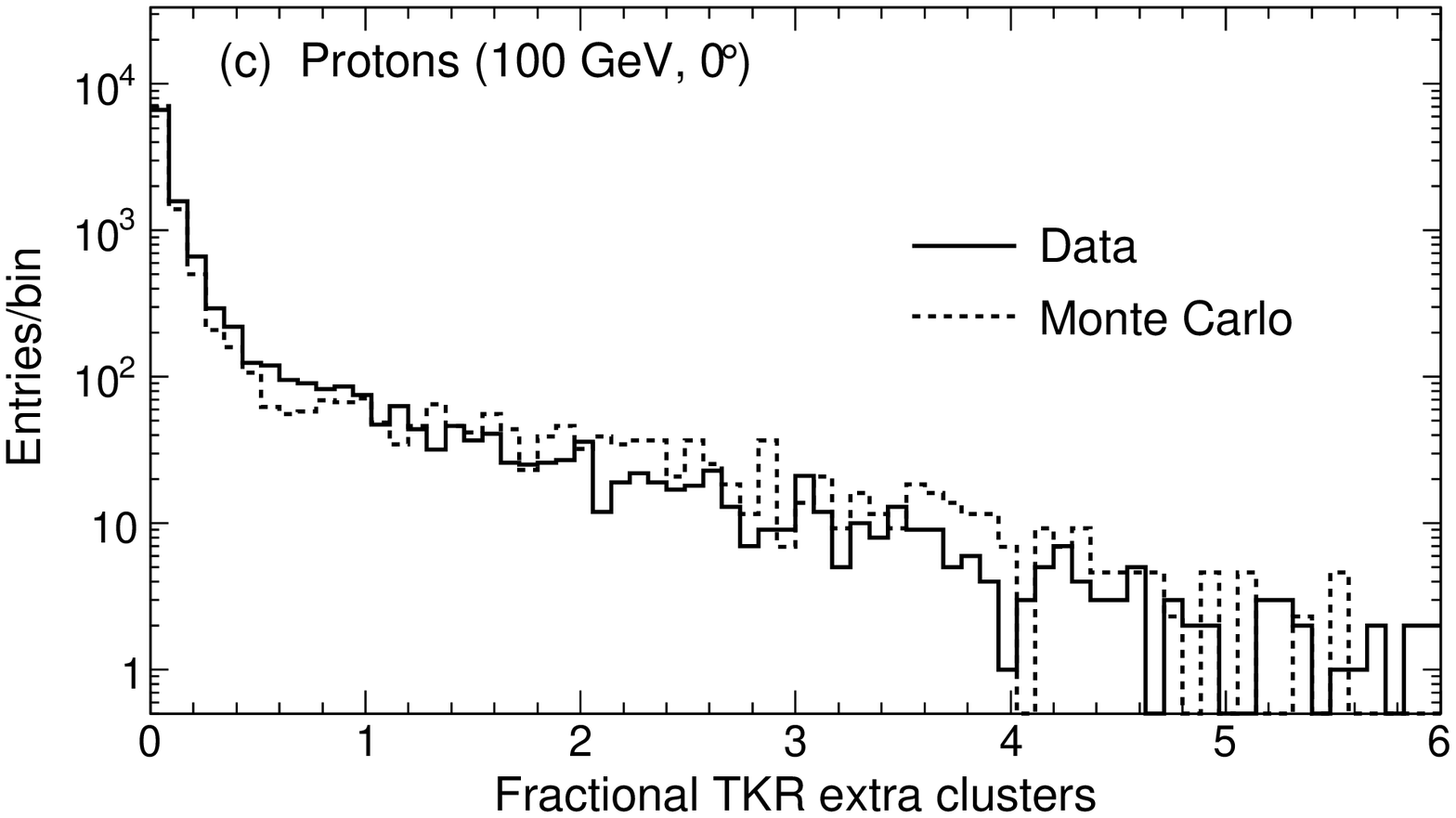}%
{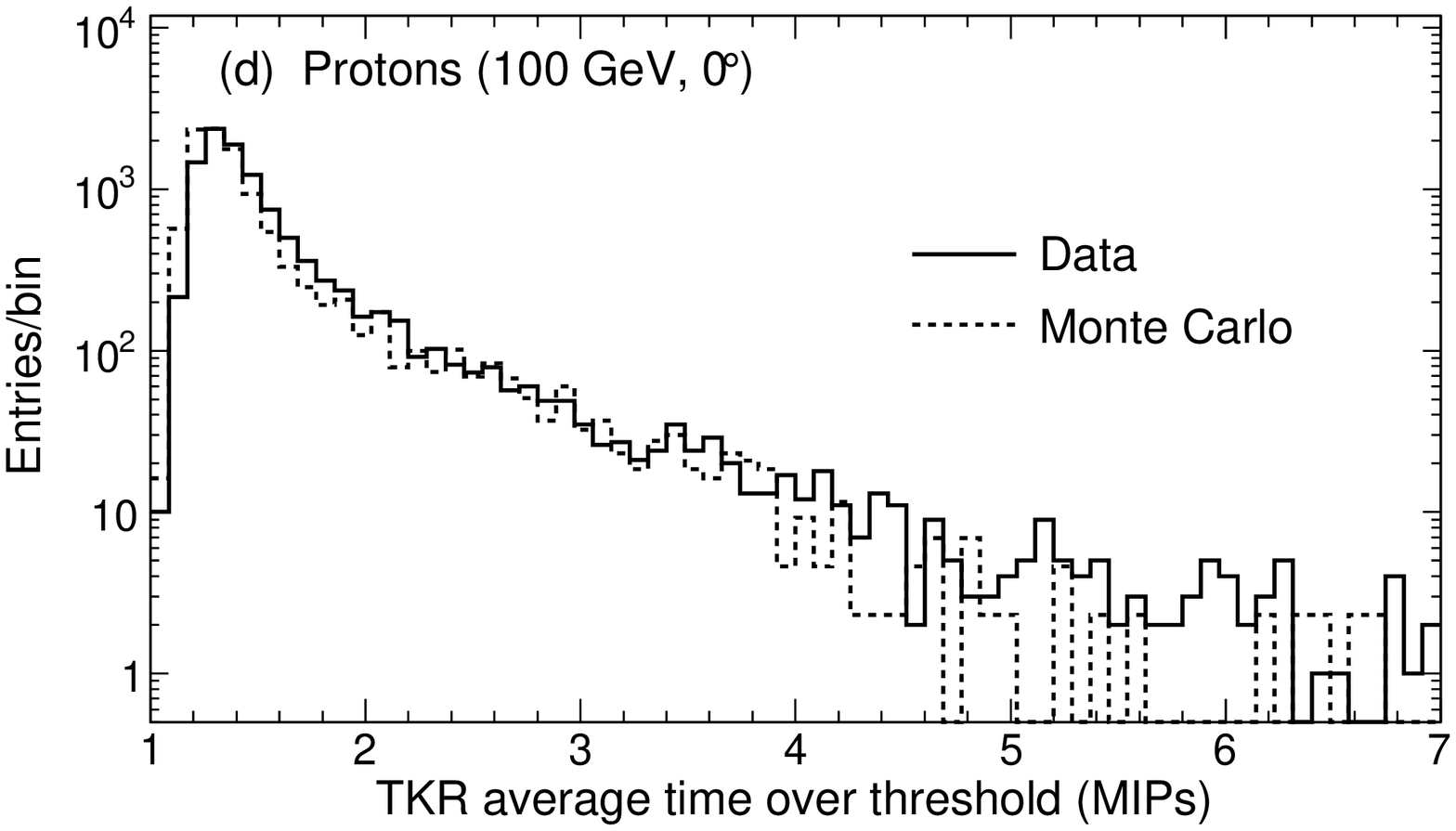}%
\caption{\label{fig:BTvariables}Comparison of beam-test data (solid line) 
and MC simulations (dashed line) for two fundamental tracker variables 
used in the electron selection: 
the number of clusters in a cone of 10 mm radius 
around the main track (left panels) and the average time over threshold 
(right panels). Both variables are shown for an electron and a proton beam. 
}
\end{figure*}

The analysis described in this paper
relies strongly on MC simulations 
for development of the event selection, 
performance parameterization, and estimation of residual background.
In order to validate the simulations, a beam-test campaign was 
performed in 2006 on a calibration unit (CU) built with flight spare modules 
integrated into a detector consisting of two complete tracker plus 
calorimeter towers, an additional third calorimeter module, 
several anticoincidence tiles, and flightlike readout electronics. 
The CU was exposed to a variety of beams of photons (up to 2.5~GeV), 
electrons \mbox{(1--300~GeV)}, hadrons ($\pi$ and $p$, a few GeV--100~GeV), 
and ions (C, Xe, 1.5~GeV/n) over 300 different instrumental configurations 
at the CERN and the GSI Helmholtz Centre for Heavy Ion Research accelerator 
complexes~\cite{BTGlastSymposium}.
Such a large data sample allows a direct comparison with
simulations over a large portion of the LAT operational phase space.

Validations studies were conducted by systematically comparing data 
taken in each experimental configuration to a simulation corresponding 
to that configuration.
Distributions of the basic quantities used for event reconstruction and
background rejection analysis, such as tracker clusters, calorimeter,
and anticoincidence detector energy 
deposits and their spatial distributions, were compared.

\begin{figure*}[!htbp]
\includegraphics[width=\twocolwidth]{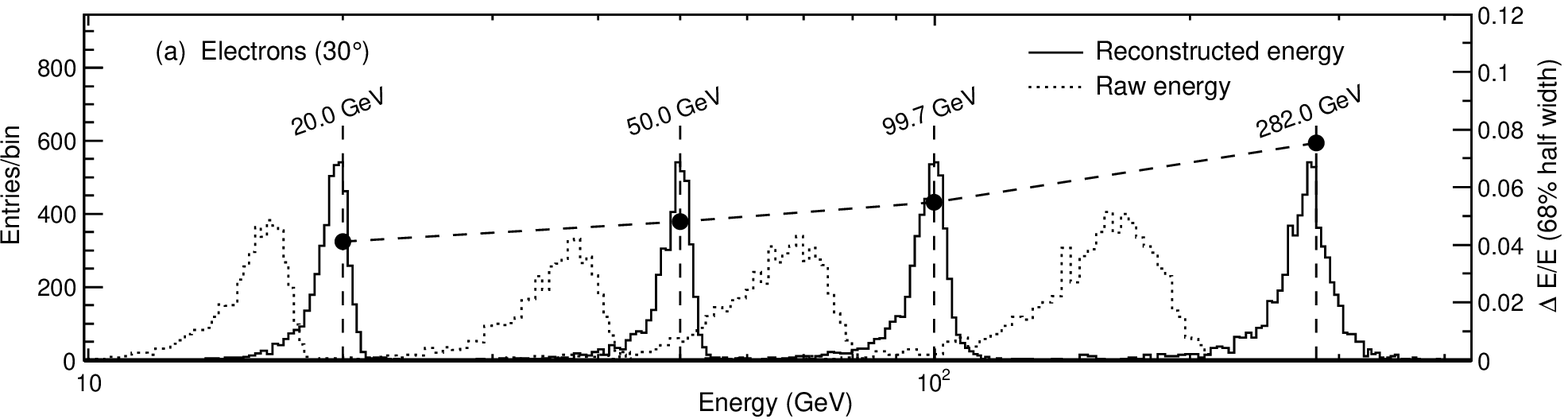}\\
\includegraphics[width=\twocolwidth]{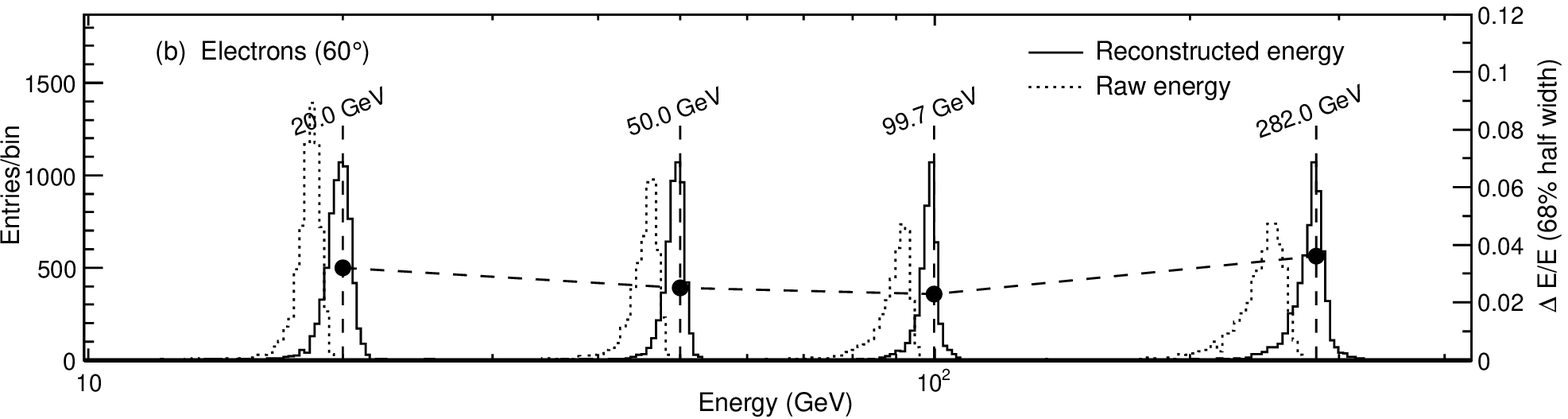}
\caption{\label{fig:BTRecon}Plots of the measured raw energy and 
reconstructed energy for different beam energies at $30^{\circ}$ (top panel) 
and $60^{\circ}$ (bottom panel). The points connected by the dashed line 
represent, for each configuration, the energy resolution (half-width of the 
68\% event containment; see Sec.~\ref{sec:energy} for details), 
which can be read on the right axis. 
The vertical dashed line represents, for each case, the nominal beam energy. 
It is clear that the leakage correction is much more pronounced 
at relatively smaller angles}
\end{figure*}

Differences were minimized after modifying the Monte Carlo simulation, 
based on the \geant\ toolkit~\cite{Geant_1}, to best match the data.
The main changes were to improve the description of the geometry and 
the materials in the instrument and along the beam lines, 
and the models describing electromagnetic (EM) 
and hadronic interactions in the detector.
Data were corrected for environmental effects that were found to affect 
the instrumental response, such as temperature drifts and beam-particle rates.

We found that EM processes are well described by the standard LHEP 
libraries~\cite{Geant_1}, 
the only exception being the \mbox{\emph{Landau-Pomeranchuk-Migdal}} effect 
(LPM,~\cite{PDG}), which was found to be inaccurately implemented. 
Based on our findings, this was fixed in the \geant\ release 
itself~\footnote{The LAT CU data were used as a benchmark for the
\geant\ EM physics classes including the LPM effect;
\geant\ releases 9.2-beta-01 and later contain the correct implementation.}.
The erroneous implementation produced a significant effect in the description 
of EM cascades at energies as low as $\sim 20$~GeV. 
The LAT is in fact sensitive to the onset of the LPM effect,
as it finely samples the longitudinal and lateral shower development.

Tuning the \geant\ simulation of hadronic interactions to the actual 
instrumental response requires choosing among the many alternative 
cross-section algorithms and interaction models that are specific 
to the energy range of interest.
\geant\ offers such flexibility through a choice of different implementations 
 from a list of possibilities~\cite{Geant_1}.
We found that the simulations that best reproduce the hadronic 
interactions recorded in the CU are obtained when using the 
Bertini libraries 
at low energies ($<20$ GeV) and the QGSP code at higher energies 
($>20$ GeV)~\cite{heikkinen,bertini}.
With such models, the agreement between data and Monte Carlo
simulations for hadronic cascades is not perfect, but appears to be
sufficient to safely estimate the residual hadronic contamination.

These codes were incorporated in both the CU and LAT simulations.
The average values of the distributions of all basic subsystem variables 
are typically reproduced by the simulations to within $\pm 5 \%$ for EM 
interactions and $\pm 10 \%$ for hadronic interactions (with maximal 
discrepancies twice as large at the limit of instrument acceptance and for 
the highest energies). 
However, essential variables such as the transverse size of showers in 
the calorimeter, the distribution of extra clusters in the tracker, and 
the average time over threshold along the best track 
(see, for example, Fig.~\ref{fig:BTvariables}) are well reproduced.
Residual  differences between the beam test data 
and the simulations are all included 
in the systematic errors evaluated according to the prescription 
in Sec.~\ref{sec:systematic}.

\begin{figure}[!bhp]
  \begin{center}
      \includegraphics[width=\onecolwidth]{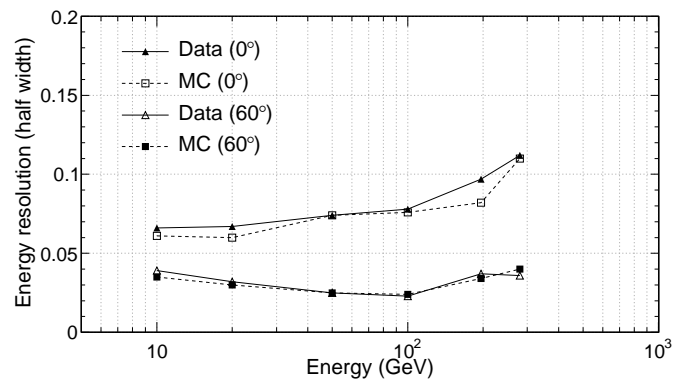}
    \caption{\label{fig:BTEResolution}Comparison of beam-test data (triangles) 
    and Monte Carlo simulations (squares) for the energy resolution 
    for electron beams entering the CU at $0^{\circ}$ and $60^{\circ}$ 
    and energies from 10 to 282 GeV.
    Lines connecting points are to guide an eye.
    }
  \end{center}
\end{figure}

Beam-test electron data were also used to validate our evaluation 
of the energy resolution.
As explained in Sec.~\ref{sec:energy}, high-energy EM showers 
are not fully contained in the LAT CU, and an evaluation of the shower 
fraction leaking from the CAL is needed to correctly reconstruct 
the shower energy.
The effect of the leakage correction in the energy reconstruction algorithm
can be seen directly in Fig.~\ref{fig:BTRecon}, 
where the raw energy deposit and the reconstructed energy distributions 
are shown for several electron beams impacting the CU. 
The energy resolution derived from the peak widths is plotted 
on the right axis.
The agreement between data and our simulations is shown 
in Fig.~\ref{fig:BTEResolution}. 

\begin{figure*}[htbp]
\fourpanel%
{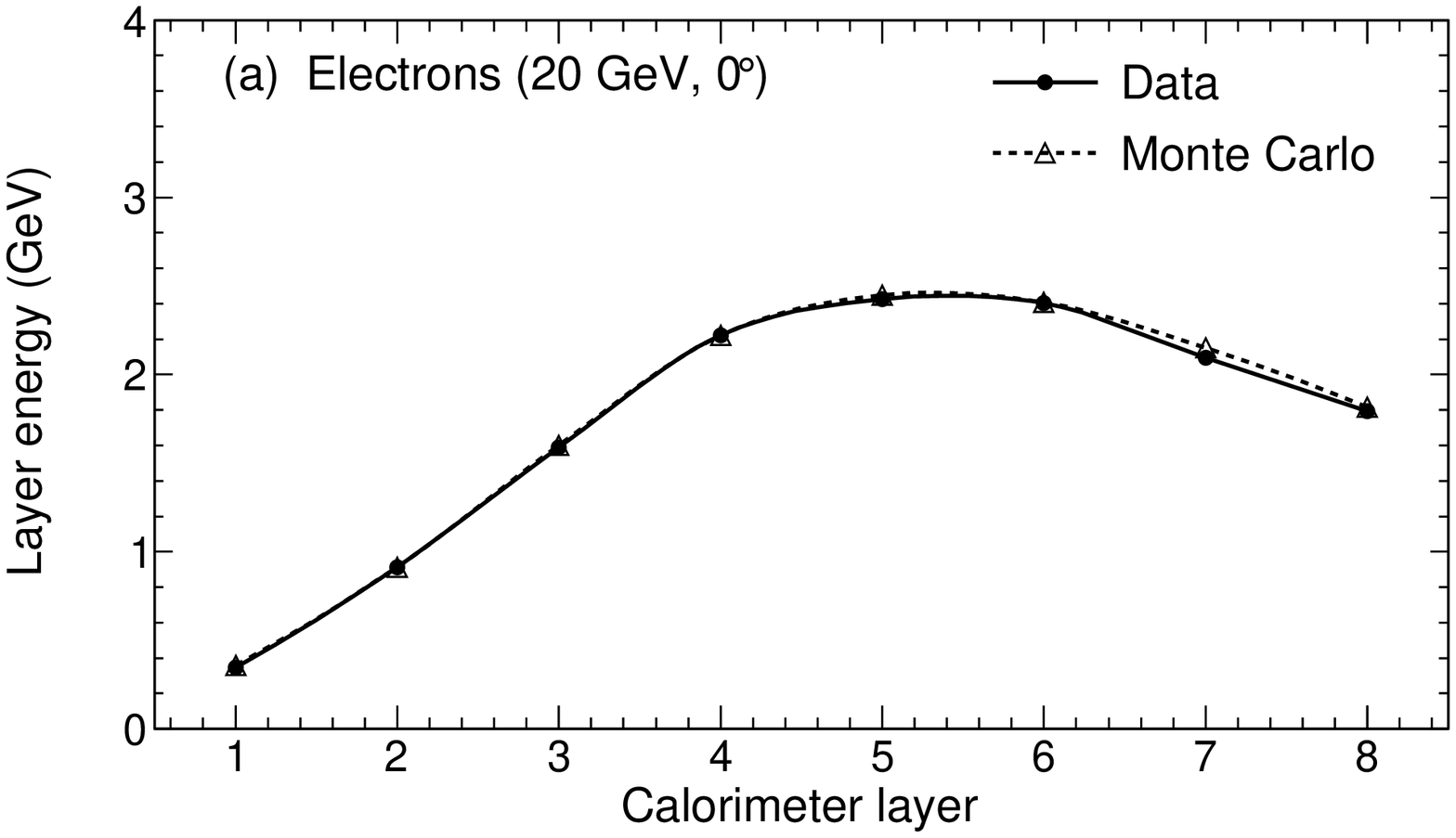}%
{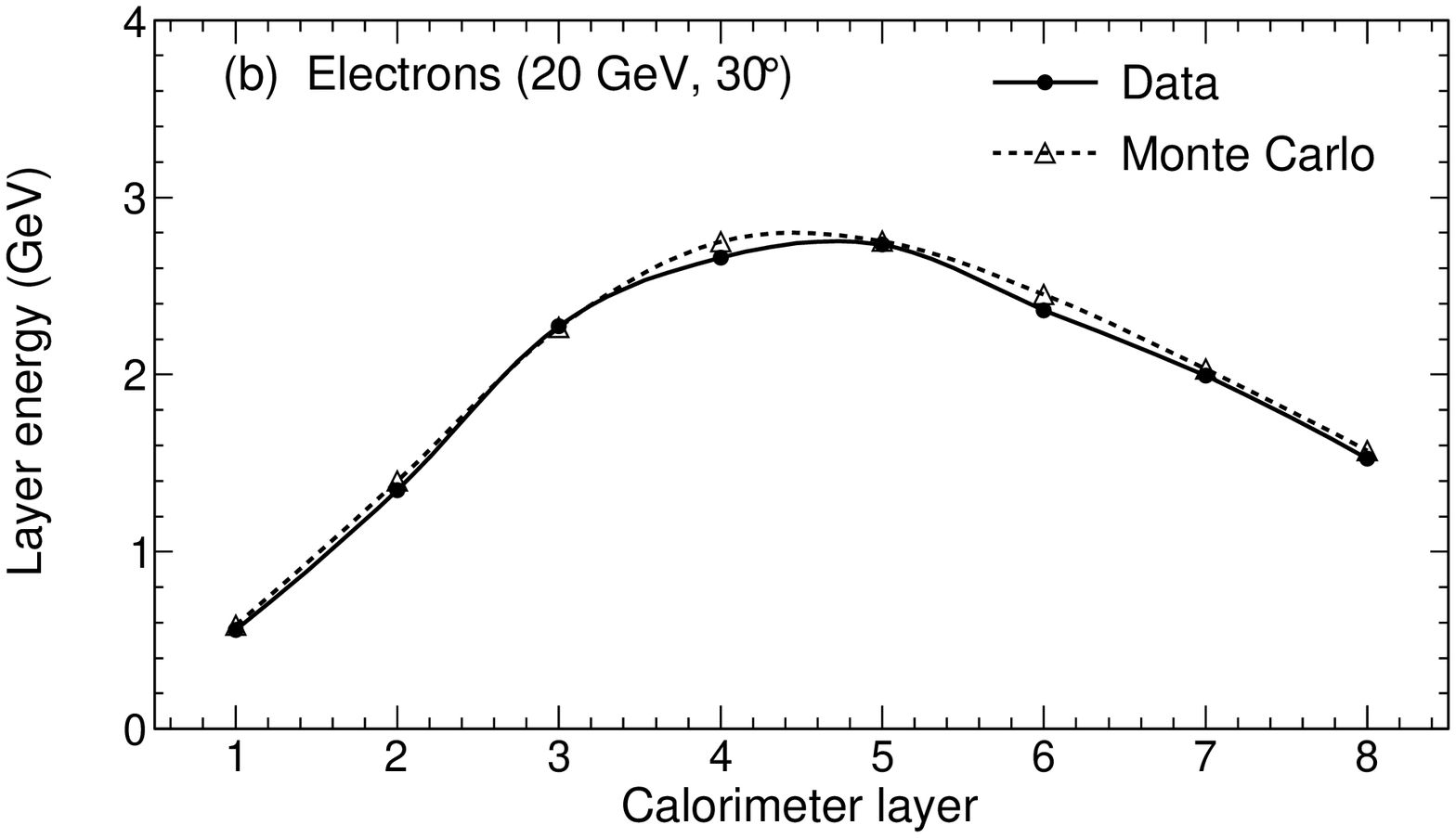}%
{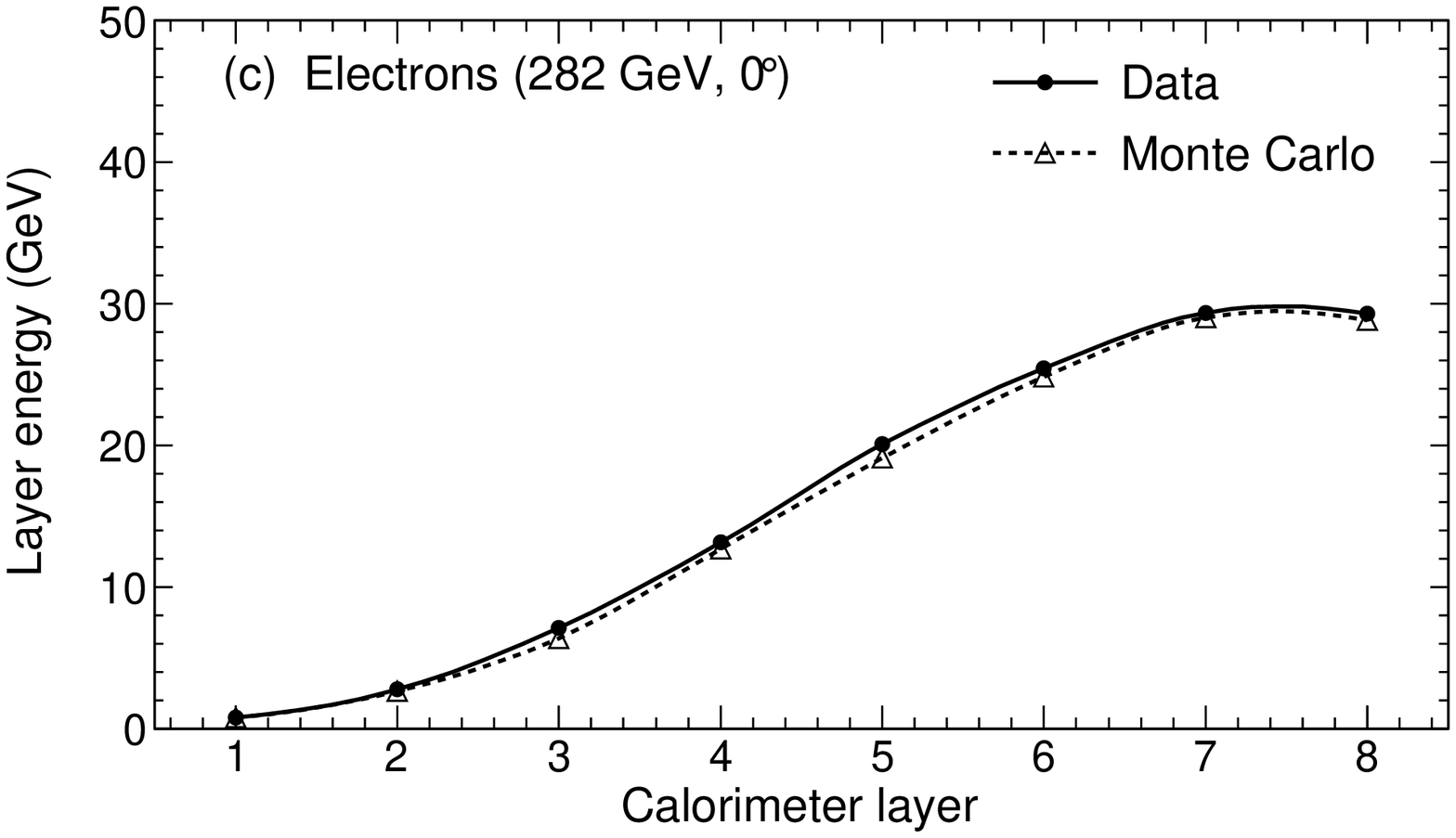}%
{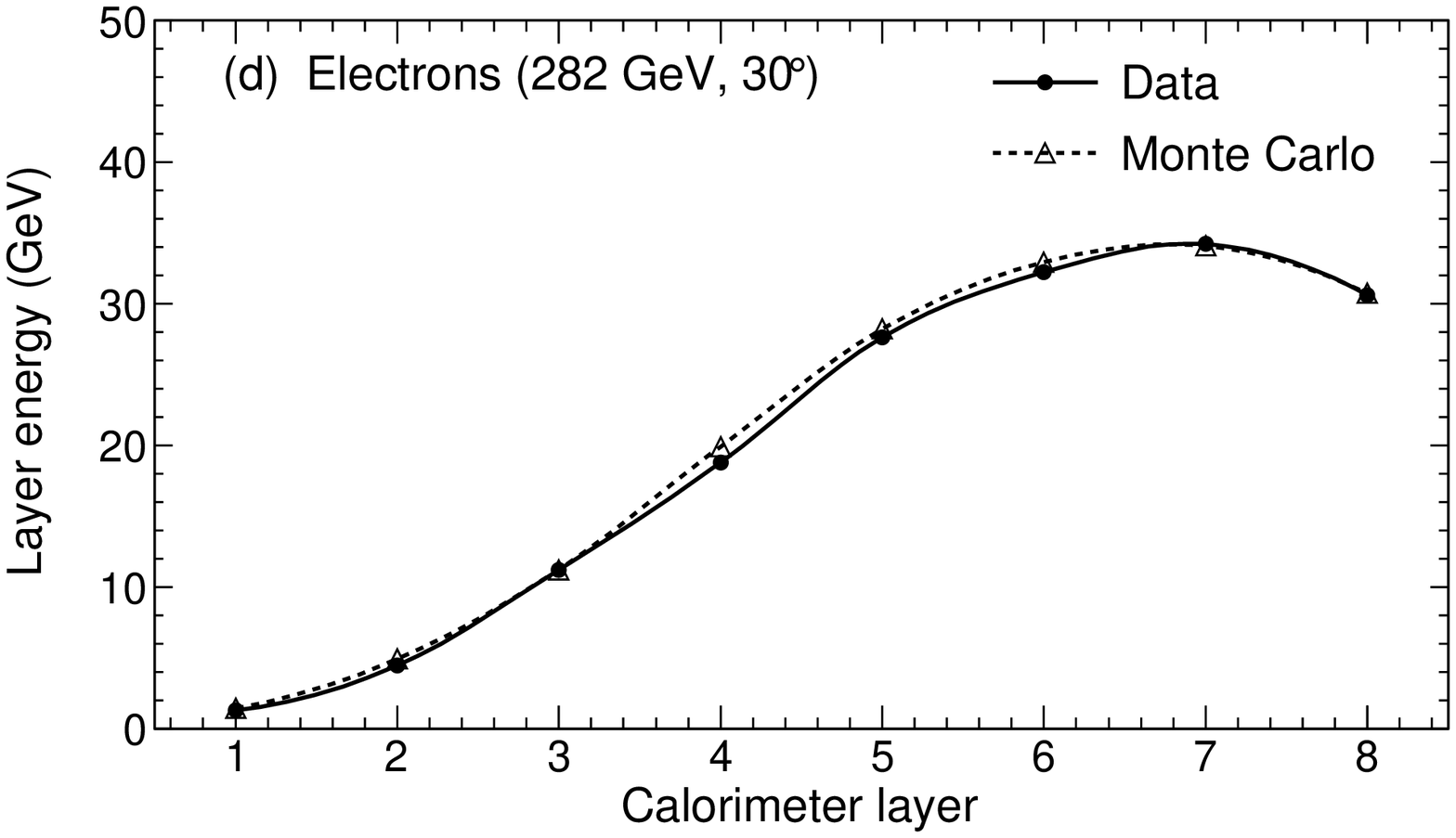} 
\caption{\label{fig:longEMshower}Comparison of beam-test data and Monte Carlo 
simulations for the longitudinal shower profiles for electron beams 
entering the CU at $0^{\circ}$ and $30^{\circ}$ and energies of 20 and 282~GeV.
}
\end{figure*}

After improving the simulation as described above,
an important residual discrepancy between the simulation and the beam test 
data was found in the raw energy deposited in the CU,
which was measured to be $9\%$ higher, on average, than predicted, 
with an asymmetric spread ranging from $-6\%$ to $+1\%$, slightly depending 
on the energy and incident angle.
This difference was corrected in beam test data using a simple scaling factor 
on the CU energy measurement, 
thus providing a good agreement between the energy deposit along 
the shower axis with the Monte Carlo simulations, as can be seen in 
Fig.~\ref{fig:longEMshower}. 

The origin of this 9\% scaling factor is unknown.
It may have to do with an imperfect calibration of the CU calorimeter modules 
or residual effects from temperature and rates at the beam test that were not 
accounted for in the data analysis. 
Further studies are now in progress with flight data.
For this reason the LAT data are not corrected with this scaling factor, 
but we include a systematic uncertainty in 
the LAT energy scale of $^{+5\%}_{-10 \%}$.

\subsection{Event selection\label{sec:selections}}

The event selection relies on the capabilities of the 
tracker, calorimeter, and anticoincidence subsystems, alone 
and in combination to discriminate between electromagnetic 
and hadronic event topologies. 
The analysis of EM showers produced in the instrument by CRE and
gamma rays is very similar.
For event reconstruction (track identification, 
energy and direction measurement, ACD analysis) 
and calculation of variables 
used in event classification we use the same reconstruction algorithms.
Although based on the same techniques, the selections 
are of course different and specific to the electron analysis. 
For example, the ACD effectively separates charged particles from photons. 
It also provides information on the topologies 
of the event useful for separating electrons from protons.   
The electron analysis covers the energy range from a few GeV to 1~TeV
while the photon analysis is currently optimized 
for the 100~MeV--300~GeV range.

Although some fraction of hadrons can have interactions that
mimic electromagnetic events, their true energies cannot be evaluated 
event by event and are underestimated by our reconstruction algorithms.
Generally, the shapes of hadronic showers differ significantly from 
EM showers. 
The most powerful separators are the comparative lateral distributions. 
Electromagnetic cascades are tightly confined, while hadronic cascades 
that leave comparable energy in the calorimeter tend to 
deposit energy over a much wider lateral region affecting 
all three detector subsystems.  
The nuclear fragments tend to leave energy far from the main trajectory 
of the particle. 
Thus hadron showers have larger transverse sizes in the calorimeter, 
larger numbers of stray tracks in the tracker, and larger energy 
deposits in more ACD tiles.

Since the phenomenology of the EM cascades and hadron interactions 
varies dramatically over the energy range of interest, 
we developed two independent event selections, one tuned for 
energies between 20 and 1000~GeV and the other for energies 
between 0.1 and 100~GeV, which we shall refer to as \highe\ and \lowe.
The \highe\ analysis takes advantage of the fact that the on-board filtering 
(event selections designed to fit the data volume into the available 
telemetry bandwidth with a minimal impact on the photon yield) 
is disengaged for events depositing 
more than 20~GeV in the calorimeter. The source of data 
for the \lowe\ selection is an unbiased sample of all trigger types, prescaled 
on-board so that one out of 250 triggered events is recorded 
without filtering.
The region of overlap in energy, between 20 and 80~GeV, allows us to 
cross-check the two independent analyses.
Above about 80~GeV the number of events in the prescaled sample 
becomes too low to be useful.

\begin{figure*}[!bht]
\fourpanel%
{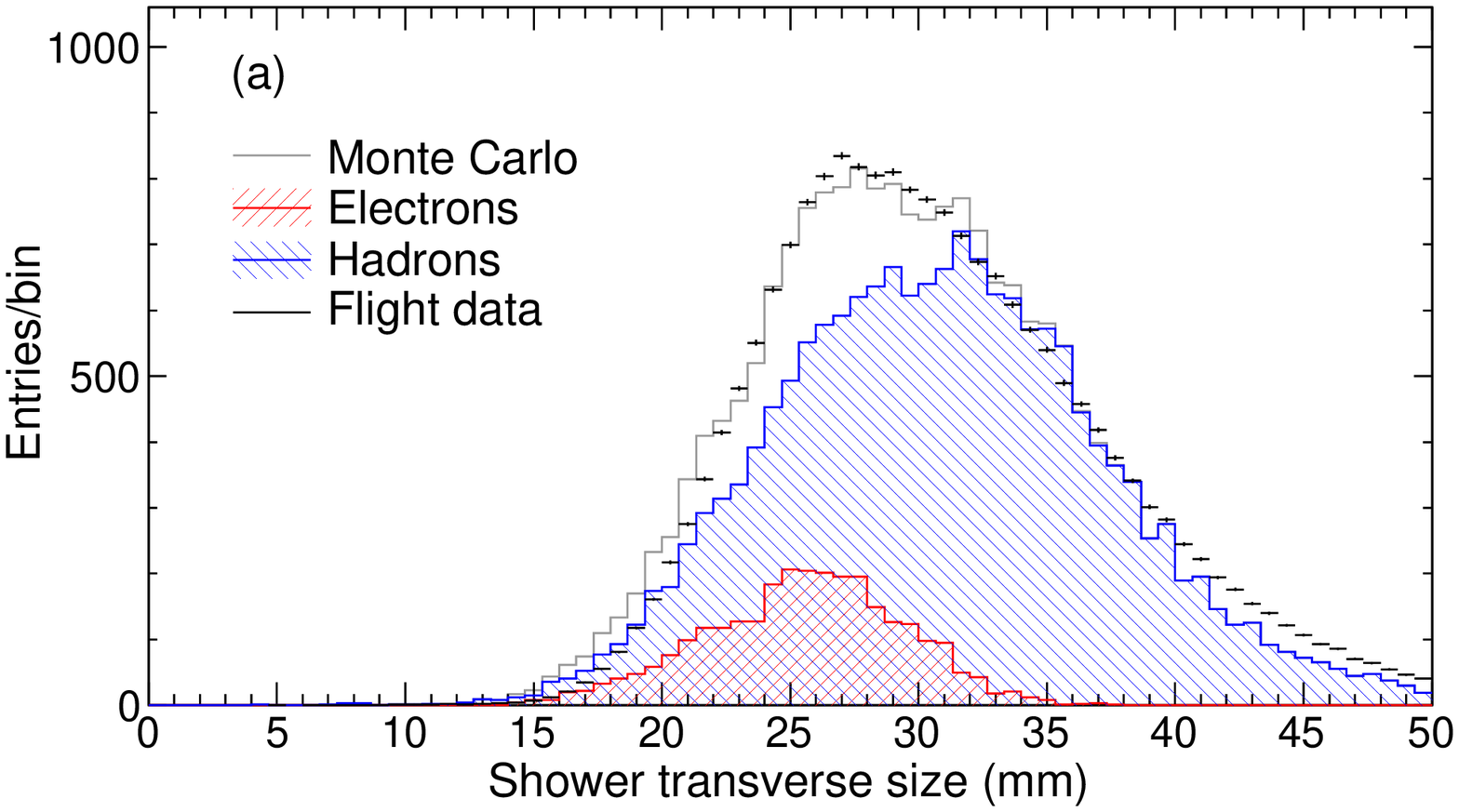}%
{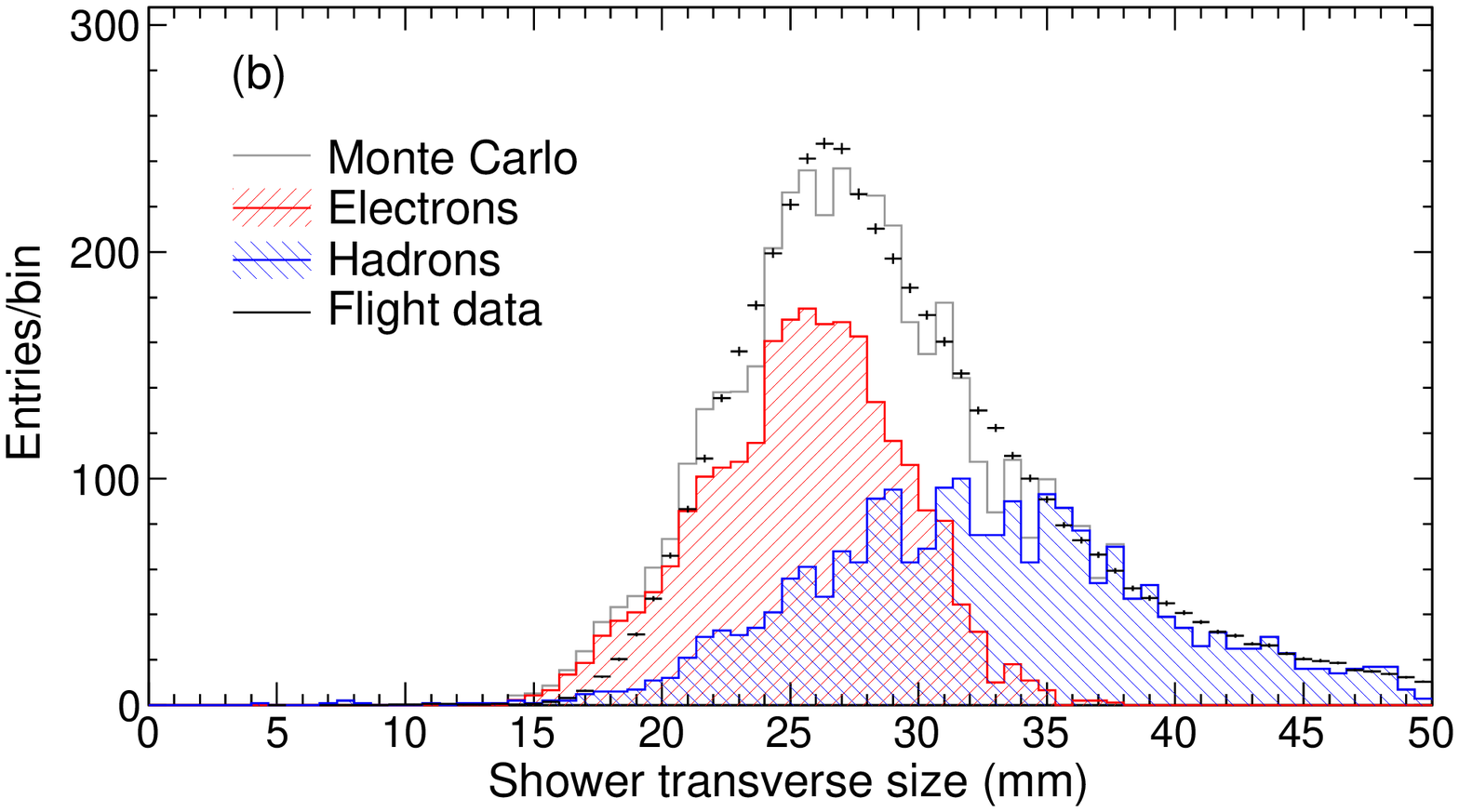}%
{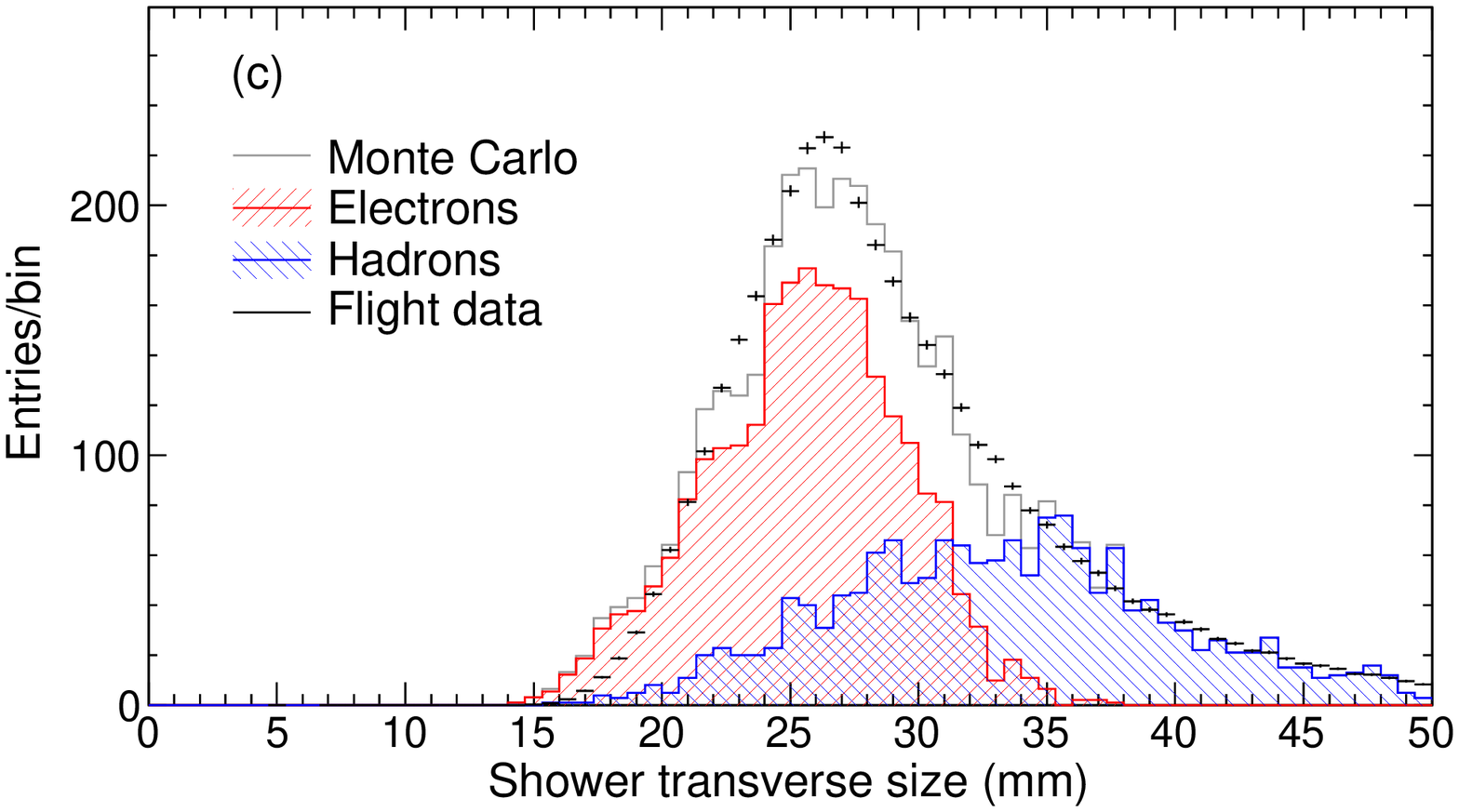}%
{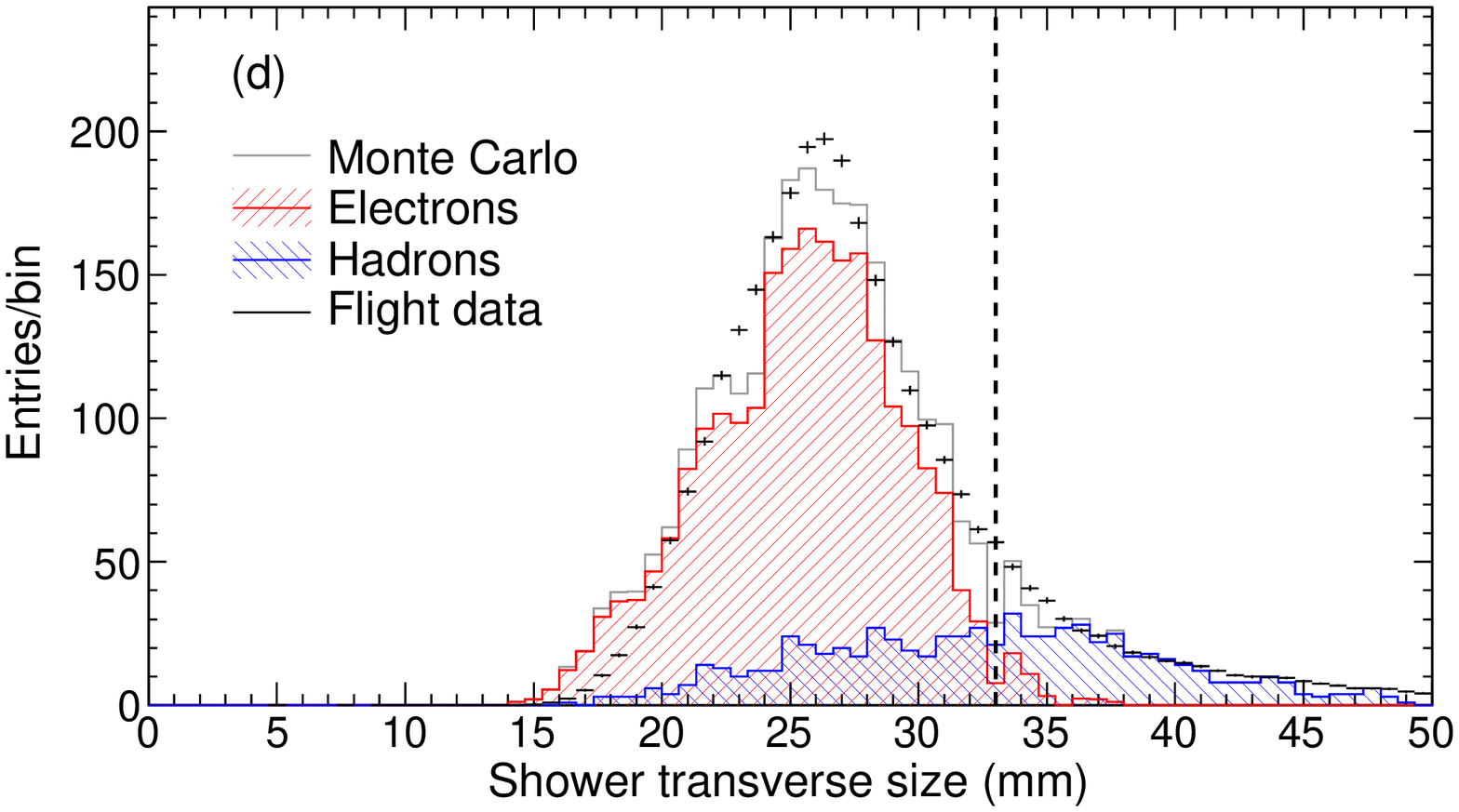}%
\caption{(color online). Distribution of the shower transverse size in the calorimeter for
the energy interval 133--210~GeV at different stages of the \highe\ selection:
(a) after the cuts on the calorimeter variables except the one
on the transverse size itself, (b) adding the selection on the tracker,
(c) on the ACD and (d) on the probability that each event is an electron
based on a classification tree analysis.
The vertical dashed line in panel (d) represents the value of the cut
on this variable.
The Monte Carlo distribution (gray line) is the sum of both the electron 
and hadron components.
The simulations have poorer statistics 
(as reflected in larger bin-to-bin fluctuations) and are scaled to the 
flight data.
}
\label{fig:CalTransRms_selection}
\end{figure*}

The event selection process must balance removal of background events 
and retaining signal events, while limiting systematic uncertainties.
We first reject those events that are badly reconstructed 
or are otherwise unusable. We require at least one reconstructed track 
and a minimum energy deposition (5~MeV for \lowe\ and 1~GeV for \highe) 
and, for \highe\ events, a pathlength longer than $7~{\rm X_0}$ 
in the calorimeter. 
We keep only events with zenith angle $<105^{\circ}$ to reduce 
the contribution from Earth albedo particles.

The next step is to select electron candidates based on the detailed 
event patterns in the calorimeter, the tracker, and the ACD subsystems.	

The calorimeter plays a central role by imaging the shower and determining 
its trajectory.  We fit both the longitudinal (for determining energy) 
and transverse shower distributions and compare them to the distributions 
expected for electromagnetic cascades.  
Figure~\ref{fig:CalTransRms_selection} shows the sequence of four successive 
cuts on the data in a single energy bin, for the transverse shower size 
in the calorimeter. This figure illustrates the difference in transverse 
shower size between electrons and hadrons, and illustrates how all three 
LAT subsystems contribute to reduce the hadron contamination.

\begin{figure*}[!thbp]
\twopanel%
{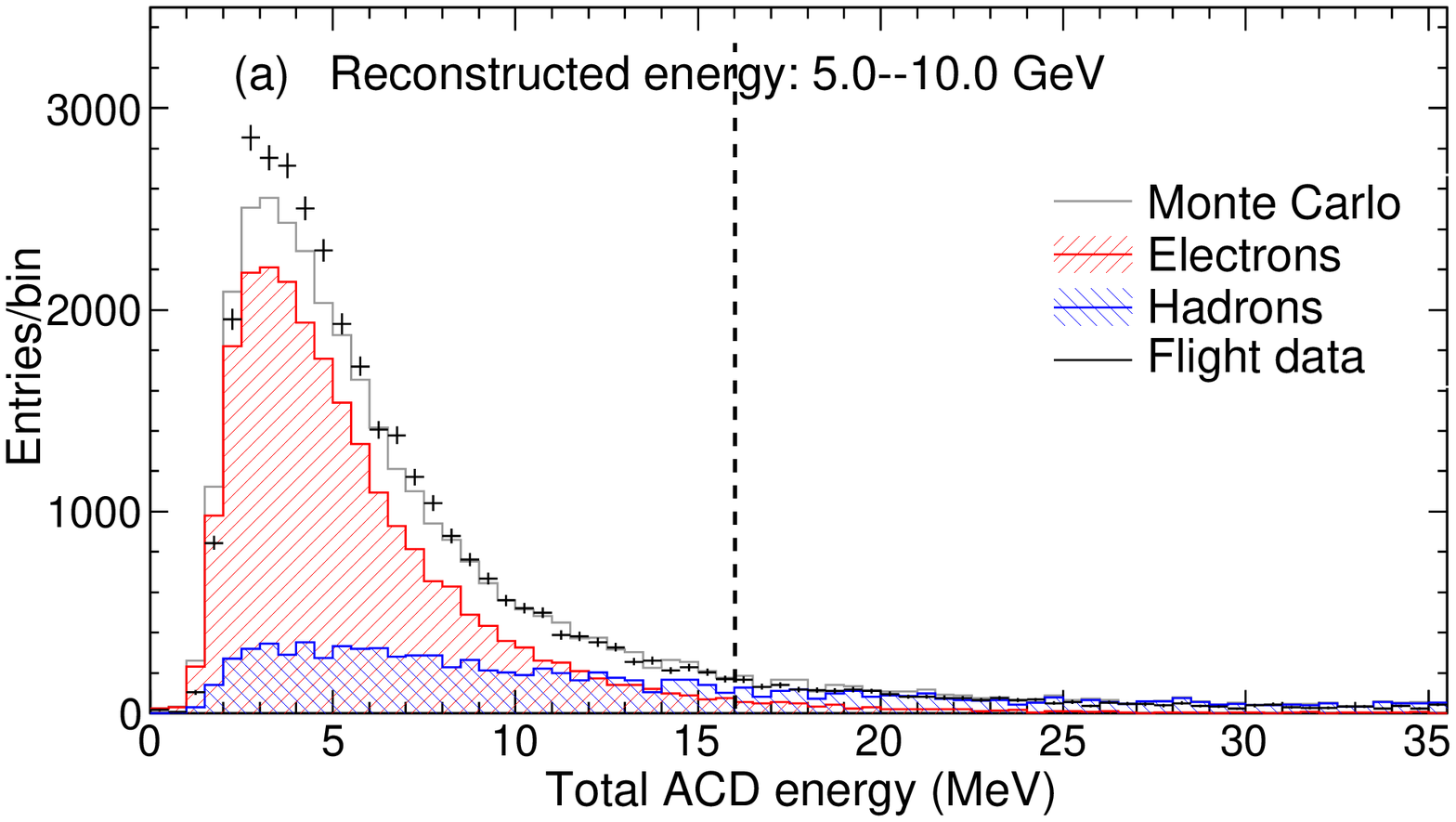}%
{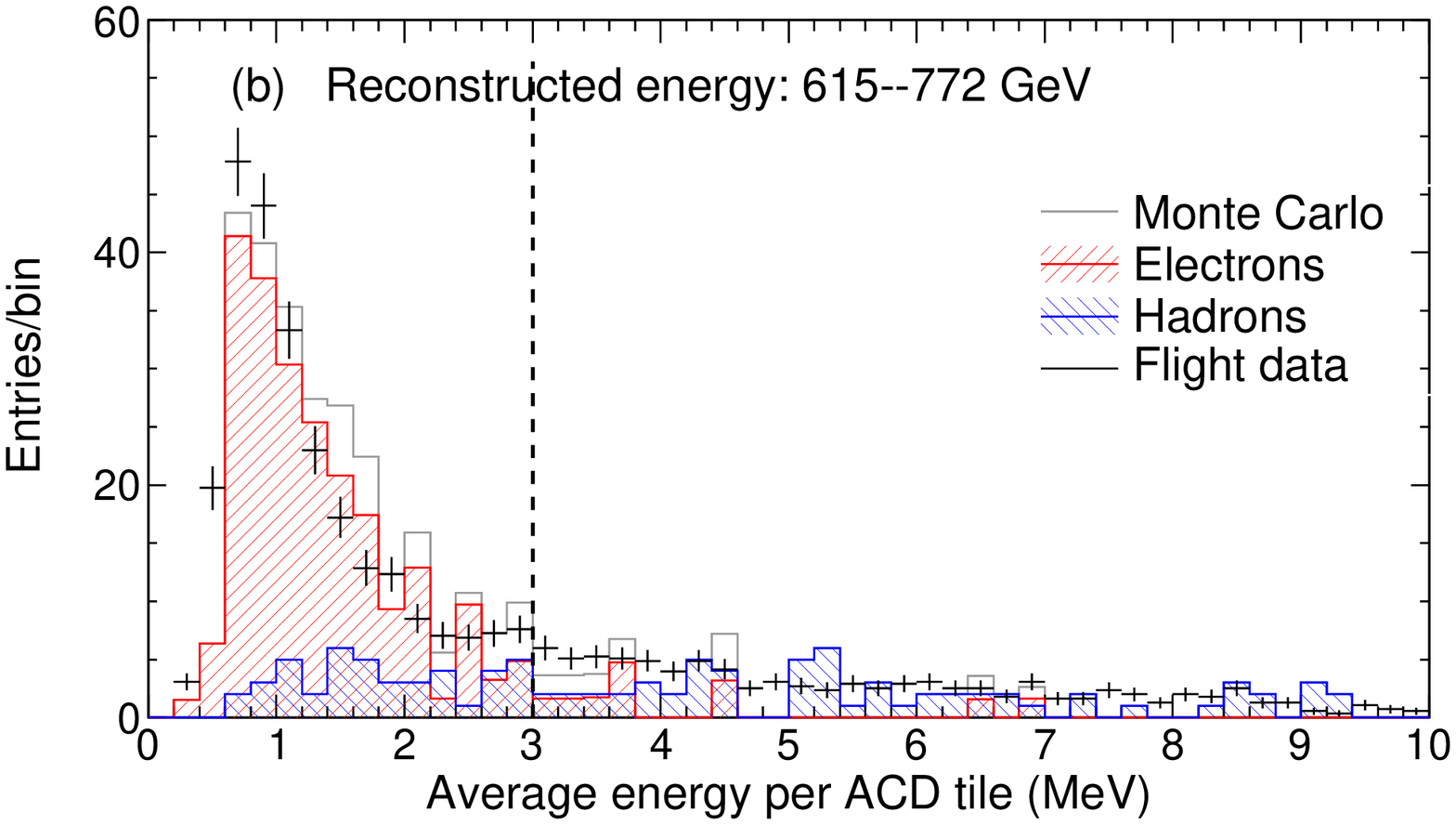}%
\caption{(color online). Distribution of (a) total energy deposition in the ACD 
used in \lowe\ selection 
and (b) average energy per ACD tile used in \highe. 
The vertical dashed line corresponds to the cut value on the variables 
in question. Both distributions are shown after the cuts on all other 
variables have been applied. 
}
\label{fig:AcdEnergy}
\end{figure*}

The tracker images the initial part of the shower.
As shown earlier in Fig.~\ref{fig:BTvariables}, electrons are selected 
by having larger energy deposition along the track 
and more clusters in the vicinity 
(within $\sim 1$~cm) of the best track, but which do not belong 
to the track itself.
As illustrated in Figs.~\ref{fig:BTvariables}(a) and~\ref{fig:BTvariables}(c),
the fraction of these extra clusters is, on average, much higher 
for energetic electrons than for protons. 
The average energy deposition in the silicon planes
(which we measure by means of the time over threshold) 
is also higher for electrons, as can be seen in 
Fig.~\ref{fig:BTvariables}(b) and~\ref{fig:BTvariables}(d).

The ACD provides part of the necessary discrimination power.
Photons are efficiently rejected using the ACD in conjunction with 
the reconstructed tracks. 
A signal in an ACD tile aligned with the selected track 
indicates that the particle crossing the LAT is charged.
Hadrons are removed by looking for energy deposition in all the
ACD tiles, mainly produced by particles backscattering from the calorimeter.
Two examples of this effect can be seen in Fig.~\ref{fig:AcdEnergy}. 
Figure~\ref{fig:AcdEnergy}(a) shows the total energy deposition
in the ACD tiles for the \lowe\ analysis; the hadrons are more likely to 
populate the high-energy tail. Figure~\ref{fig:AcdEnergy}(b) shows 
the average energy per tile in the \highe\ analysis; it is significantly 
higher for hadrons than for electrons, due to backsplash from nuclear cascades.

A classification tree (CT) analysis~\footnote{The reader can refer 
to~\cite{datamining} for a comprehensive review of the use of 
data mining and machine learning techniques in astrophysics.}
provides the remaining hadron rejection power necessary 
for the CRE spectrum measurement. 

\begin{figure*}[htbp]
\fourpanel%
{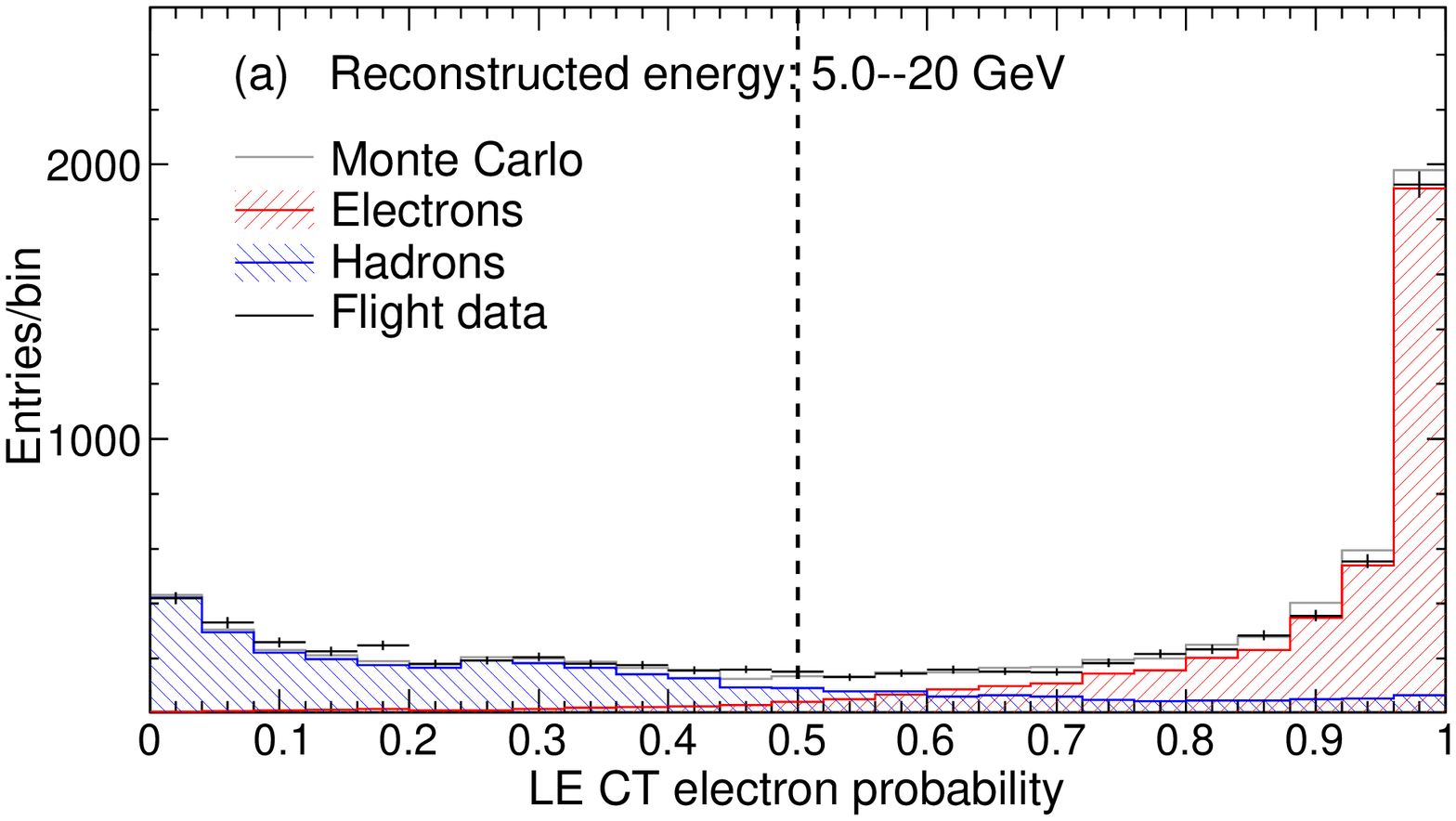}%
{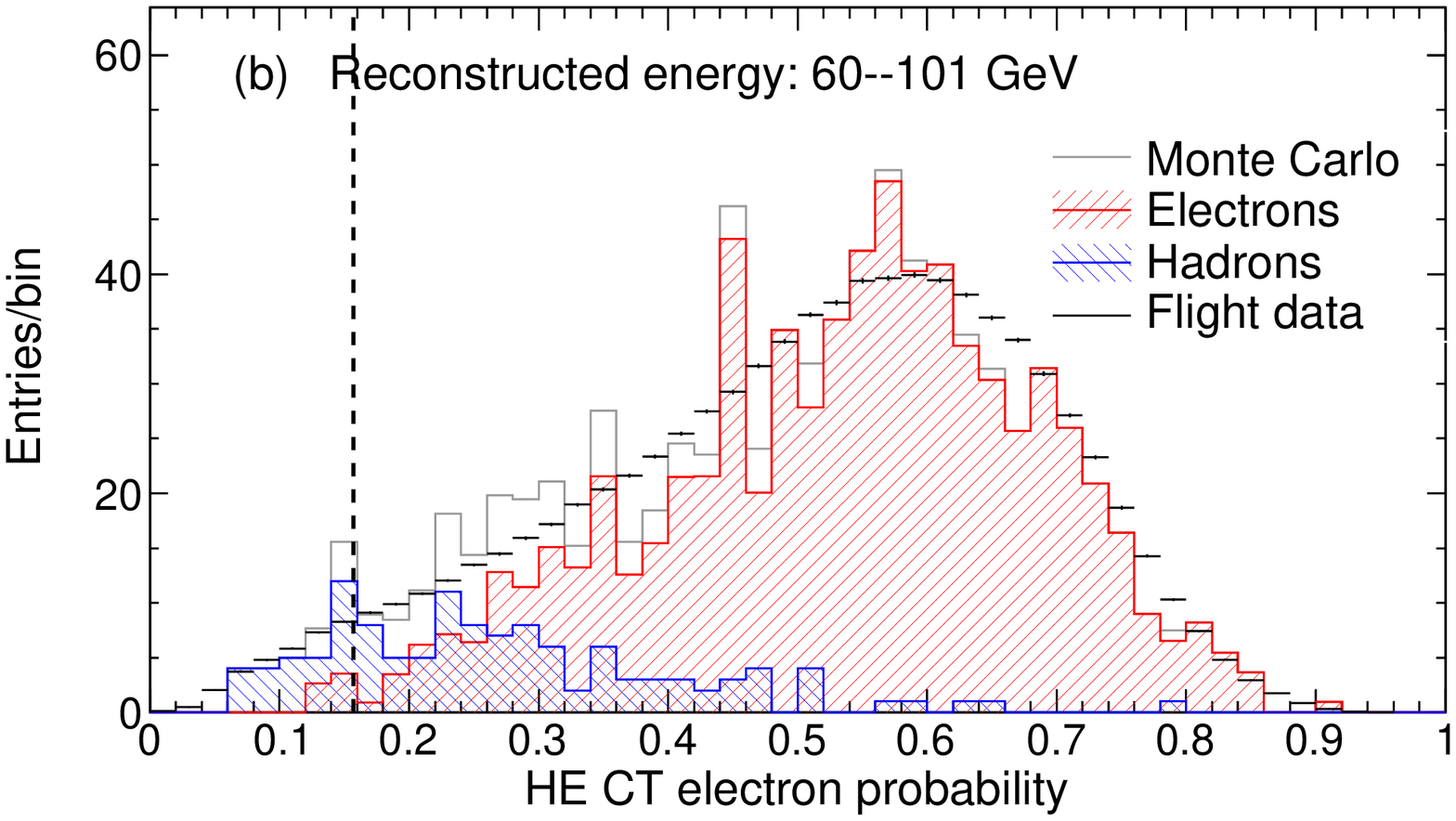}%
{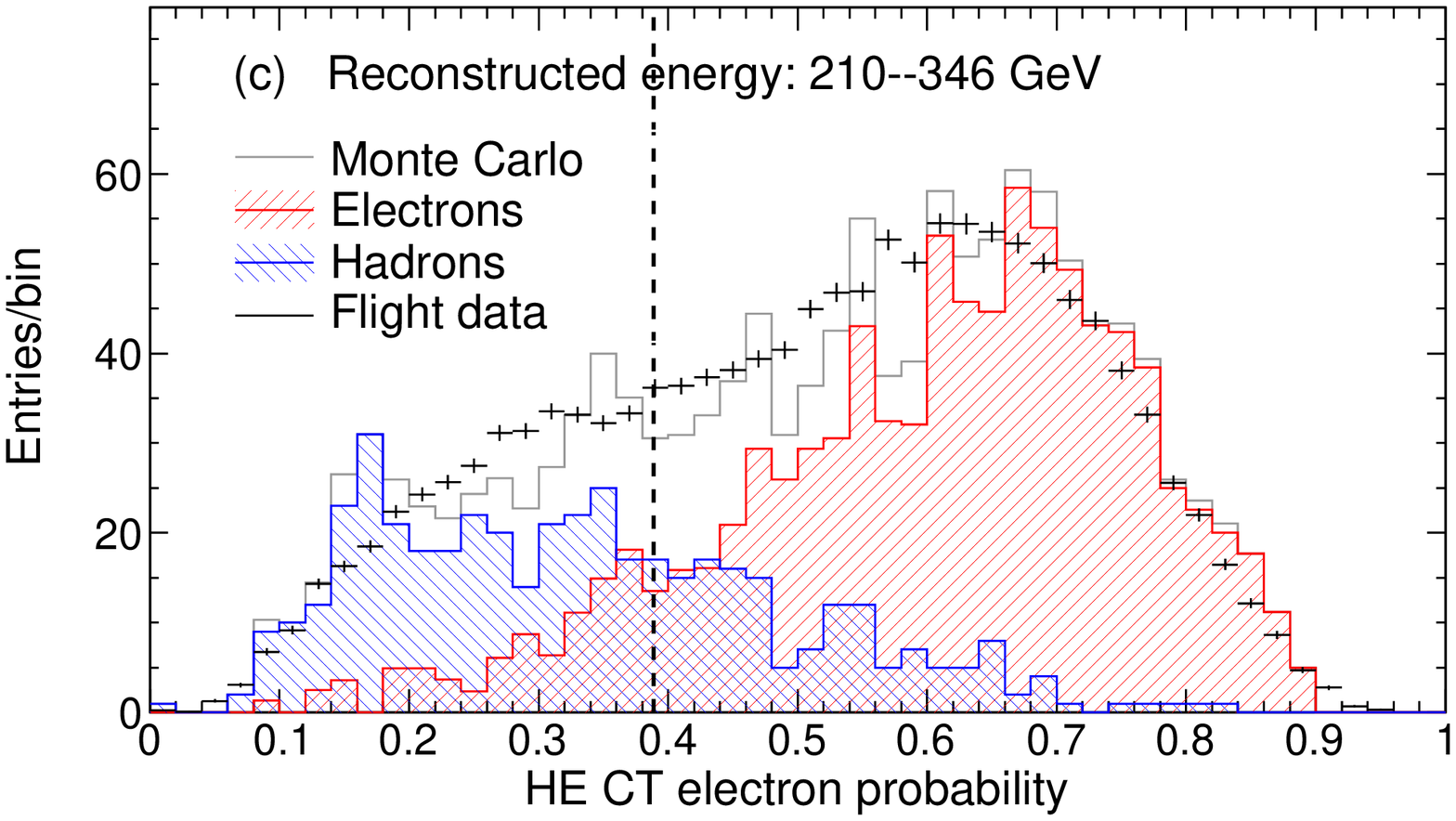}%
{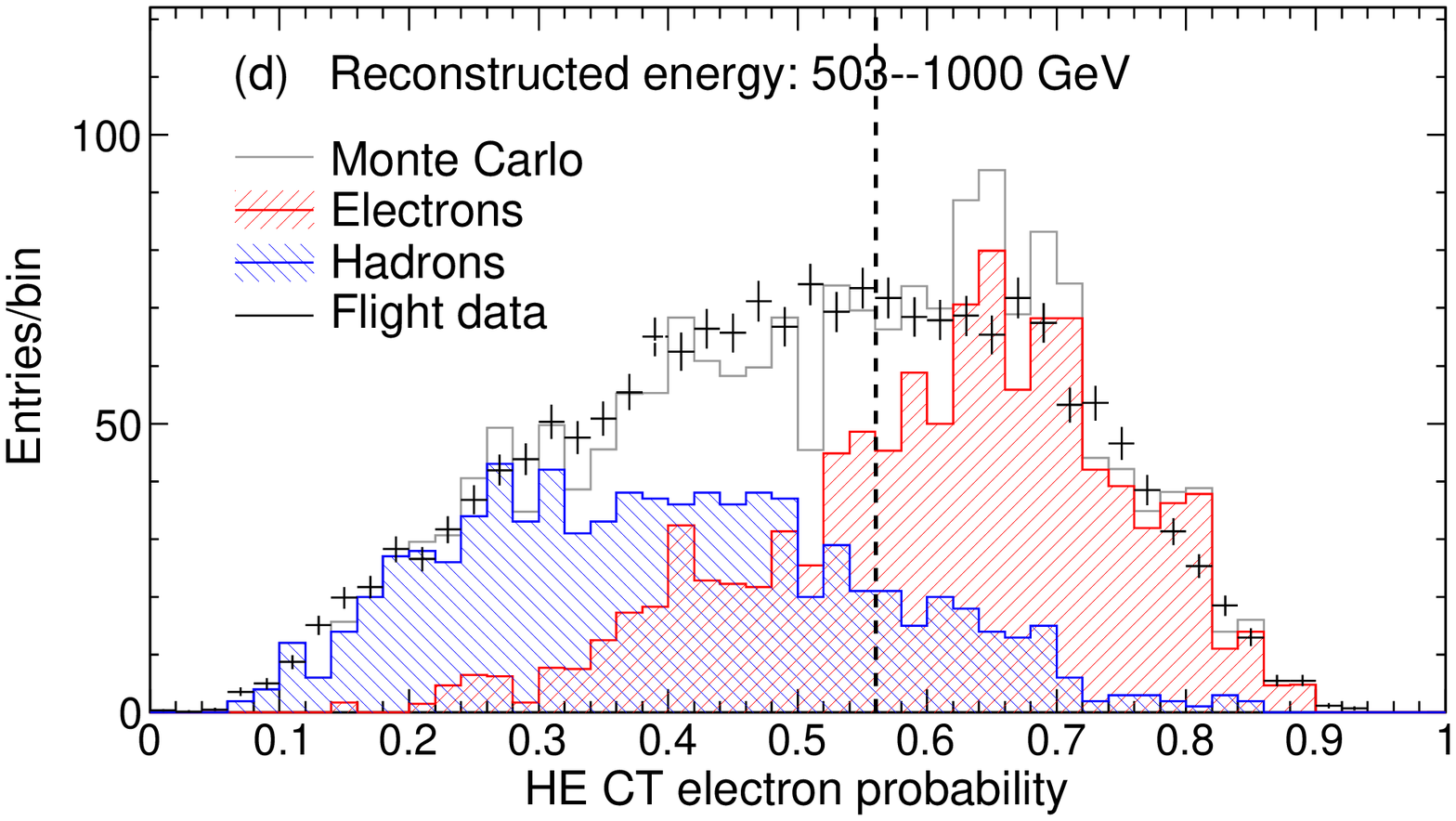}%
\caption{(color online). Distribution of CT-predicted probability (a) for \lowe\ analysis  
and (b), (c), and (d) for \highe\ analysis in different energy intervals. 
Monte Carlo generated distributions are compared with flight distributions. 
The cut value is a continuous function of energy and is represented
by the vertical dashed line in each panel.
The distributions are shown after the cuts on all other 
variables have been applied.
}
\label{fig:CTBProb_energy}
\end{figure*}

We identified the quantities (variables) derived from the event 
reconstructions that are most sensitive 
to the differences between electromagnetic and hadronic event topologies.
For example, the multiplicity of tracks and the extra hits outside of 
reconstructed tracks is useful for rejecting interacting hadrons.
Variables mapping the shower 
development in the calorimeter are also important.  
The CTs are trained using simulated events 
and, for each event, predict the probability that the event is an electron.
The cut that we have adopted on the resulting CT-predicted electron 
probability is energy dependent.
For \highe\ analysis, a higher probability is required as energy increases.
These cuts give us a set of candidate electron events 
with a residual contamination of hadrons that cannot be removed on an 
event-by-event basis. The remaining contamination must be estimated 
using the simulations and will be discussed 
in Sec.~\ref{sec:contamination}. 

Though the simulations are the starting point for 
the event selection, we systematically compare them with the 
flight data as illustrated in 
Figs.~\ref{fig:CalTransRms_selection}--\ref{fig:CTBProb_energy}.
The input energy spectra for all the particles are those included 
in the model of energetic particles in the Fermi orbit 
(Sec.~\ref{sec:background}), with the exception of the electrons. 
For the electrons we use instead a power-law spectrum 
that fits our previous publication~\cite{PRL}. 
For any single variable we use the signal and proton background 
distributions at the very end of the selection chain 
(after the cuts on all the other variables have been applied) to quantify 
the additional rejection power provided by that particular variable.
Any variables for which the data-MC agreement was not 
satisfactory were not used in any part of the selection. 

The procedure used to characterize the discrepancies between data and
Monte Carlo and quantify the associated systematic uncertainties 
will be described in Sec.~\ref{sec:systematic}. 
We stress, however, that there is a good qualitative agreement 
(both in terms of the shapes of the distributions and in terms of 
the relative weights of the electron and hadron populations) 
in all the energy bins and at all the stages of the selection. 
This is a good indication of the self-consistency of the analysis and 
that both the CR flux model and detector simulation 
adequately reproduce the data.

\subsection{Energy reconstruction}\label{sec:energy}

As mentioned in the previous section, the electron energy reconstruction 
is performed using the algorithms developed for the photon analysis~\cite{LAT}.
These algorithms are based on comprehensive simulations and
validated with the beam test data~\cite{BTGlastSymposium}.

The total depth of the LAT, including both the tracker and the calorimeter,
is 10.1~X$_0$ on axis. The average amount of material
traversed by the candidate electrons, integrated over the instrument field of
view, is 12.5~X$_0$.
However, for electromagnetic cascades $\gtrsim 100$~GeV a significant 
fraction of the energy is not contained in the calorimeter.
Here, the calorimeter shower imaging capability is crucial in order
to correct for the energy leakage from the sides and the back of the
calorimeter and through the gaps between calorimeter modules.

The event reconstruction is an iterative process~\cite{LAT}.
The best track provides the reference axis for the analysis of the
shower in the calorimeter. The energy reconstruction is completed only after
the particle tracks are identified and fitted. 
Following this procedure, each single event
is fed into three different energy reconstruction algorithms:
\begin{itemize}
\item[a)] a parametric correction method based on the energy centroid depth 
along the shower axis in the calorimeter in combination with the total 
energy absorbed --- valid over the entire energy range for the LAT;
\item[b)] a maximum likelihood fit, based on the correlation between 
the total deposited energy, the energy deposited in the last layer of the 
calorimeter
and the number of tracker hits --- valid up to 300~GeV; and
\item[c)] a three-dimensional fit to the shower profile, taking into
account the longitudinal and transverse development --- valid above 1~GeV.
\end{itemize}
For each event the best energy reconstruction method is
then selected by means of a CT analysis similar to that
described in Sec.~\ref{sec:selections}. The classifier is trained on a
Monte Carlo data sample and exploits all the available topological information
to infer which energy estimate is closest to the true
energy for the particular event being processed.
The final stage of the energy analysis, again based on a set of CTs, 
provides an estimate of the quality of the energy reconstruction,
which we explicitly use in the analysis to reject events 
with poorly measured energy.

\begin{figure*}[hbtp]

\twopanel%
{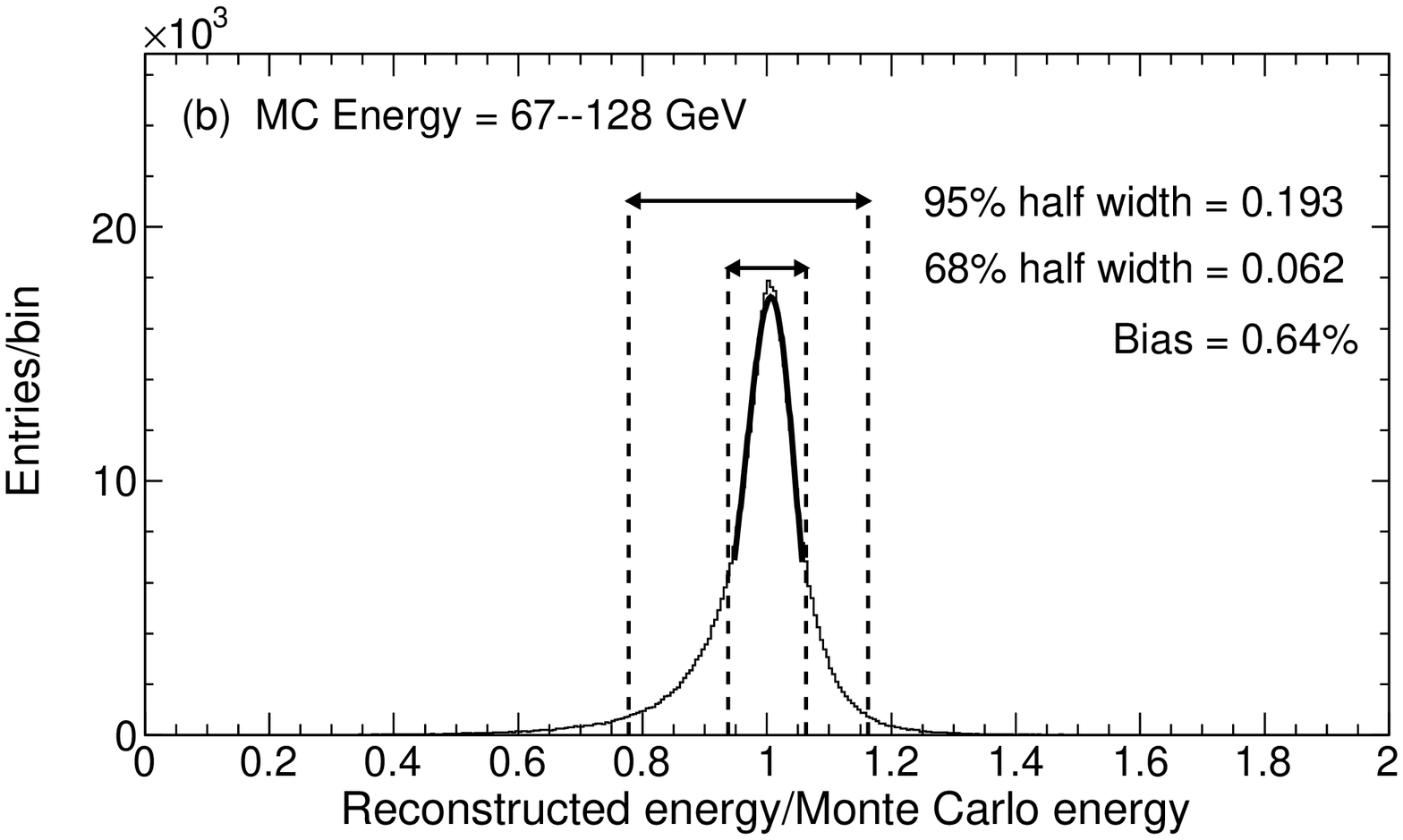}%
{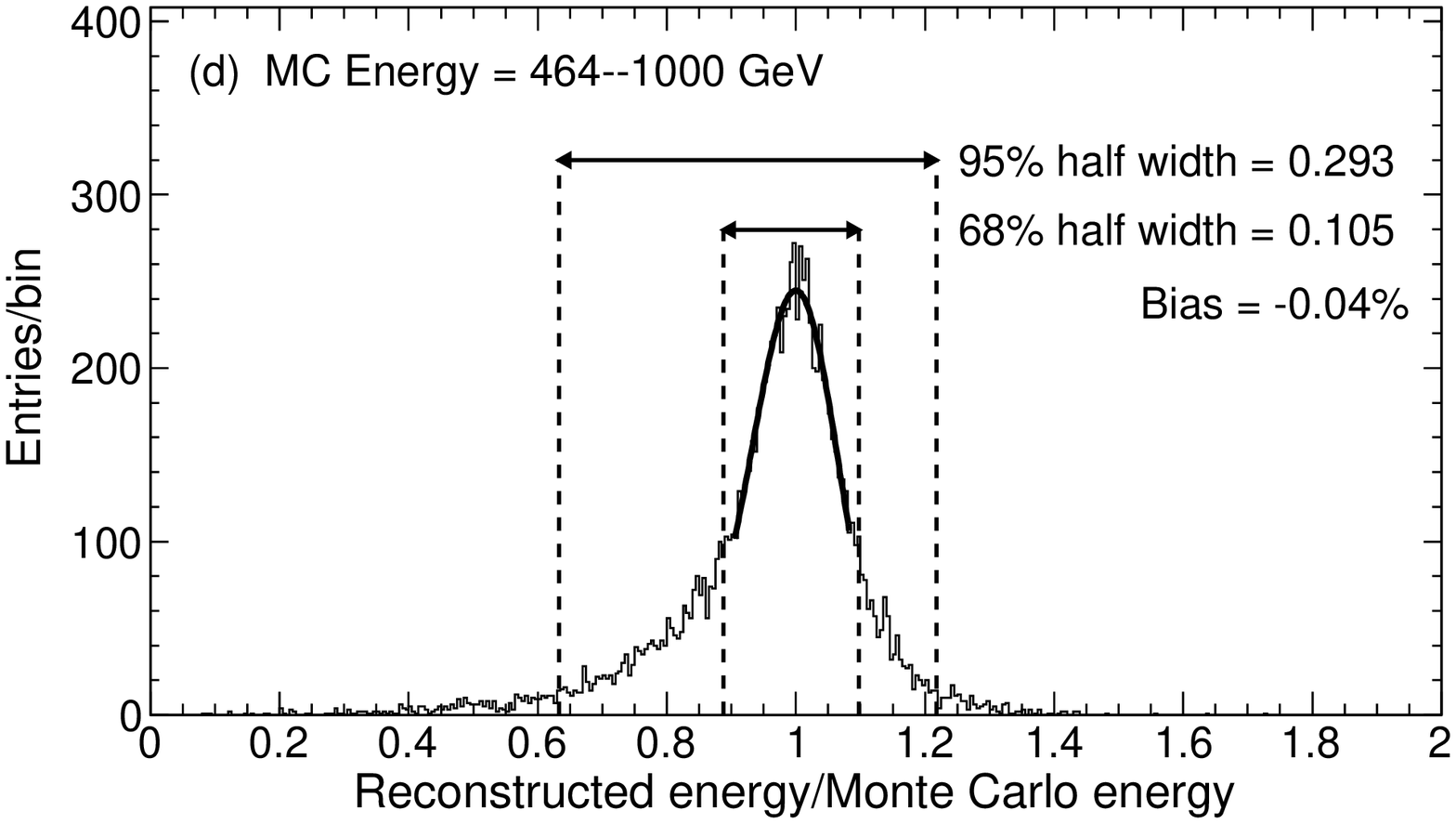}%

\caption{\label{fig:EnergyResCalc}
Energy dispersion distributions (after the \highe\ selection cuts have been
applied) in two sample bins.
The 68\% and the 95\% containment windows, defining the energy resolution, are
represented by the horizontal double arrows. The fractional bias is defined as
the deviation from unity of the position of the most probable value of the
log-normal function used for fitting.}
\end{figure*}

\begin{figure*}[htbp]
\twopanelsmaller%
{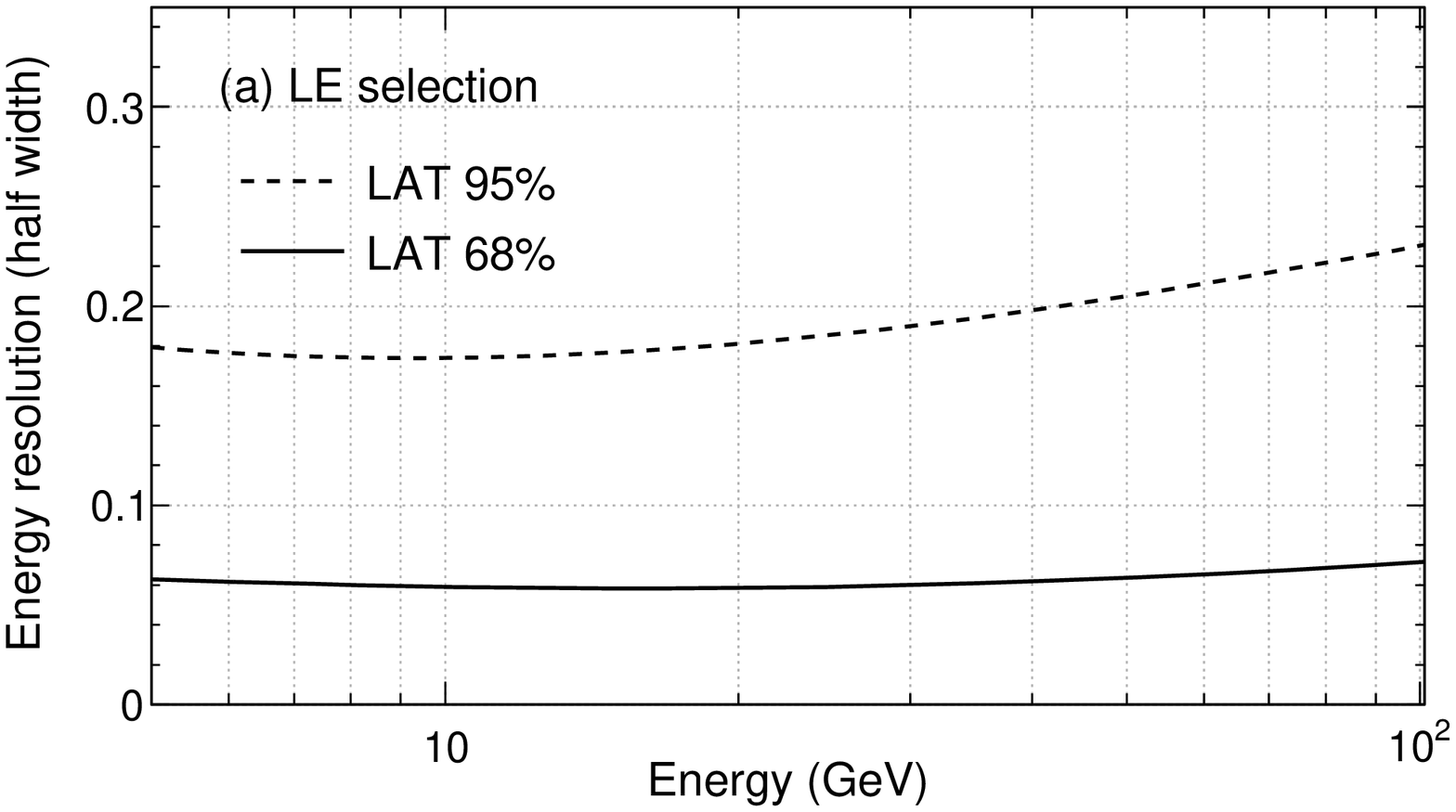}%
{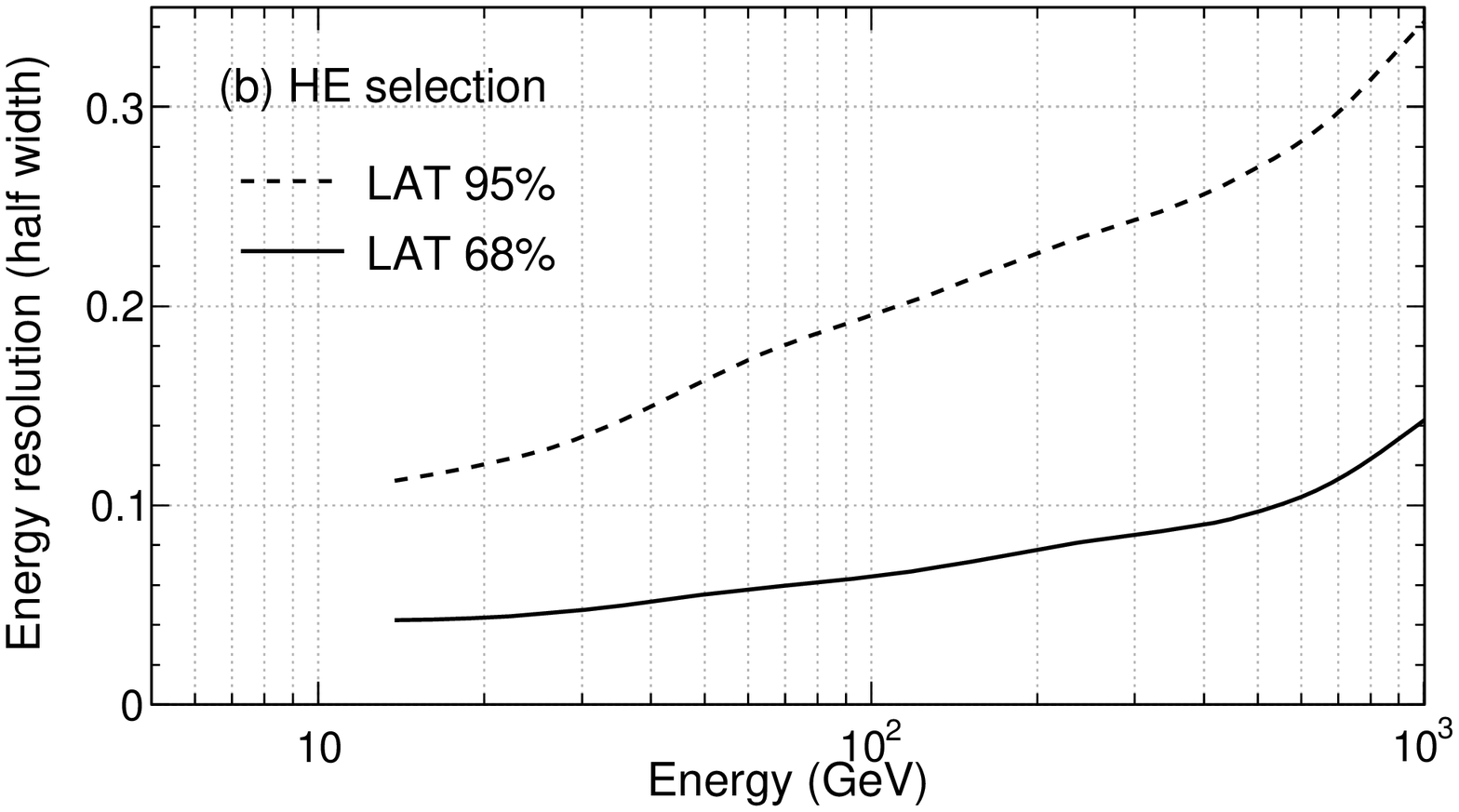}%
\caption{\label{fig:EResolution}
Energy resolution (half-width of the $68\%$ and $95\%$ energy dispersion 
containment windows) for the \lowe\ (left panel) 
and the \highe\ (right panel) analysis.
}
\end{figure*}

At high energies (and especially above 300~GeV, where the
likelihood fit is no longer available), the three-dimensional fit to the shower
profile is the reconstruction method chosen for the vast majority of events. 
This method takes into account the saturation of the calorimeter readout 
electronics that occurs at $\sim 70$~GeV for an individual crystal.
However above $\sim 1$~TeV the number of saturated crystals 
increases quickly, requiring a more complex correction.
This is beyond the scope of the current
paper and will be addressed in subsequent publications. 
We therefore limit ourselves to events with energies $< 1$~TeV.

The performance of the energy reconstruction algorithm has been characterized
across the whole energy range of interest using Monte Carlo simulations
of an isotropic 1/E electron flux.
We divided the energy range into 6 partially overlapping bins per
decade and quantified the bias and the resolution in each bin, based on
the resulting energy dispersion distributions (defined as
the ratio between the reconstructed energy and the true energy, as shown in
Fig.~\ref{fig:EnergyResCalc}).

The energy dispersion distribution in each energy window is fitted with a 
log-normal function 
and the bias is calculated as the deviation of the most probable
value of the fit function from 1. 
This bias is smaller than $1\%$ over the entire phase space explored.
We characterize the energy resolution by quoting the half-width of the
smallest window containing $68\%$ and $95\%$ of the events in the energy
dispersion distributions. Those windows are graphically indicated in
Fig.~\ref{fig:EnergyResCalc} and correspond to 1 and 2 sigma, respectively,
in the ideal case of a Gaussian response.
The energy resolution corresponding to a $68\%$ half-width
containment is about $6\%$ at 7~GeV and increases as the energy increases,
reaching $15\%$ at 1~TeV as shown in Fig.~\ref{fig:EResolution}.
The $95\%$ containment is useful to quantify the 
tails of the distribution and is within a factor 3 of the $68\%$ containment.
The deviation with respect to a Gaussian distribution is mainly due
to a higher probability to underestimate the energy than to overestimate it
and is reflected in the low-energy tails in Fig.~\ref{fig:EnergyResCalc}.
We verified, with our simulations, that the energy response 
does not generate any discontinuity 
that could create spurious features in the spectrum.

\section{Spectral analysis}\label{sec:spec_analysis}%
\subsection{Instrument acceptance \label{sec:geomfactor}}

The instrument acceptance for electrons, 
or effective  geometric factor (EGF),  is defined as a product 
of the instrument field of view and its effective area.
To calculate the EGF we use a Monte Carlo simulation
of an isotropic electron spectrum 
with a power-law index of $\Gamma = 1$ (the same simulation
described in Sec.~\ref{sec:energy}).
In this case,
\begin{equation}                                           
{\rm EGF}_i = A \times \frac{N^{\rm pass}_i}{N^{\rm gen}_i}
\end{equation}
where $N^{\rm gen}_i$ and $N^{\rm pass}_i$ are, respectively, 
the number of generated 
events and the number of events surviving the selection cuts
in the $i^{th}$ energy bin.
The normalization constant $A$ depends on the area and the solid angle
over which the events have been generated.

\begin{figure}[htbp]
  \begin{center}
      \includegraphics[width=\onecolwidth]{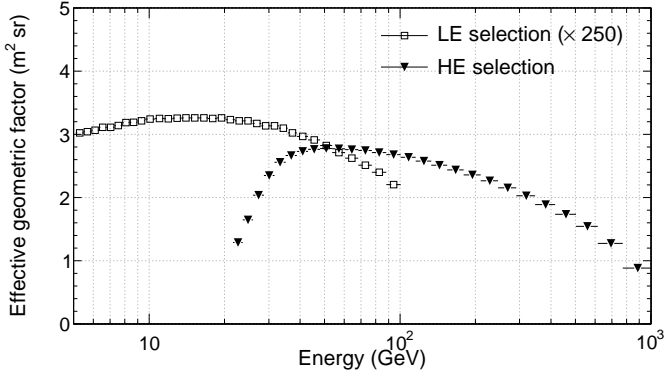}
    \caption{\label{fig:GeomFactor}Effective geometric factor 
    for \lowe\ (squares, multiplied by a factor 250) and \highe\ events (triangles)}
  \end{center}
\end{figure}

The EGF for \lowe\ and \highe\ 
events is shown in Fig.~\ref{fig:GeomFactor}.
The \highe\ EGF has a peak value of $\sim2.8~\rm{m}^2 \rm{sr}$  
at an energy $\rm{E}\sim 50$~GeV. 
The falloff below 50~GeV is due to the on-board 
filtering (see Sec.~\ref{sec:selections}), while the decrease 
for energies above 50~GeV is due to the energy dependence of the 
event selection.
The \lowe\ EGF in Fig.~\ref{fig:GeomFactor} has been multiplied by a factor
of 250 for graphical clarity.
For energies below 30~GeV its value is almost constant while for higher
energies it decreases rapidly. This effect is due to the fact that the \lowe\
event selection is optimized for relatively low energies. 
The statistical error on the EGF is less than $\sim 1\%$ for 
each energy bin of the reconstructed spectrum for both \lowe\ and \highe.

\subsection{Correction for residual contamination} \label{sec:contamination}

\begin{figure}[!bhtp]
  \begin{center}
    \includegraphics[width=\onecolwidth]{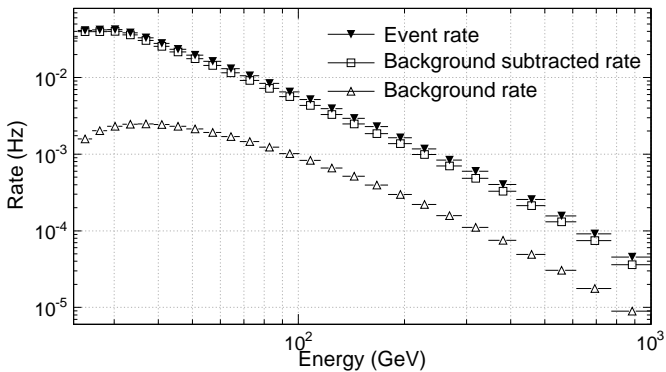}
    \caption{\label{fig:contamination} 
    Flight rate of electron candidates after \highe\ selection
    (inverted triangles), corresponding simulated rate of hadron
    events (open triangles) and resulting rate of electrons (open squares)
    after background subtraction. }
  \end{center} 
\end{figure}

We estimate the contamination in each energy bin 
by applying the selection cuts to the on-orbit simulation
to determine the rate of remaining background events 
(protons and heavy nuclei). 
To correct for the contamination, this rate is subtracted 
from that of the flight electron candidates (shown 
in Fig.~\ref{fig:contamination} for the \highe\ analysis).
With this procedure the contribution to the systematic uncertainty 
due to the residual contamination depends on
the energy spectra for hadrons
in our simulation and not on the one for electrons 
(see Sec.~\ref{sec:systematic}).
The contamination (defined as the ratio between simulated residual 
hadron rate and total event rate) ranges 
from $\sim 4$\% at 20~GeV to $\sim 20$\% at 1~TeV 
for the \highe\ selection while for \lowe\ is $\sim 10$\% at 7~GeV increasing 
with energy up to $\sim 18$\% at 80~GeV.
Note that the \lowe\ analysis (which is independent of the \highe\ analysis) 
deals with large variation in the event topology, especially at its low-energy 
end; also it was optimized for the efficiency for electrons to compensate 
for lower input statistics. It is reflected in slightly higher 
(but still under 20\%) residual hadron contamination.

The number of simulated events generated 
was chosen to keep statistical fluctuation on the background rate
small compared to the systematic uncertainties. 

As a cross-check, we also carried out Monte Carlo simulations 
using only protons with spectral index $\Gamma = 1$, 
thereby enriching the sample statistics with high-energy events.  
After applying the \highe\ selection cuts, we determine the rate of residual 
proton events corresponding to a spectral index of 1, and reweight it to 
derive the residual proton event rate corresponding to the real CR proton 
spectral index of 2.76.
We add 5\% to this rate in order to take into account 
the contribution of heavy nuclei, mainly helium, which was not simulated.
The resulting rate agrees within statistical errors with that obtained using 
the on-orbit flux model.

As mentioned in Sec.~\ref{sec:selections},
the ACD is very effective in removing gamma-ray initiated events.
To check the gamma-ray contamination in our electron candidate sample,
we use the all-sky average gamma-ray flux measured by the LAT
and extrapolate it over the energy range of interest.
We then convolve it with the effective geometric factor for gamma rays 
after electron selection cuts
to obtain the rate of remaining gamma-ray events.
The ratio of this rate to the measured event rate
provides an estimate of the gamma contamination,
which remains below 0.1\% over the whole energy range.

\subsection{Spectral reconstruction}\label{sec:spectrum}
Once we have the rate of electrons, the spectrum is found
by dividing the event rate by the EGF (described in Sec.~\ref{sec:geomfactor})
and the width of the energy interval.
The energy dispersion (which causes events to migrate to adjacent bins)
is taken into account by unfolding the 
background-subtracted rate with a technique based on Bayes' 
theorem~\cite{unfolding}. The event migration is calculated using a 
matrix based on the energy dispersion obtained from 
simulations. We found that this correction
is less than 5\% in all the energy intervals.

The reconstruction procedure is similar for both the
\lowe\ and the \highe\ analyses. The former is more complicated
due to the presence of the Earth's magnetic field.
In fact, for energies below $\sim 20$~GeV we need to consider 
the shielding effect of the geomagnetic field as characterized by the 
cutoff rigidity. 
The lowest allowed primary-electron energy is strongly dependent on
geomagnetic position and decreases with increasing 
geomagnetic latitude. 
For the orbit of Fermi, the cutoff ranges between about 6 and 15~GeV.
\begin{figure}[!htb]
\includegraphics[width=\onecolwidth]{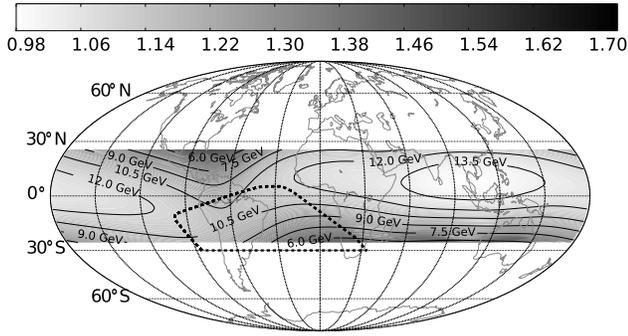}
\caption{Map of McIlwain L values for the Fermi orbit. Overlaid in contours are
the corresponding values for vertical cutoff rigidity. These values were
calculated using the $10^{\rm th}$ generation IGRF model~\cite{IGRF}, which is 
valid outside of the South Atlantic Anomaly (represented by the dashed
black line in the figure).}
\label{fig:McIlwainLVsRigidity}
\end{figure}

As recognized in~\cite{SmartShea},
the McIlwain L~\footnote{The McIlwain L parameter 
is a geomagnetic coordinate defined as the distance in Earth radii 
from the center of the Earth's titled, off-center, equivalent 
dipole to the equatorial crossing of a field line.}
parameter is particularly convenient for characterizing cutoff rigidities and
has been used for selecting data in the \lowe\ analysis. 
Figure~\ref{fig:McIlwainLVsRigidity} illustrates the distribution of the
McIlwain L parameter for the Fermi orbit. 
We want to stress here that the contours shown in 
Fig.~\ref{fig:McIlwainLVsRigidity} are the vertical cutoff rigidities 
based on the International Geomagnetic Reference Field (IGRF) 
model~\cite{IGRF} and are intended 
for illustrative purposes only. 
Our analysis does not depend in any way on the vertical
cutoff values from this model.

\begin{figure}[!htb]
\includegraphics[width=\onecolwidth]{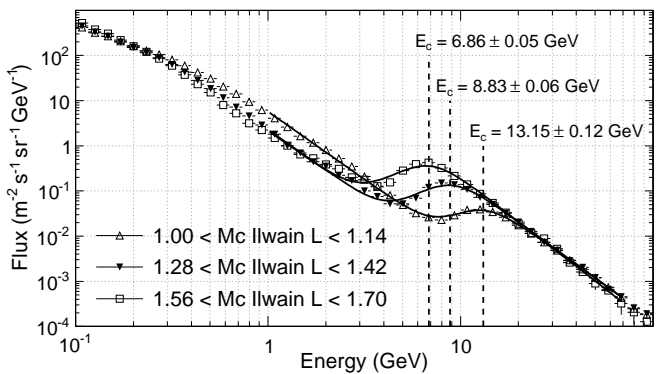}
\caption{The measured electron flux in three McIlwain L bins. 
For each bin the fit of the flux with equation~\ref{eq:spectrum_function} 
and the resulting estimated cutoff rigidity, $E_c$, is shown. As 
described in the text, $E_c$ decreases for larger values of McIlwain L.}
\label{fig:FluxFitted}
\end{figure}

Each McIlwain L interval has an associated
cutoff; we determine $E_c$ by parameterizing the shape of the CRE spectrum as
\begin{equation}\label{eq:spectrum_function}
\frac{{\rm d}N}{{\rm d}E} = c_{\rm s}E^{-\Gamma_{\rm s}} +
\frac{c_{\rm p}E^{-\Gamma_{\rm p}}}{1 + (E/E_c)^{-6}}
\end{equation}
where $c_{\rm s}$ and $c_{\rm p}$ are the normalization constants for the 
secondary (albedo) and primary components of the spectrum while 
$\Gamma_{\rm s}$ and $\Gamma_{\rm p}$ are their spectral indexes. 
Figure~\ref{fig:FluxFitted} illustrates how we determine $E_c$ using 
Eq.~\ref{eq:spectrum_function} for three McIlwain L intervals.

As can be seen in Fig.~\ref{fig:FluxFitted}, 
the transition to cutoff is smoothed out due to the complexity 
of the particle orbits in the Earth's magnetosphere. 
Therefore, we increase $E_c$ by $15\%$ to arrive at an effective 
minimum energy of the primary electron flux not affected 
by the Earth's magnetic field. 
To verify that this increase is sufficient, we have performed a series of 
tests to quantify the changes in the flux level as a function of this parameter
and found that the final spectrum does not vary significantly for values 
greater than 15\%.
We split the \lowe\ data sample into 10 intervals of McIlwain L parameter.
For each energy bin we use the interval of McIlwain L parameter 
whose effective minimum energy is lower than the energy in question.
This procedure is illustrated in
Fig.~\ref{fig:FluxOverlap}, where the electron spectrum is shown together 
with the McIlwain L intervals from which the flux was measured.

The electron flux below the geomagnetic cutoff is due to secondary electrons 
produced in the Earth atmosphere including reentrance albedo. 
Discussion of the spectrum below the cutoff is beyond the scope of this paper.

\subsection{Assessment of systematic uncertainties}\label{sec:systematic}
\begin{figure*}[htbp]
\twopanel%
{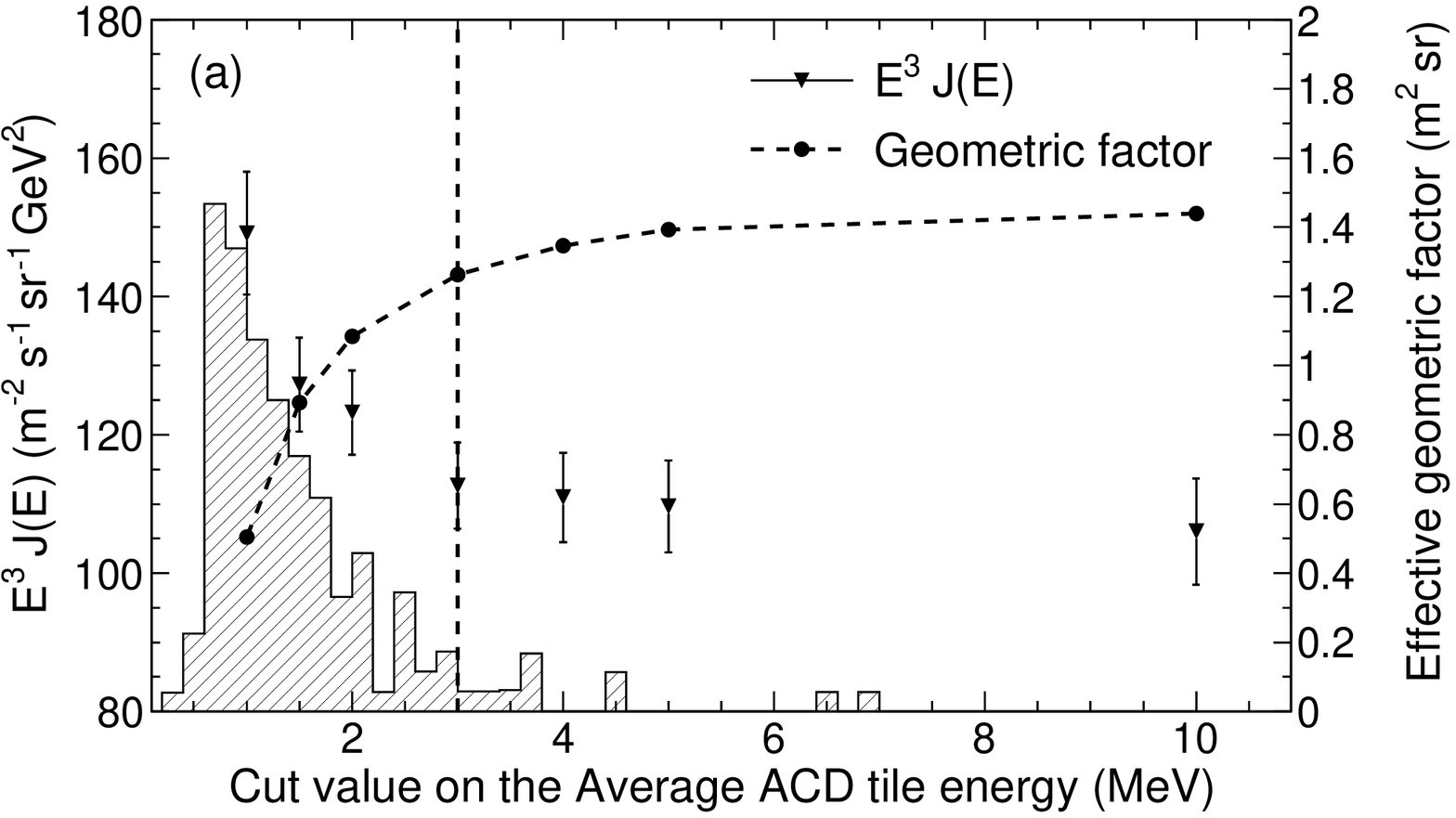}%
{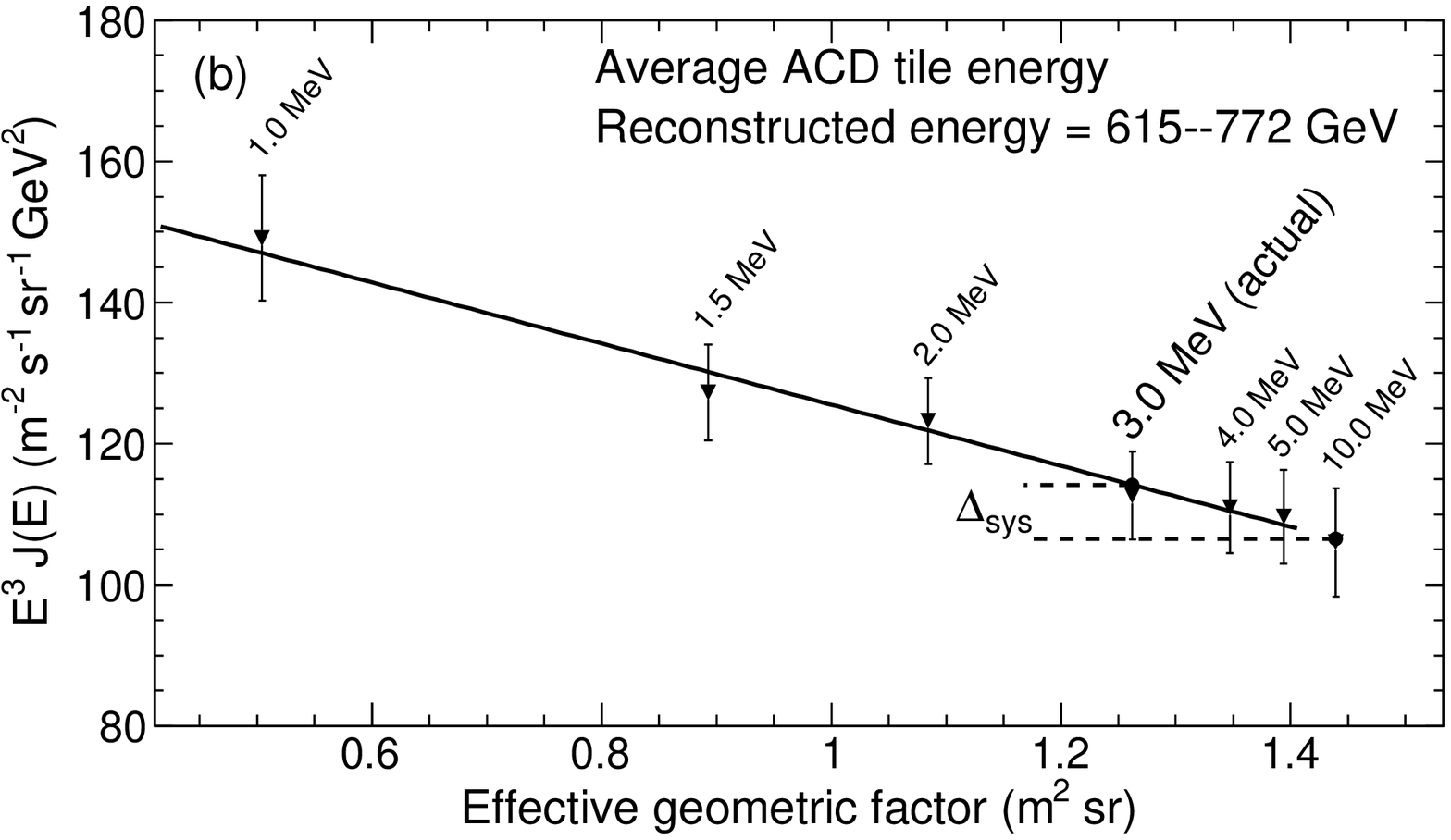}%
\caption{\label{fig:flux_scan}
Effect of cut on the average energy release per ACD tile for 
the energy interval 615 to 772 GeV.
Panel (a) shows how the measured flux (triangles, scale on left axis) 
and the effective geometric factor (circles, scale on right axis) 
depend on the cut value. The vertical dashed line indicates the value used. 
For reference, the Monte Carlo distribution of the average ACD tile energy 
is shown. Panel (b) shows the measured flux vs 
the effective geometric factor for the cut values indicated.
}
\end{figure*}

The imperfect knowledge of the EGF constitutes one of
the main sources of systematic uncertainty. This is a direct consequence of the
fact that the simulations we use for the evaluation of the EGF cannot 
perfectly reproduce the topological variables used in the electron
selection. Differences between data and simulation may affect the flux
measurement also through the subtraction of the hadronic background, but
this contribution is relatively easier to keep under control because the
contamination itself is always under 20\%.
In order to characterize the agreement between simulations and data, and assess
the effect of the residual discrepancies, we systematically studied the
variations of the measured flux induced by changes in the selection cuts
around the optimal values. If the agreement were perfect the flux would not
depend on the cut values. However, this is in general not true and 
such changes translate into systematically higher or lower flux values.

Consider a variable for which we wish to know the effect of changing 
cut values. We first apply all other cuts, and then vary the cut value 
on this variable and study the effects.
The procedure we used is illustrated in Fig.~\ref{fig:flux_scan} for one
variable in one energy bin. Panel~\ref{fig:flux_scan}(a) 
shows how the geometric factor and the
measured flux depend on the cut value. In this particular
case a harsher cut translates into a systematically higher flux. This can be
qualitatively understood by looking at the comparison between data and 
Monte Carlo simulation for the distribution of the average energy per ACD
tile shown in Fig.~\ref{fig:AcdEnergy}(b).
The distribution of this quantity in our simulation is slightly shifted
toward higher energies with respect to the flight data and
therefore, for any given cut, we effectively tend to underestimate the EGF
(i.e., overestimate the flux).
It is important to note that this variable is directly related to the 
topology of the backsplash in the ACD, which is extremely hard to simulate, 
especially at very high energies.

We found that the scatter plot of the measured flux vs the geometric factor,
as shown in Fig.~\ref{fig:flux_scan}(b), can be fitted reasonably well with
a straight line in all the cases we encountered (the slope returned by the fit
being directly related to the agreement between Monte Carlo and flight data).
The fit function is used to determine the difference between the flux
at the selected cut value and that measured when the cut is loose enough that
the variable under study no longer contributes to the selection.
We take this difference [$\Delta_{\rm{sys}}$ in 
Fig.~\ref{fig:flux_scan}(b)] 
as the estimate of the systematic effect introduced by
the variable itself. For the case illustrated in Fig.~\ref{fig:flux_scan} it
is 7\% and represents the largest single contribution, among all the selection
variables, to the total systematic error in this energy bin.

The method described here is sensitive to differential discrepancies around
the cut values for both signal and background and allows us to map them to the
actual measured spectrum. It has been performed separately for each energy
bin and each selection variable (setting the cuts for all the other variables
to the optimal values). Positive (negative) contributions, corresponding to
variables for which the slope of the fit is negative (positive) are summed up
in quadrature separately to provide an asymmetric bracketing of systematic
uncertainty.

The error on the absolute normalization of the background flux (predominantly
protons) constitutes an additional source of systematic uncertainty.
We conservatively assumed a constant value of 20\%, which is properly weighted
with the residual contamination (Sec.~\ref{sec:contamination}).

The uncertainty in the absolute energy scale of the detector is also a
significant contribution to the systematic error on the measurement.
Assuming that this uncertainty $\Delta s/s$ is energy-independent (as the
results of our beam test indicate) it translates into a rigid shift of the
overall spectrum. 
For a given spectral index $\Gamma$ the vertical component of this shift is
given by $(\Gamma - 1)\Delta s/s$ (i.e., is 20\% for an uncertainty of 10\% on
the absolute energy scale and a spectral index $\Gamma=3$).

The simulated data sample used for the evaluation of the geometric factor
and the residual contamination is large enough that any effect due to
statistical fluctuations is negligible in both the \lowe\ and the \highe\
analysis. This is not true for the analysis with sampled statistics presented
in Sec.~\ref{sec:longpath}.

\subsection{Cross-check using events with long path in the instrument}\label{sec:longpath}
 
In order to cross-check the impact of the energy resolution on the measured
spectrum, we performed a dedicated analysis in which we selected events
with the longest path lengths (at least~$12~{\rm X}_0$) in the calorimeter.
We further select events that do
not cross any of the boundary gaps between calorimeter tower modules and that
have sufficient track length (at least~$1~{\rm X}_0$) in the tracker 
for a good direction reconstruction.
For the event sample defined by these three requirements the average amount
of material traversed is $\sim 16~{\rm X}_0$
\begin{figure}[!bhtp]
\includegraphics[width=\onecolwidth]{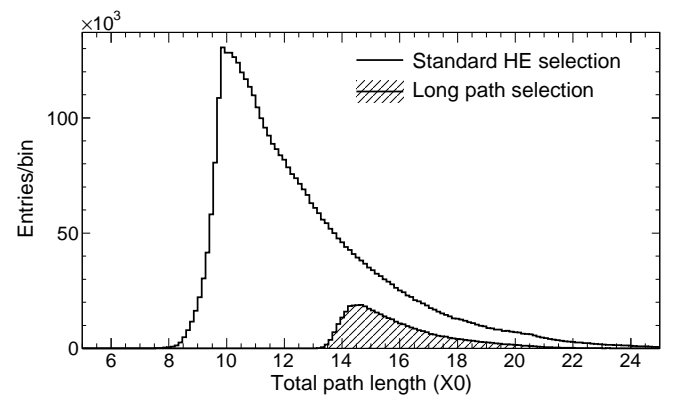}
\caption{\label{fig:PathLong}Distribution of the amount of material traversed
by the candidate electrons passing the long path selection, compared with that
for the entire data sample used in the standard analysis (the sharp edge
at $\sim10~{\rm X}_0$ in the latter reflects the total thickness of
the instrument on-axis). Note the difference in the number of events.}
\end{figure}
\begin{figure}[!thbp]
\includegraphics[width=\onecolwidth]{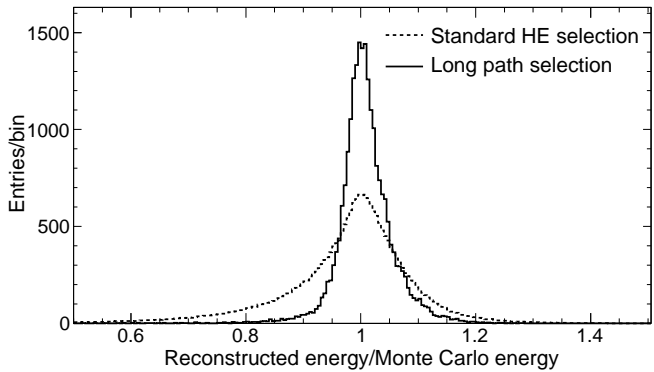}
\caption{\label{fig:EnergyDispLong}Energy dispersion distribution in the energy
range 242--458~GeV for the long-path selection (solid line) 
and the standard \highe\ analysis (dashed line).}
\end{figure}
(see Fig.~\ref{fig:PathLong}), ensuring that the shower maximum is well
contained in the calorimeter up to at least 1~TeV (the average depth of the 
shower maximum for electrons at this energy is $10.9~{\rm X}_0$). 
Correspondingly the instrument acceptance decreases to $\sim 5\%$ of that 
achieved in the standard analysis described in the previous sections.

\begin{figure}[!hbtp]
\includegraphics[width=\onecolwidth]{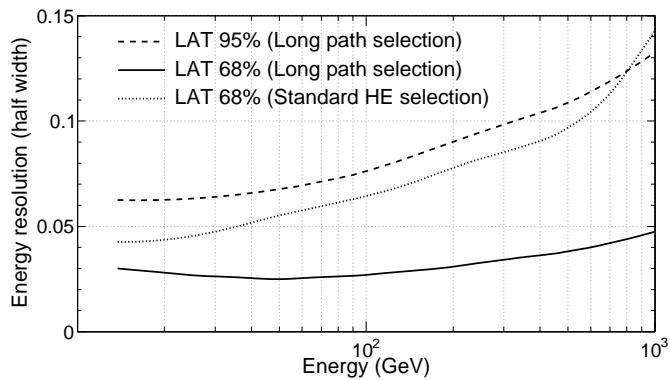}
\caption{\label{fig:EnResComp}Energy resolution for the long-path selection 
analysis. The half-width of the 68\% containment window for the \highe\
analysis, which is comparable with that of the 95\% window for the more
restrictive analysis, is overlaid for reference.}
\end{figure}

As illustrated in Figs.~\ref{fig:EnergyDispLong} and~\ref{fig:EnResComp},
the energy resolution for events passing this restrictive selection is
significantly better than that presented in Sec.~\ref{sec:energy}
for the full analysis.
The energy dispersion distributions are much narrower and symmetric, with no
prominent low-energy tails. The energy resolution (half-width of the 68\%
containment window) is around 3\% at 100~GeV and increases to approximately
5\% at~1 TeV.

\begin{figure}[htbp]
\includegraphics[width=\onecolwidth]{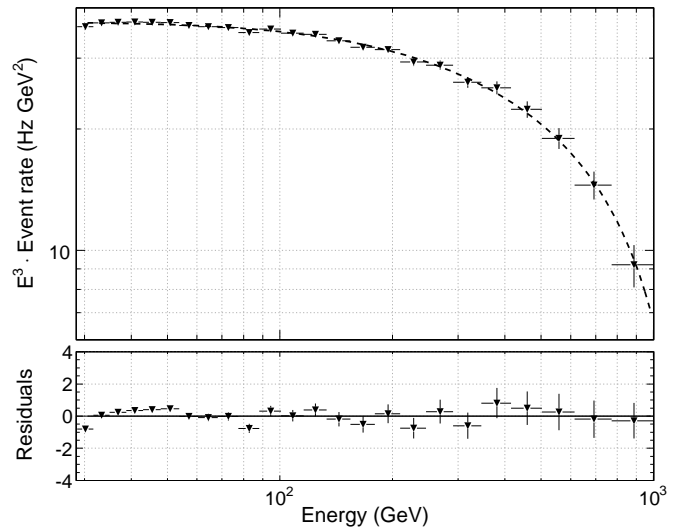}
\caption{\label{fig:CountRate}Count rate multiplied by $E^3$ for long-path
selection. The bottom panel shows residuals from the smooth function fit
(dashed line in the top panel).}
\end{figure}

Figure~\ref{fig:CountRate} shows the event rate (multiplied by $E^3$) for the
long path length selection. There is no evidence of any significant spectral
feature. The dashed line is a fit with a smooth function; the residuals of the
fit are plotted in the bottom panel.

\dblfloatsatbottom
\squeezetable
\begin{table*}[bhtp!] 
\centering
 \caption{Number of events, residual hadronic contamination, 
  flux $J_{\rm E}$ and minimum McIlwain L value for \lowe\ analysis.
  Statistical error is followed by systematic error 
  (see Sec.~\ref{sec:systematic}). 
  Residual contamination is defined as the ratio between 
  hadronic background rate and measured event rate.
  } 	 
\label{table:LESelection}
\begin{tabular}{l r c r c}
\hline\hline
Energy~(GeV) & Counts & Residual contamination & 
$J_{\rm E}$ (GeV$^{-1}$ s$^{-1}$ m$^{-2}$ sr$^{-1}$) & McIlwain~L~$>$\\
\hline 	 
6.8--7.3 & 109 & 0.11 & $(54.6 \pm 7.5^{+7.9}_{-3.9})\cdot 10^{-2}$ & 1.72\\
7.3--7.8 & 532 & 0.07 & $(44.3 \pm 2.5^{+6.3}_{-3.0})\cdot 10^{-2}$ & 1.67\\
7.8--8.4 & 1425 & 0.09 & $(34.1 \pm 1.3^{+4.6}_{-2.3})\cdot 10^{-2}$ & 1.6\\
8.4--9.0 & 2777 & 0.11 & $(264 \pm 8.0^{+35}_{-18})\cdot 10^{-3}$ & 1.56\\
9.0--9.7 & 3885 & 0.08 & $(226 \pm 5.6^{+29}_{-15})\cdot 10^{-3}$ & 1.51\\
\phantom{0}9.7--10.6 & 5648 & 0.09 & $(171 \pm 3.7^{+22}_{-11})\cdot 10^{-3}$ & 1.46\\
10.6--11.5 & 5300 & 0.10 & $(131 \pm 3.0^{+16}_{-8})\cdot 10^{-3}$ & 1.42\\
11.5--12.4 & 4409 & 0.08 & $(101 \pm 2.3^{+12}_{-6})\cdot 10^{-3}$ & 1.42\\
12.4--13.5 & 6742 & 0.08 & $(75.8 \pm 1.5^{+8.6}_{-4.4})\cdot 10^{-3}$ & 1.28\\
13.5--14.6 & 5880 & 0.07 & $(62.3 \pm 1.3^{+6.8}_{-3.4})\cdot 10^{-3}$ & 1.28\\
14.6--15.8 & 9857 & 0.08 & $(457 \pm 8.3^{+48}_{-25})\cdot 10^{-4}$ & 1.14\\
15.8--17.2 & 8527 & 0.09 & $(363 \pm 7.0^{+37}_{-20})\cdot 10^{-4}$ & 1.14\\
17.2--18.6 & 7189 & 0.07 & $(281 \pm 5.5^{+27}_{-14})\cdot 10^{-4}$ & 1.14\\
18.6--20.2 & 6102 & 0.10 & $(217 \pm 4.7^{+21}_{-11})\cdot 10^{-4}$ & 1.14\\
20.2--21.9 & 9361 & 0.10 & $(168 \pm 3.2^{+15}_{-8})\cdot 10^{-4}$ & 1.0\\
21.9--23.8 & 7883 & 0.10 & $(132 \pm 2.7^{+11}_{-6})\cdot 10^{-4}$ & 1.0\\
23.8--25.8 & 6639 & 0.10 & $(105.2 \pm 2.2^{+8.6}_{-4.8})\cdot 10^{-4}$ & 1.0\\
25.8--28.0 & 5674 & 0.12 & $(80.4 \pm 1.9^{+6.4}_{-4.0})\cdot 10^{-4}$ & 1.0\\
28.0--30.4 & 4781 & 0.10 & $(63.3 \pm 1.5^{+4.7}_{-2.8})\cdot 10^{-4}$ & 1.0\\
30.4--32.9 & 4234 & 0.11 & $(52.5 \pm 1.3^{+3.7}_{-2.3})\cdot 10^{-4}$ & 1.0\\
32.9--35.7 & 3411 & 0.13 & $(38.7 \pm 1.1^{+2.6}_{-1.8})\cdot 10^{-4}$ & 1.0\\
35.7--38.8 & 2899 & 0.13 & $(297 \pm 9.3^{+19}_{-13})\cdot 10^{-5}$ & 1.0\\
38.8--43.1 & 2948 & 0.14 & $(222 \pm 6.9^{+14}_{-9})\cdot 10^{-5}$ & 1.0\\
43.1--48.0 & 2325 & 0.16 & $(153.7 \pm 5.6^{+9.3}_{-7.4})\cdot 10^{-5}$ & 1.0\\
48.0--53.7 & 1955 & 0.17 & $(113.9 \pm 4.5^{+6.5}_{-6.5})\cdot 10^{-5}$ & 1.0\\
53.7--60.4 & 1527 & 0.14 & $(79.6 \pm 3.3^{+3.8}_{-3.9})\cdot 10^{-5}$ & 1.0\\
60.4--68.2 & 1172 & 0.15 & $(53.8 \pm 2.6^{+2.7}_{-2.6})\cdot 10^{-5}$ & 1.0\\
68.2--77.4 & 901 & 0.18 & $(35.7 \pm 2.1^{+2.1}_{-1.5})\cdot 10^{-5}$ & 1.0\\
\hline\hline
\end{tabular} 
\end{table*}

A complete assessment of the systematic uncertainties related to the
discrepancies between data and simulations for this subset of data
(as discussed in Sec.~\ref{sec:systematic} for the full analysis) 
would require us, in this case, to undertake more
complex simulations with about 20 times as many events; 
this is not possible at this time. 
It is reasonable to assume that such uncertainties are of the same order of
magnitude as those quoted for the \highe\ events.
Because this source of systematic errors comes from the analysis 
of data sets of very different size,
\begin{figure}[!bhtp]
\includegraphics[width=\onecolwidth]{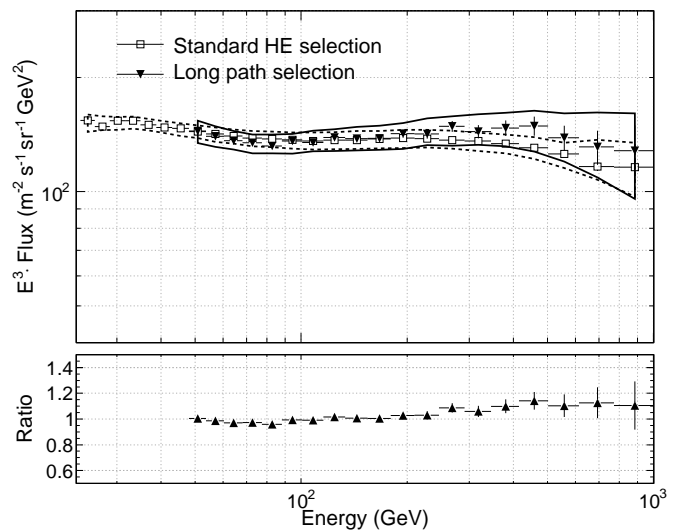}
\caption{\label{fig:SpectrumLong}Comparison of the spectra obtained with
the long-path selection and the standard \highe\ selection. 
The continuous lines represent the systematic uncertainties
for the long-path analysis and the dashed lines for the standard analysis. 
The bottom panel shows the ratio of the two spectra.}
\end{figure} 
with one being only $\sim 5\%$ of the other, we can assume that 
they are substantially independent.
\begin{figure}[!tbp]
  \includegraphics[width=\onecolwidth]{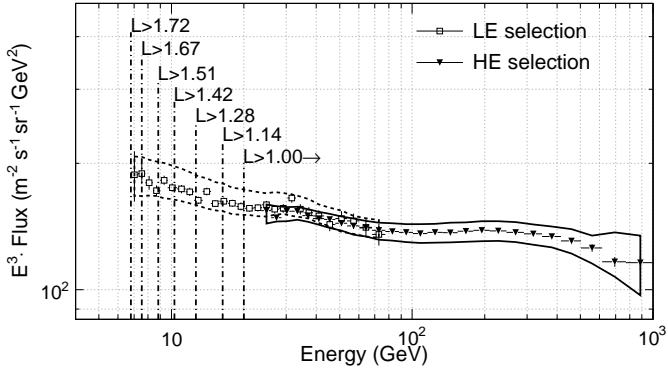}
  \caption{Cosmic-ray electron spectra as measured by Fermi LAT for 1 yr of
    observations for \lowe\ events (squares) and \highe\ events (triangles). 
    The continuous lines represent the systematic uncertainties.
    The two spectra agrees within systematic errors in the overlap region
    between 20~GeV and 80~GeV.}
\label{fig:FluxOverlap}
\end{figure} 

However this assumption is not critical for our purposes, because the
systematic uncertainties in the evaluation of the EGF and the residual
contamination (which again are connected to the limited size of the  
simulated event samples) are significantly larger, here, 
and in fact constitute the dominant contribution.
Figure~\ref{fig:SpectrumLong} shows the consistency, within the systematic
errors, between the spectrum obtained with the standard analysis and 
that obtained with the long-path selection. This confirms that the energy
resolution quoted in Sec.~\ref{sec:energy} is indeed sufficient for the
measurement and does not have any significant effect on the spectrum.
     
\section{Result and discussion}\label{sec:result} 

We analyzed data collected in nominal sky survey mode 
from 4 August 2008 to 4 August 2009, for 
a total live time of about 265 days.
The event sample after the selection is composed of 
$1.24\times 10^5$ events in the \lowe\ range and 
$7.8\times 10^6 $ events in the \highe\ range. 
For the latter analysis, the energy bins were chosen to be the
full width of 68\% containment of the energy dispersion, 
evaluated at the bin center. 
The resulting electron spectra are shown in Fig.~\ref{fig:FluxOverlap}
for the two selections. They agree within systematic errors in
the overlap region between 20~GeV and 80~GeV.

\begin{figure}[htbp]
  \includegraphics[width=\linewidth]{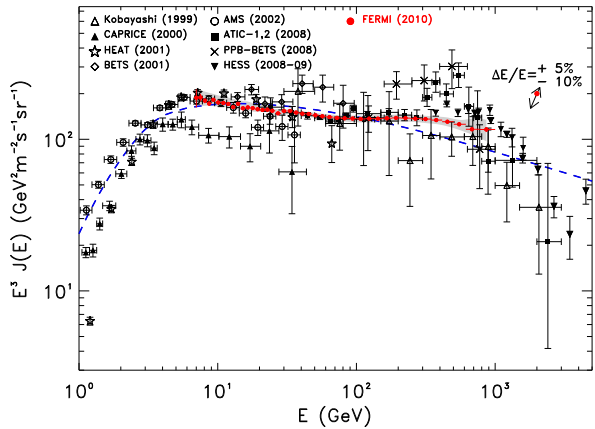}
  \caption{\label{fig:Spectrum} (color). Cosmic-ray electron spectrum as measured 
    by Fermi LAT for 1 yr of observations - shown by filled circles, 
    along with other recent high-energy results. 
    The \lowe\ spectrum is used to extend the \highe\ analysis at low energy.
    Systematic errors are shown by the gray band. 
    The range of the spectrum rigid shift implied by a shift of the 
    absolute energy is shown by the arrow in the upper right corner. 
    Dashed line shows the model based on pre-Fermi 
    results~\cite{2004ApJ...613..962S}.
    Data from other experiments are: Kobayashi~\cite{Kobayashi1999}, 
    CAPRICE~\cite{CAPRICE}, HEAT~\cite{HEAT2001}, BETS~\cite{BETS}, 
    AMS~\cite{ams}, ATIC~\cite{ATIC}, PPB-BETS~\cite{PPB}, and
    HESS~\cite{HESS,HESS2}.
    Note that the AMS and CAPRICE data are for $e^-$ only.}
\end{figure}

Numerical values are given in Table~\ref{table:LESelection} for \lowe\ 
and in Table~\ref{table:HESelection} for \highe\ events.  
Note that the \lowe\ part of the spectrum has 
poorer statistical precision due to the 1:250 on-board prescale. 
Figure~\ref{fig:Spectrum} shows the LAT spectrum 
along with other recent experiments and with a CR 
propagation model based on pre-Fermi data~\cite{2004ApJ...613..962S}.

The CR electron spectrum reported in this paper and shown 
in Fig.~\ref{fig:Spectrum} is essentially the same as that published 
in~\cite{PRL} for the energy above 20~GeV, but with twice the data volume.
Within the systematic errors (shown by the gray band 
in Fig.~\ref{fig:Spectrum}) 
the entire spectrum from 7~GeV to 1~TeV can be fitted by a power law 
with spectral index in the interval 3.03--3.13 (best fit 3.08), 
similar to that given in~\cite{PRL}.
The spectrum is significantly harder (flatter) than that reported 
by previous experiments. 
The cross-check analysis using events with long paths in the instrument 
confirms the absence of any evident feature in the $e^+ + e^-$ spectrum 
from 50~GeV to 1~TeV, as originally reported in~\cite{PRL}.

Below $\sim 50$~GeV the electron spectrum is consistent with 
previous experiments and does not indicate any flattening at low energies. 
This  may be compared with previous experiments that made measurements over 
the last solar cycle with an opposite polarity of the solar magnetic field 
(e.g.~\cite{ams, CAPRICE}), and which indicate that 
a significant flattening occurs only below $\sim 6$~GeV. 

\begin{figure}[hbtp]
\centering
\includegraphics[width=\linewidth]{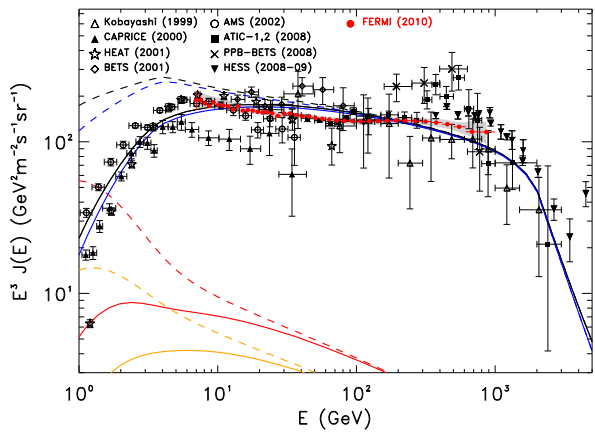}
\caption{(color). The $e^+ + e^-$ spectrum computed with the conventional GALPROP
  model~\cite{ptuskin} (shown by solid black line) is compared with the 
  Fermi LAT (red filled circles) and other experimental data.
  This model adopts an injection spectral index $\Gamma = 1.6/2.5$ below/above
  4~GeV, and a steepening $\Gamma = 5$ above 2~TeV. 
  Blue lines show $e^-$ spectrum only.
  The solar modulation was treated using the force-field approximation with 
  $\Phi=550$~MV.
  The dashed/solid lines show the before modulation/modulated spectra.
  Secondary $e^+$ (red lines) and $e^-$ (orange lines) are calculated using 
  the formalism from~\cite{MS1998}.  
 }
\label{fig:conventional_fig5}
\end{figure}

To fit the high-energy part of the Fermi LAT spectrum and to agree with 
the HESS data, a conventional propagation model requires 
an injection power-law index $\alpha \simeq 2.5$ above $\sim 4$~GeV 
and a cutoff at $\sim 2$~TeV.
However, while providing good agreement with the high-energy part of 
the spectrum, a model with a single power-law injection index fails 
to reproduce the low-energy data. 
To obtain an agreement with all the available data at low energies we need
the injection spectrum $\alpha \sim 1.5 - 2.0$ below $\sim 4$~GeV 
and a modulation parameter in the range $\Phi = 400 - 600$~MV. 
The latter was set to match proton spectrum at low energy during 
the first year of Fermi LAT operation~\cite{2009NuPhS.190..293C}.
An example of such a calculation using GALPROP 
code~\cite{SM1998} is shown in Fig.~\ref{fig:conventional_fig5}. 
This model includes spatial Kolmogorov diffusion with spectral index 
$\delta = 0.33$ and diffusive reacceleration characterized by
an Alfv\'en speed $v_A = 30$~km/s; the halo height was 4~kpc. 
Energy losses by inverse Compton scattering and synchrotron radiation were 
computed as a function of energy and position.
Secondary electrons and positrons from CR proton and helium interactions 
with interstellar gas make a significant contribution to the total leptons 
flux, especially at low energies. 
These secondary particle fluxes were computed for the same GALPROP model 
as for the primary electrons as described in~\cite{MS1998} 
and references therein.
This model is essentially a conventional one with distributed reacceleration,
described in~\cite{ptuskin}. For more information on CR and their propagation 
in the interstellar medium see e.g. a recent review~\cite{ARNPS}. 

\renewcommand{\tabcolsep}{7pt}
\squeezetable
\begin{table}[hbp!] 	 
  \centering 	 
  \caption{Number of events, residual hadronic contamination 
    and flux $J_{\rm E}$ for \highe\ analysis.
    Statistical error is followed by systematic error 
    (see Sec.~\ref{sec:systematic}). 
    Residual contamination is defined as the ratio between 
    hadronic background rate and measured event rate.} 
  \label{table:HESelection}     
  \begin{tabular}{c r c r} 	 
    \hline\hline 	 
    Energy  & Counts & Residual & $J_{\rm E}$ \hspace*{1.2cm}\\
    (GeV) &  & contamination & (GeV$^{-1}$ s$^{-1}$ m$^{-2}$ sr$^{-1}$)\\
    \hline 
    23.6--26.0 & 944~264 & 0.04 & $(1020 \pm 1.2^{+50}_{-54}) \cdot 10^{-5}$\\
    26.0--28.7 & 958~983 & 0.05 & $(735 \pm 0.9^{+30}_{-33}) \cdot 10^{-5}$\\
    28.7--31.7 & 967~571 & 0.05 & $(566 \pm 0.6^{+20}_{-22}) \cdot 10^{-5}$\\
    31.7--35.0 & 880~243 & 0.06 & $(420 \pm 0.5^{+13}_{-16}) \cdot 10^{-5}$\\
    35.0--38.8 & 754~385 & 0.08 & $(302 \pm 0.4^{+9}_{-11}) \cdot 10^{-5}$\\
    38.8--43.1 & 638~368 & 0.09 & $(2180 \pm 3.0^{+71}_{-83}) \cdot 10^{-6}$\\
    43.1--48.0 & 534~109 & 0.10 & $(1577 \pm 2.4^{+55}_{-65}) \cdot 10^{-6}$\\
    48.0--53.7 & 447~219 & 0.11 & $(1110 \pm 1.9^{+38}_{-46}) \cdot 10^{-6}$\\
    53.7--60.4 & 371~444 & 0.12 & $(775 \pm 1.4^{+29}_{-38}) \cdot 10^{-6}$\\
    60.4--68.2 & 297~616 & 0.13 & $(536 \pm 1.1^{+21}_{-24}) \cdot 10^{-6}$\\
    68.2--77.4 & 241~956 & 0.14 & $(365 \pm 0.9^{+14}_{-19}) \cdot 10^{-6}$\\
    77.4--88.1 & 191~926 & 0.15 & $(247 \pm 0.7^{+11}_{-12}) \cdot 10^{-6}$\\
    88.1--101 & 148~899 & 0.16 & $(1659 \pm 5.1^{+77}_{-84}) \cdot 10^{-7}$\\
    101--116 & 118~212 & 0.16 & $(1097 \pm 3.8^{+52}_{-59}) \cdot 10^{-7}$\\
    116--133 & 89~641 & 0.17 & $(725 \pm 2.9^{+37}_{-38}) \cdot 10^{-7}$\\
    133--154 & 67~146 & 0.18 & $(470 \pm 2.2^{+25}_{-25}) \cdot 10^{-7}$\\
    154--180 & 52~453 & 0.17 & $(303 \pm 1.6^{+15}_{-16}) \cdot 10^{-7}$\\
    180--210 & 37~352 & 0.18 & $(192 \pm 1.2^{+10}_{-10}) \cdot 10^{-7}$\\
    210--246 & 26~807 & 0.19 & $(1198 \pm 9.0^{+68}_{-62}) \cdot 10^{-8}$\\
    246--291 & 19~150 & 0.19 & $(728 \pm 6.5^{+41}_{-37}) \cdot 10^{-8}$\\
    291--346 & 13~648 & 0.19 & $(434 \pm 4.6^{+23}_{-24}) \cdot 10^{-8}$\\
    346--415 & 9183 & 0.19 & $(251 \pm 3.2^{+14}_{-14}) \cdot 10^{-8}$\\
    415--503 & 5845 & 0.19 & $(1407 \pm 22^{+95}_{-92}) \cdot 10^{-9}$\\
    503--615 & 3577 & 0.20 & $(754 \pm 15^{+56}_{-57}) \cdot 10^{-9}$\\
    615--772 & 2092 & 0.19 & $(370 \pm 10^{+54}_{-39}) \cdot 10^{-9}$\\
    \phantom{0}772--1000 & 1039 & 0.20 & $(179 \pm 6.9^{+32}_{-25}) \cdot 10^{-9}$\\
    \hline\hline
  \end{tabular}     
\end{table}

We note that the force-field treatment~\cite{Gleeson&Axford}, 
used in our calculation to evaluate the effect of solar modulation, 
is  approximate and does not take into account many important effects, such as 
the configuration of the heliospheric magnetic field and drift
effects which lead to the charge-sign dependence 
(e.g.~\cite{clem,potg,Potgieter}).
In addition, the value of the modulation potential $\Phi$ depends on the 
assumed interstellar particle spectra, and thus other combinations
of parameters are also possible. Ultimately the interstellar spectrum 
of CREs can be tested using the LAT observations of the 
Galactic diffuse gamma-ray emission where the inverse Compton component is
dominating the gas component at medium to high Galactic 
latitudes~\cite{LatDiffuse}.

\begin{figure}[hbtp]
\centering
\includegraphics[width=\linewidth]{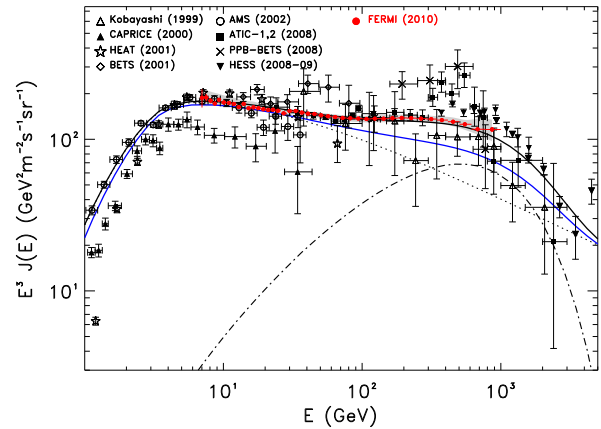}
\caption{(color). The $e^+ + e^-$ spectrum (solid line) computed with the conventional 
  GALPROP model~\cite{ptuskin} but with a different injection spectrum: 
  an injection index $\Gamma=1.6/2.7$ below/above 4~GeV (dotted line). 
  An additional component with an injection index $\Gamma=1.5$ and 
  exponential cutoff is shown by the dashed line.
  Blue line shows $e^-$ spectrum only. 
  Secondary $e^+$ and $e^-$ are treated as in 
  Fig.~\ref{fig:conventional_fig5}.
  Fermi-LAT data points are shown by red filled circles. 
 }
\label{fig:extra_source_fig9}
\end{figure}

The Fermi LAT measured spectrum suggests some spectral flattening at 
70--200~GeV and a noticeable excess above 200~GeV as compared to our 
power-law spectral fit. These gentle features of the spectrum can be 
explained within a conventional model by adjusting the injection spectra.

Another possibility that provides a good overall agreement with our spectrum 
is the introduction of an additional leptonic component with a hard spectrum 
(Fig.~\ref{fig:extra_source_fig9}). Such an additional component is motivated
by the rise in the positron fraction reported by PAMELA~\cite{Pamela}.
Recent papers have suggested different models for this component. The data
can accommodate a contribution from nearby sources (such as pulsars) or from
the annihilation of dark matter particles (see e.g.~\cite{Dario} 
for a comprehensive list of references). 
The features may also be explained by other
astrophysical effects (\cite{2010ApJ...710..236S,Blasi:2009hv} 
and others).
Further discussion of these many models, as well as an interpretation 
of low-energy data with more realistic models for heliospheric propagation, 
is beyond the scope of this paper.

\begin{acknowledgments}
The Fermi LAT Collaboration acknowledges generous ongoing support
from a number of agencies and institutes that have supported both the
development and the operation of the LAT as well as scientific data analysis.
These include the National Aeronautics and Space Administration and the
Department of Energy in the United States; the Commissariat \`a l'Energie Atomique
and the Centre National de la Recherche Scientifique/Institut National de Physique
Nucl\'eaire et de Physique des Particules in France; the Agenzia Spaziale Italiana
and the Istituto Nazionale di Fisica Nucleare in Italy; the Ministry of Education,
Culture, Sports, Science, and Technology (MEXT), High Energy Accelerator Research
Organization (KEK), and Japan Aerospace Exploration Agency (JAXA) in Japan; and
the K.~A.~Wallenberg Foundation, the Swedish Research Council, and the
Swedish National Space Board in Sweden.
Additional support for science analysis during the operations phase is gratefully
acknowledged from the Istituto Nazionale di Astrofisica in Italy and the Centre National d'\'Etudes Spatiales in France.
We would like to thank
the INFN GRID Data Centers of Pisa, Trieste, and CNAF-Bologna;
the DOE SLAC National Accelerator Laboratory Computing Division
and the CNRS/IN2P3 Computing Center (CC-IN2P3---Lyon/Villeurbanne) in partnership with CEA/DSM/Irfu
for their strong support in performing the massive simulations necessary for this work. 
J. Conrad is Fellow of the Royal Swedish Academy of Sciences, funded by 
a grant from the K. A. Wallenberg Foundation.
L. Tibaldo is partially supported by the International Doctorate on 
Astroparticle Physics (IDAPP) program.

\end{acknowledgments}



\begin{thebibliography}{47}
\expandafter\ifx\csname natexlab\endcsname\relax\def\natexlab#1{#1}\fi
\expandafter\ifx\csname bibnamefont\endcsname\relax
  \def\bibnamefont#1{#1}\fi
\expandafter\ifx\csname bibfnamefont\endcsname\relax
  \def\bibfnamefont#1{#1}\fi
\expandafter\ifx\csname citenamefont\endcsname\relax
  \def\citenamefont#1{#1}\fi
\expandafter\ifx\csname url\endcsname\relax
  \def\url#1{\texttt{#1}}\fi
\expandafter\ifx\csname urlprefix\endcsname\relax\def\urlprefix{URL }\fi
\providecommand{\bibinfo}[2]{#2}
\providecommand{\eprint}[2][]{\url{#2}}

\bibitem[{\citenamefont{{Atwood} et~al.}(2009)\citenamefont{{Atwood}, {Abdo},
  {Ackermann}, {Althouse}, {Anderson}, {Axelsson}, {Baldini}, {Ballet}, {Band},
  {Barbiellini} et~al.}}]{LAT}
\bibinfo{author}{\bibfnamefont{W.~B.} \bibnamefont{{Atwood}}},
  \bibinfo{author}{\bibfnamefont{A.~A.} \bibnamefont{{Abdo}}},
  \bibinfo{author}{\bibfnamefont{M.}~\bibnamefont{{Ackermann}}},
  \bibinfo{author}{\bibfnamefont{W.}~\bibnamefont{{Althouse}}},
  \bibinfo{author}{\bibfnamefont{B.}~\bibnamefont{{Anderson}}},
  \bibinfo{author}{\bibfnamefont{M.}~\bibnamefont{{Axelsson}}},
  \bibinfo{author}{\bibfnamefont{L.}~\bibnamefont{{Baldini}}},
  \bibinfo{author}{\bibfnamefont{J.}~\bibnamefont{{Ballet}}},
  \bibinfo{author}{\bibfnamefont{D.~L.} \bibnamefont{{Band}}},
  \bibinfo{author}{\bibfnamefont{G.}~\bibnamefont{{Barbiellini}}},
  \bibnamefont{et~al.}, \bibinfo{journal}{Astrophys. J.} \textbf{\bibinfo{volume}{697}},
  \bibinfo{pages}{1071} (\bibinfo{year}{2009}).

\bibitem[{\citenamefont{{Abdo} et~al.}(2009{\natexlab{a}})\citenamefont{{Abdo},
  {Ackermann}, {Ajello}, {Atwood}, {Axelsson}, {Baldini}, {Ballet},
  {Barbiellini}, {Bastieri}, {Battelino} et~al.}}]{PRL}
\bibinfo{author}{\bibfnamefont{A.~A.} \bibnamefont{{Abdo}}},
  \bibinfo{author}{\bibfnamefont{M.}~\bibnamefont{{Ackermann}}},
  \bibinfo{author}{\bibfnamefont{M.}~\bibnamefont{{Ajello}}},
  \bibinfo{author}{\bibfnamefont{W.~B.} \bibnamefont{{Atwood}}},
  \bibinfo{author}{\bibfnamefont{M.}~\bibnamefont{{Axelsson}}},
  \bibinfo{author}{\bibfnamefont{L.}~\bibnamefont{{Baldini}}},
  \bibinfo{author}{\bibfnamefont{J.}~\bibnamefont{{Ballet}}},
  \bibinfo{author}{\bibfnamefont{G.}~\bibnamefont{{Barbiellini}}},
  \bibinfo{author}{\bibfnamefont{D.}~\bibnamefont{{Bastieri}}},
  \bibinfo{author}{\bibfnamefont{M.}~\bibnamefont{{Battelino}}},
  \bibnamefont{et~al.}, \bibinfo{journal}{Phys. Rev. Lett.}
  \textbf{\bibinfo{volume}{102}}, \bibinfo{pages}{181101}
  (\bibinfo{year}{2009}{\natexlab{a}}).

\bibitem[{\citenamefont{{Nishimura} et~al.}(1980)\citenamefont{{Nishimura},
  {Fujii}, {Taira}, {Aizu}, {Hiraiwa}, {Kobayashi}, {Niu}, {Ohta}, {Golden},
  and {Koss}}}]{nishimura}
\bibinfo{author}{\bibfnamefont{J.}~\bibnamefont{{Nishimura}}},
  \bibinfo{author}{\bibfnamefont{M.}~\bibnamefont{{Fujii}}},
  \bibinfo{author}{\bibfnamefont{T.}~\bibnamefont{{Taira}}},
  \bibinfo{author}{\bibfnamefont{E.}~\bibnamefont{{Aizu}}},
  \bibinfo{author}{\bibfnamefont{H.}~\bibnamefont{{Hiraiwa}}},
  \bibinfo{author}{\bibfnamefont{T.}~\bibnamefont{{Kobayashi}}},
  \bibinfo{author}{\bibfnamefont{K.}~\bibnamefont{{Niu}}},
  \bibinfo{author}{\bibfnamefont{I.}~\bibnamefont{{Ohta}}},
  \bibinfo{author}{\bibfnamefont{R.~L.} \bibnamefont{{Golden}}},
  \bibnamefont{and} \bibinfo{author}{\bibfnamefont{T.~A.}
  \bibnamefont{{Koss}}}, \bibinfo{journal}{Astrophys. J.} \textbf{\bibinfo{volume}{238}},
  \bibinfo{pages}{394} (\bibinfo{year}{1980}).

\bibitem[{\citenamefont{{Barwick} et~al.}(1998)\citenamefont{{Barwick},
  {Beatty}, {Bower}, {Chaput}, {Coutu}, {de Nolfo}, {DuVernois}
  et~al.}}]{barwick}
\bibinfo{author}{\bibfnamefont{S.~W.} \bibnamefont{{Barwick}}},
  \bibinfo{author}{\bibfnamefont{J.~J.} \bibnamefont{{Beatty}}},
  \bibinfo{author}{\bibfnamefont{C.~R.} \bibnamefont{{Bower}}},
  \bibinfo{author}{\bibfnamefont{C.~J.} \bibnamefont{{Chaput}}},
  \bibinfo{author}{\bibfnamefont{S.}~\bibnamefont{{Coutu}}},
  \bibinfo{author}{\bibfnamefont{G.~A.} \bibnamefont{{de Nolfo}}},
  \bibinfo{author}{\bibfnamefont{M.~A.} \bibnamefont{{DuVernois}}},
  \bibnamefont{et~al.}, \bibinfo{journal}{Astrophys. J.} \textbf{\bibinfo{volume}{498}},
  \bibinfo{pages}{779} (\bibinfo{year}{1998}).

\bibitem[{\citenamefont{{M\"{u}ller}}(2001)}]{mueller}
\bibinfo{author}{\bibfnamefont{D.}~\bibnamefont{{M\"{u}ller}}},
  \bibinfo{journal}{Adv. Space Res.} \textbf{\bibinfo{volume}{27}},
  \bibinfo{pages}{659} (\bibinfo{year}{2001}).

\bibitem[{\citenamefont{{Aharonian} et~al.}(1995)\citenamefont{{Aharonian},
  {Atoyan}, and {Voelk}}}]{ahar}
\bibinfo{author}{\bibfnamefont{F.~A.} \bibnamefont{{Aharonian}}},
  \bibinfo{author}{\bibfnamefont{A.~M.} \bibnamefont{{Atoyan}}},
  \bibnamefont{and} \bibinfo{author}{\bibfnamefont{H.~J.}
  \bibnamefont{{Voelk}}}, \bibinfo{journal}{Astron. Astrophys.}
  \textbf{\bibinfo{volume}{294}}, \bibinfo{pages}{L41} (\bibinfo{year}{1995}).

\bibitem[{\citenamefont{{Chang} et~al.}(2008)\citenamefont{{Chang}, {Adams},
  {Ahn}, {Bashindzhagyan}, {Christl}, {Ganel}, {Guzik}, {Isbert}, {Kim},
  {Kuznetsov} et~al.}}]{ATIC}
\bibinfo{author}{\bibfnamefont{J.}~\bibnamefont{{Chang}}},
  \bibinfo{author}{\bibfnamefont{J.~H.} \bibnamefont{{Adams}}},
  \bibinfo{author}{\bibfnamefont{H.~S.} \bibnamefont{{Ahn}}},
  \bibinfo{author}{\bibfnamefont{G.~L.} \bibnamefont{{Bashindzhagyan}}},
  \bibinfo{author}{\bibfnamefont{M.}~\bibnamefont{{Christl}}},
  \bibinfo{author}{\bibfnamefont{O.}~\bibnamefont{{Ganel}}},
  \bibinfo{author}{\bibfnamefont{T.~G.} \bibnamefont{{Guzik}}},
  \bibinfo{author}{\bibfnamefont{J.}~\bibnamefont{{Isbert}}},
  \bibinfo{author}{\bibfnamefont{K.~C.} \bibnamefont{{Kim}}},
  \bibinfo{author}{\bibfnamefont{E.~N.} \bibnamefont{{Kuznetsov}}},
  \bibnamefont{et~al.}, \bibinfo{journal}{Nature (London)}
  \textbf{\bibinfo{volume}{456}}, \bibinfo{pages}{362} (\bibinfo{year}{2008}).

\bibitem[{\citenamefont{{Torii} et~al.}(2008)\citenamefont{{Torii}, {Yamagami},
  {Tamura}, {Yoshida}, {Kitamura}, {Anraku}, {Chang}, {Ejiri}, {Iijima},
  {Kadokura} et~al.}}]{PPB}
\bibinfo{author}{\bibfnamefont{S.}~\bibnamefont{{Torii}}},
  \bibinfo{author}{\bibfnamefont{T.}~\bibnamefont{{Yamagami}}},
  \bibinfo{author}{\bibfnamefont{T.}~\bibnamefont{{Tamura}}},
  \bibinfo{author}{\bibfnamefont{K.}~\bibnamefont{{Yoshida}}},
  \bibinfo{author}{\bibfnamefont{H.}~\bibnamefont{{Kitamura}}},
  \bibinfo{author}{\bibfnamefont{K.}~\bibnamefont{{Anraku}}},
  \bibinfo{author}{\bibfnamefont{J.}~\bibnamefont{{Chang}}},
  \bibinfo{author}{\bibfnamefont{M.}~\bibnamefont{{Ejiri}}},
  \bibinfo{author}{\bibfnamefont{I.}~\bibnamefont{{Iijima}}},
  \bibinfo{author}{\bibfnamefont{A.}~\bibnamefont{{Kadokura}}},
  \bibnamefont{et~al.}, \bibinfo{journal}{arXiv:0809.0760}.

\bibitem[{\citenamefont{{Aharonian} et~al.}(2008)\citenamefont{{Aharonian},
  {Akhperjanian}, {Barres de Almeida}, {Bazer-Bachi}, {Becherini}, {Behera},
  {Benbow}, {Bernl{\"o}hr}, {Boisson}, {Bochow} et~al.}}]{HESS}
\bibinfo{author}{\bibfnamefont{F.}~\bibnamefont{{Aharonian}}},
  \bibinfo{author}{\bibfnamefont{A.~G.} \bibnamefont{{Akhperjanian}}},
  \bibinfo{author}{\bibfnamefont{U.}~\bibnamefont{{Barres de Almeida}}},
  \bibinfo{author}{\bibfnamefont{A.~R.} \bibnamefont{{Bazer-Bachi}}},
  \bibinfo{author}{\bibfnamefont{Y.}~\bibnamefont{{Becherini}}},
  \bibinfo{author}{\bibfnamefont{B.}~\bibnamefont{{Behera}}},
  \bibinfo{author}{\bibfnamefont{W.}~\bibnamefont{{Benbow}}},
  \bibinfo{author}{\bibfnamefont{K.}~\bibnamefont{{Bernl{\"o}hr}}},
  \bibinfo{author}{\bibfnamefont{C.}~\bibnamefont{{Boisson}}},
  \bibinfo{author}{\bibfnamefont{A.}~\bibnamefont{{Bochow}}},
  \bibnamefont{et~al.}, \bibinfo{journal}{Phys. Rev. Lett.}
  \textbf{\bibinfo{volume}{101}}, \bibinfo{pages}{261104}
  (\bibinfo{year}{2008}).

\bibitem[{\citenamefont{{Aharonian} et~al.}(2009)\citenamefont{{Aharonian},
  {Akhperjanian}, {Anton}, {Barres de Almeida}, {Bazer-Bachi}, {Becherini},
  {Behera}, {Bernl{\"o}hr}, {Bochow}, {Boisson} et~al.}}]{HESS2}
\bibinfo{author}{\bibfnamefont{F.}~\bibnamefont{{Aharonian}}},
  \bibinfo{author}{\bibfnamefont{A.~G.} \bibnamefont{{Akhperjanian}}},
  \bibinfo{author}{\bibfnamefont{G.}~\bibnamefont{{Anton}}},
  \bibinfo{author}{\bibfnamefont{U.}~\bibnamefont{{Barres de Almeida}}},
  \bibinfo{author}{\bibfnamefont{A.~R.} \bibnamefont{{Bazer-Bachi}}},
  \bibinfo{author}{\bibfnamefont{Y.}~\bibnamefont{{Becherini}}},
  \bibinfo{author}{\bibfnamefont{B.}~\bibnamefont{{Behera}}},
  \bibinfo{author}{\bibfnamefont{K.}~\bibnamefont{{Bernl{\"o}hr}}},
  \bibinfo{author}{\bibfnamefont{A.}~\bibnamefont{{Bochow}}},
  \bibinfo{author}{\bibfnamefont{C.}~\bibnamefont{{Boisson}}},
  \bibnamefont{et~al.}, \bibinfo{journal}{Astron. Astrophys.} \textbf{\bibinfo{volume}{508}},
  \bibinfo{pages}{561} (\bibinfo{year}{2009}).

\bibitem[{\citenamefont{{Adriani} et~al.}(2009)\citenamefont{{Adriani},
  {Barbarino}, {Bazilevskaya}, {Bellotti}, {Boezio}, {Bogomolov}, {Bonechi},
  {Bongi}, {Bonvicini}, {Bottai} et~al.}}]{Pamela}
\bibinfo{author}{\bibfnamefont{O.}~\bibnamefont{{Adriani}}},
  \bibinfo{author}{\bibfnamefont{G.~C.} \bibnamefont{{Barbarino}}},
  \bibinfo{author}{\bibfnamefont{G.~A.} \bibnamefont{{Bazilevskaya}}},
  \bibinfo{author}{\bibfnamefont{R.}~\bibnamefont{{Bellotti}}},
  \bibinfo{author}{\bibfnamefont{M.}~\bibnamefont{{Boezio}}},
  \bibinfo{author}{\bibfnamefont{E.~A.} \bibnamefont{{Bogomolov}}},
  \bibinfo{author}{\bibfnamefont{L.}~\bibnamefont{{Bonechi}}},
  \bibinfo{author}{\bibfnamefont{M.}~\bibnamefont{{Bongi}}},
  \bibinfo{author}{\bibfnamefont{V.}~\bibnamefont{{Bonvicini}}},
  \bibinfo{author}{\bibfnamefont{S.}~\bibnamefont{{Bottai}}},
  \bibnamefont{et~al.}, \bibinfo{journal}{Nature (London)}
  \textbf{\bibinfo{volume}{458}}, \bibinfo{pages}{607} (\bibinfo{year}{2009}).

\bibitem[{\citenamefont{{Moskalenko} and {Strong}}(1998)}]{MS1998}
\bibinfo{author}{\bibfnamefont{I.~V.} \bibnamefont{{Moskalenko}}}
  \bibnamefont{and} \bibinfo{author}{\bibfnamefont{A.~W.}
  \bibnamefont{{Strong}}}, \bibinfo{journal}{Astrophys. J.}
  \textbf{\bibinfo{volume}{493}}, \bibinfo{pages}{694} (\bibinfo{year}{1998}).

\bibitem[{\citenamefont{{Kobayashi} et~al.}(2004)\citenamefont{{Kobayashi},
  {Komori}, {Yoshida}, and {Nishimura}}}]{kob}
\bibinfo{author}{\bibfnamefont{T.}~\bibnamefont{{Kobayashi}}},
  \bibinfo{author}{\bibfnamefont{Y.}~\bibnamefont{{Komori}}},
  \bibinfo{author}{\bibfnamefont{K.}~\bibnamefont{{Yoshida}}},
  \bibnamefont{and}
  \bibinfo{author}{\bibfnamefont{J.}~\bibnamefont{{Nishimura}}},
  \bibinfo{journal}{Astrophys. J.} \textbf{\bibinfo{volume}{601}}, \bibinfo{pages}{340}
  (\bibinfo{year}{2004}).

\bibitem[{\citenamefont{{Grasso} et~al.}(2009)\citenamefont{{Grasso},
  {Profumo}, {Strong}, {Baldini}, {Bellazzini}, {Bloom}, {Bregeon}, {di
  Bernardo}, {Gaggero}, {Giglietto} et~al.}}]{Dario}
\bibinfo{author}{\bibfnamefont{D.}~\bibnamefont{{Grasso}}},
  \bibinfo{author}{\bibfnamefont{S.}~\bibnamefont{{Profumo}}},
  \bibinfo{author}{\bibfnamefont{A.~W.} \bibnamefont{{Strong}}},
  \bibinfo{author}{\bibfnamefont{L.}~\bibnamefont{{Baldini}}},
  \bibinfo{author}{\bibfnamefont{R.}~\bibnamefont{{Bellazzini}}},
  \bibinfo{author}{\bibfnamefont{E.~D.} \bibnamefont{{Bloom}}},
  \bibinfo{author}{\bibfnamefont{J.}~\bibnamefont{{Bregeon}}},
  \bibinfo{author}{\bibfnamefont{G.}~\bibnamefont{{di Bernardo}}},
  \bibinfo{author}{\bibfnamefont{D.}~\bibnamefont{{Gaggero}}},
  \bibinfo{author}{\bibfnamefont{N.}~\bibnamefont{{Giglietto}}},
  \bibnamefont{et~al.}, \bibinfo{journal}{Astropart. Phys.}
  \textbf{\bibinfo{volume}{32}}, \bibinfo{pages}{140} (\bibinfo{year}{2009}).

\bibitem[{\citenamefont{{Ormes} et~al.}(1997)}]{Durban}
\bibinfo{author}{\bibfnamefont{J.~F.} \bibnamefont{{Ormes}}}
  \bibnamefont{et~al.}, in \emph{\bibinfo{booktitle}{Proceedings of the 25th
  International Cosmic Ray Conference}}, edited by
  \bibinfo{editor}{\bibnamefont{{M. S. Potgieter, C. Raubenheimer, and D. J.
  van der Walt}}} (\bibinfo{publisher}{{Potchefstroom University, Transvaal,
  South Africa}}, \bibinfo{year}{1997}), vol.~\bibinfo{volume}{5}, pp.
  \bibinfo{pages}{73}.

\bibitem[{\citenamefont{{Moiseev} et~al.}(2008)\citenamefont{{Moiseev},
  {Ormes}, and {Moskalenko}}}]{Alex}
\bibinfo{author}{\bibfnamefont{A.~A.} \bibnamefont{{Moiseev}}},
  \bibinfo{author}{\bibfnamefont{J.~F.} \bibnamefont{{Ormes}}},
  \bibnamefont{and} \bibinfo{author}{\bibfnamefont{I.~V.}
  \bibnamefont{{Moskalenko}}}, in \emph{\bibinfo{booktitle}{{Proceedings of the
  30th International Cosmic Ray Conference}}}, edited by
  \bibinfo{editor}{\bibnamefont{{Rogelio Caballero, Juan Carlos D'Olivo,
  Gustavo Medina-Tanco, Lukas Nellen, Federico A. S\'anchez, Jos\'e F.
  Vald\'es-Galicia}}} (\bibinfo{publisher}{{Universidad Nacional Aut\'onoma de
  M\'exico, Mexico City, Mexico}}, \bibinfo{year}{2008}),
  vol.~\bibinfo{volume}{2}, pp. \bibinfo{pages}{449},
  \eprint{arXiv:astro-ph/0706.0882}.

\bibitem[{\citenamefont{{Baldini} et~al.}(2006)\citenamefont{{Baldini}, {Brez},
  {Himel}, {Hirayama}, {Johnson}, {Kroeger}, {Latronico}, {Minuti}, {Nelson},
  {Rando} et~al.}}]{2006ITNS...53..466B}
\bibinfo{author}{\bibfnamefont{L.}~\bibnamefont{{Baldini}}},
  \bibinfo{author}{\bibfnamefont{A.}~\bibnamefont{{Brez}}},
  \bibinfo{author}{\bibfnamefont{T.}~\bibnamefont{{Himel}}},
  \bibinfo{author}{\bibfnamefont{M.}~\bibnamefont{{Hirayama}}},
  \bibinfo{author}{\bibfnamefont{R.~P.} \bibnamefont{{Johnson}}},
  \bibinfo{author}{\bibfnamefont{W.}~\bibnamefont{{Kroeger}}},
  \bibinfo{author}{\bibfnamefont{L.}~\bibnamefont{{Latronico}}},
  \bibinfo{author}{\bibfnamefont{M.}~\bibnamefont{{Minuti}}},
  \bibinfo{author}{\bibfnamefont{D.}~\bibnamefont{{Nelson}}},
  \bibinfo{author}{\bibfnamefont{R.}~\bibnamefont{{Rando}}},
  \bibnamefont{et~al.}, \bibinfo{journal}{IEEE Trans. Nucl. Sci.}
  \textbf{\bibinfo{volume}{53}}, \bibinfo{pages}{466} (\bibinfo{year}{2006}).

\bibitem[{\citenamefont{{Abdo} et~al.}(2009{\natexlab{b}})\citenamefont{{Abdo},
  {Ackermann}, {Ajello}, {Ampe}, {Anderson}, {Atwood}, {Axelsson}, {Bagagli},
  {Baldini}, {Ballet} et~al.}}]{Eduardo}
\bibinfo{author}{\bibfnamefont{A.~A.} \bibnamefont{{Abdo}}},
  \bibinfo{author}{\bibfnamefont{M.}~\bibnamefont{{Ackermann}}},
  \bibinfo{author}{\bibfnamefont{M.}~\bibnamefont{{Ajello}}},
  \bibinfo{author}{\bibfnamefont{J.}~\bibnamefont{{Ampe}}},
  \bibinfo{author}{\bibfnamefont{B.}~\bibnamefont{{Anderson}}},
  \bibinfo{author}{\bibfnamefont{W.~B.} \bibnamefont{{Atwood}}},
  \bibinfo{author}{\bibfnamefont{M.}~\bibnamefont{{Axelsson}}},
  \bibinfo{author}{\bibfnamefont{R.}~\bibnamefont{{Bagagli}}},
  \bibinfo{author}{\bibfnamefont{L.}~\bibnamefont{{Baldini}}},
  \bibinfo{author}{\bibfnamefont{J.}~\bibnamefont{{Ballet}}},
  \bibnamefont{et~al.}, \bibinfo{journal}{Astropart. Phys.}
  \textbf{\bibinfo{volume}{32}}, \bibinfo{pages}{193}
  (\bibinfo{year}{2009}{\natexlab{b}}).

\bibitem[{\citenamefont{{Aguilar} et~al.}(2002)\citenamefont{{AMS
  Collaboration}, {Aguilar}, {Alcaraz}, {Allaby}, {Alpat}, {Ambrosi},
  {Anderhub}, {Ao}, {Arefiev}, {Azzarello} et~al.}}]{ams}
  \bibinfo{author}{\bibfnamefont{M.}~\bibnamefont{{Aguilar}}},
  \bibinfo{author}{\bibfnamefont{J.}~\bibnamefont{{Alcaraz}}},
  \bibinfo{author}{\bibfnamefont{J.}~\bibnamefont{{Allaby}}},
  \bibinfo{author}{\bibfnamefont{B.}~\bibnamefont{{Alpat}}},
  \bibinfo{author}{\bibfnamefont{G.}~\bibnamefont{{Ambrosi}}},
  \bibinfo{author}{\bibfnamefont{H.}~\bibnamefont{{Anderhub}}},
  \bibinfo{author}{\bibfnamefont{L.}~\bibnamefont{{Ao}}},
  \bibinfo{author}{\bibfnamefont{A.}~\bibnamefont{{Arefiev}}},
  \bibinfo{author}{\bibfnamefont{P.}~\bibnamefont{{Azzarello}}},
  \bibnamefont{et~al.}, 
\bibinfo{author}{\bibnamefont{{(AMS Collaboration)}}},
\bibinfo{journal}{Phys.~Rep.}
  \textbf{\bibinfo{volume}{366}}, \bibinfo{pages}{331} (\bibinfo{year}{2002}).

\bibitem[{\citenamefont{{Haino} et~al.}(2004)\citenamefont{{Haino}, {Sanuki},
  {Abe}, {Anraku}, {Asaoka}, {Fuke}, {Imori}, {Itasaki}, {Maeno}, {Makida}
  et~al.}}]{bess}
\bibinfo{author}{\bibfnamefont{S.}~\bibnamefont{{Haino}}},
  \bibinfo{author}{\bibfnamefont{T.}~\bibnamefont{{Sanuki}}},
  \bibinfo{author}{\bibfnamefont{K.}~\bibnamefont{{Abe}}},
  \bibinfo{author}{\bibfnamefont{K.}~\bibnamefont{{Anraku}}},
  \bibinfo{author}{\bibfnamefont{Y.}~\bibnamefont{{Asaoka}}},
  \bibinfo{author}{\bibfnamefont{H.}~\bibnamefont{{Fuke}}},
  \bibinfo{author}{\bibfnamefont{M.}~\bibnamefont{{Imori}}},
  \bibinfo{author}{\bibfnamefont{A.}~\bibnamefont{{Itasaki}}},
  \bibinfo{author}{\bibfnamefont{T.}~\bibnamefont{{Maeno}}},
  \bibinfo{author}{\bibfnamefont{Y.}~\bibnamefont{{Makida}}},
  \bibnamefont{et~al.}, \bibinfo{journal}{Phys. Lett. B}
  \textbf{\bibinfo{volume}{594}}, \bibinfo{pages}{35} (\bibinfo{year}{2004})

\bibitem[{\citenamefont{{Agostinelli} et~al.}(2003)\citenamefont{{Agostinelli},
  {Allison}, {Amako}, {Apostolakis}, {Araujo}, {Arce}, {Asai}, {Axen},
  {Banerjee}, {Barrand} et~al.}}]{Geant_1}
\bibinfo{author}{\bibfnamefont{S.}~\bibnamefont{{Agostinelli}}},
  \bibinfo{author}{\bibfnamefont{J.}~\bibnamefont{{Allison}}},
  \bibinfo{author}{\bibfnamefont{K.}~\bibnamefont{{Amako}}},
  \bibinfo{author}{\bibfnamefont{J.}~\bibnamefont{{Apostolakis}}},
  \bibinfo{author}{\bibfnamefont{H.}~\bibnamefont{{Araujo}}},
  \bibinfo{author}{\bibfnamefont{P.}~\bibnamefont{{Arce}}},
  \bibinfo{author}{\bibfnamefont{M.}~\bibnamefont{{Asai}}},
  \bibinfo{author}{\bibfnamefont{D.}~\bibnamefont{{Axen}}},
  \bibinfo{author}{\bibfnamefont{S.}~\bibnamefont{{Banerjee}}},
  \bibinfo{author}{\bibfnamefont{G.}~\bibnamefont{{Barrand}}},
  \bibnamefont{et~al.}, \bibinfo{journal}{Nucl. Instrum. Methods
  Phys. Res., Sect. A} \textbf{\bibinfo{volume}{506}}, \bibinfo{pages}{250}
  (\bibinfo{year}{2003}).

\bibitem[{\citenamefont{{Sgr{\`o}} et~al.}(2009)\citenamefont{{Sgr{\`o}},
  {Bregeon}, {Baldini}, and {for the FERMI LAT collaboration}}}]{carmelo}
\bibinfo{author}{\bibfnamefont{C.}~\bibnamefont{{Sgr{\`o}}}},
  \bibinfo{author}{\bibfnamefont{J.}~\bibnamefont{{Bregeon}}},
  \bibinfo{author}{\bibfnamefont{L.}~\bibnamefont{{Baldini}}}
  \bibinfo{author}{\bibnamefont{{(for the Fermi LAT
  collaboration)}}}, in \emph{\bibinfo{booktitle}{{Proceedings of 31st
  International Cosmic Ray Conference}}} \bibinfo{publisher}{{{\L}\'{o}d\'{z},
  Poland}}, \bibinfo{year}{2009}, \eprint{arXiv:0907.0385}.

\bibitem[{\citenamefont{{Zuccon} et~al.}(2003)\citenamefont{{Zuccon},
  {Bertucci}, {Alpat}, {Ambrosi}, {Battiston}, {Battistoni}, {Burger},
  {Caraffini}, {Cecchi}, {di Masso} et~al.}}]{zuccon}
\bibinfo{author}{\bibfnamefont{P.}~\bibnamefont{{Zuccon}}},
  \bibinfo{author}{\bibfnamefont{B.}~\bibnamefont{{Bertucci}}},
  \bibinfo{author}{\bibfnamefont{B.}~\bibnamefont{{Alpat}}},
  \bibinfo{author}{\bibfnamefont{G.}~\bibnamefont{{Ambrosi}}},
  \bibinfo{author}{\bibfnamefont{R.}~\bibnamefont{{Battiston}}},
  \bibinfo{author}{\bibfnamefont{G.}~\bibnamefont{{Battistoni}}},
  \bibinfo{author}{\bibfnamefont{W.~J.} \bibnamefont{{Burger}}},
  \bibinfo{author}{\bibfnamefont{D.}~\bibnamefont{{Caraffini}}},
  \bibinfo{author}{\bibfnamefont{C.}~\bibnamefont{{Cecchi}}},
  \bibinfo{author}{\bibfnamefont{L.}~\bibnamefont{{di Masso}}},
  \bibnamefont{et~al.}, \bibinfo{journal}{Astropart. Phys.}
  \textbf{\bibinfo{volume}{20}}, \bibinfo{pages}{221} (\bibinfo{year}{2003}).

\bibitem[{\citenamefont{{Baldini} et~al.}(2007)\citenamefont{{Baldini},
  {Barbiellini}, {Bellazzini}, {Bogart}, {Bogaert}, {Bonamente}, {Bregeon},
  {Brez}, {Brigida}, {Borgland} et~al.}}]{BTGlastSymposium}
\bibinfo{author}{\bibfnamefont{L.}~\bibnamefont{{Baldini}}},
  \bibinfo{author}{\bibfnamefont{G.}~\bibnamefont{{Barbiellini}}},
  \bibinfo{author}{\bibfnamefont{R.}~\bibnamefont{{Bellazzini}}},
  \bibinfo{author}{\bibfnamefont{J.~R.} \bibnamefont{{Bogart}}},
  \bibinfo{author}{\bibfnamefont{G.}~\bibnamefont{{Bogaert}}},
  \bibinfo{author}{\bibfnamefont{E.}~\bibnamefont{{Bonamente}}},
  \bibinfo{author}{\bibfnamefont{J.}~\bibnamefont{{Bregeon}}},
  \bibinfo{author}{\bibfnamefont{A.}~\bibnamefont{{Brez}}},
  \bibinfo{author}{\bibfnamefont{M.}~\bibnamefont{{Brigida}}},
  \bibinfo{author}{\bibfnamefont{A.~W.} \bibnamefont{{Borgland}}},
  \bibnamefont{et~al.}, in \emph{\bibinfo{booktitle}{The First GLAST
  Symposium}}, edited by \bibinfo{editor}{\bibnamefont{{S.~Ritz, P.~Michelson,
  \& C.~A.~Meegan}}} (\bibinfo{year}{2007}), vol. \bibinfo{volume}{921} of
  \emph{\bibinfo{series}{American Institute of Physics Conference Series}}, pp.
  \bibinfo{pages}{190--204}.

\bibitem[{\citenamefont{{Nakamura} et~al.}(2010)}]{PDG}
\bibinfo{author}{\bibfnamefont{K.}~\bibnamefont{{Nakamura}}}
  \bibnamefont{et~al.}, \bibinfo{journal}{{J. Phys. G}}
  \textbf{\bibinfo{volume}{37}}, \bibinfo{pages}{075021}
  (\bibinfo{year}{2010}), \urlprefix\url{http://pdg.lbl.gov}.

\bibitem[{\citenamefont{{Heikkinen}}(2005)}]{heikkinen}
\bibinfo{author}{\bibfnamefont{A.}~\bibnamefont{{Heikkinen}}}, in
  \emph{\bibinfo{booktitle}{The Monte Carlo Method: Versatility Unbounded in a
  Dynamic Computing World}} (\bibinfo{publisher}{American Nuclear Society},
  \bibinfo{year}{2005}).

\bibitem[{ber()}]{bertini}
\bibinfo{author}{\geant\ Collaboration}, \bibinfo{journal}{\geant\ Physics List},
\urlprefix\href{http://geant4.cern.ch/support/proc_mod_catalog/physics_lists/physicsLists.shtml}{%
\texttt{http://geant4.cern.ch/support/proc\_mod\_catalog/}} \\
\href{http://geant4.cern.ch/support/proc_mod_catalog/physics_lists/physicsLists.shtml}{%
\texttt{physics\_lists/ physicsLists.shtml}}.

\bibitem[{\citenamefont{{Ball} and {Brunner}}(2009)}]{datamining}
\bibinfo{author}{\bibfnamefont{N.~M.} \bibnamefont{{Ball}}} \bibnamefont{and}
  \bibinfo{author}{\bibfnamefont{R.~J.} \bibnamefont{{Brunner}}},
  \bibinfo{journal}{Int. J. Mod. Phys. D}, \textbf{\bibinfo{volume}{19}}, \bibinfo{pages}{1049},
(\bibinfo{year}{2010}), \eprint{arXiv:0906.2173}.

\bibitem[{\citenamefont{{D'Agostini}}(1995)}]{unfolding}
\bibinfo{author}{\bibfnamefont{G.}~\bibnamefont{{D'Agostini}}},
  \bibinfo{journal}{Nucl. Instrum. Methods Phys. Res., Sect. A}
  \textbf{\bibinfo{volume}{362}}, \bibinfo{pages}{487} (\bibinfo{year}{1995}).

\bibitem[{\citenamefont{{Smart} and {Shea}}(2005)}]{SmartShea}
\bibinfo{author}{\bibfnamefont{D.~F.} \bibnamefont{{Smart}}} \bibnamefont{and}
  \bibinfo{author}{\bibfnamefont{M.~A.} \bibnamefont{{Shea}}},
  \bibinfo{journal}{Adv. Space Res.} \textbf{\bibinfo{volume}{36}},
  \bibinfo{pages}{2012} (\bibinfo{year}{2005}).

\bibitem[{IGR()}]{IGRF}
\bibinfo{author}{International Association of Geomagnetism and Aeronomy}, 
\bibinfo{journal}{International Geomagnetic Reference Field},
  \urlprefix\url{http://www.ngdc.noaa.gov/IAGA/vmod/igrf.html}.



\bibitem[{\citenamefont{{Strong} et~al.}(2004)\citenamefont{{Strong},
  {Moskalenko}, and {Reimer}}}]{2004ApJ...613..962S}
\bibinfo{author}{\bibfnamefont{A.~W.} \bibnamefont{{Strong}}},
  \bibinfo{author}{\bibfnamefont{I.~V.} \bibnamefont{{Moskalenko}}},
  \bibnamefont{and} \bibinfo{author}{\bibfnamefont{O.}~\bibnamefont{{Reimer}}},
  \bibinfo{journal}{Astrophys. J.} \textbf{\bibinfo{volume}{613}}, \bibinfo{pages}{962}
  (\bibinfo{year}{2004}).

\bibitem[{\citenamefont{{Boezio} et~al.}(2000)\citenamefont{{Boezio},
  {Carlson}, {Francke}, {Weber}, {Suffert}, {Hof}, {Menn}, {Simon}, {Stephens},
  {Bellotti} et~al.}}]{CAPRICE}
\bibinfo{author}{\bibfnamefont{M.}~\bibnamefont{{Boezio}}},
  \bibinfo{author}{\bibfnamefont{P.}~\bibnamefont{{Carlson}}},
  \bibinfo{author}{\bibfnamefont{T.}~\bibnamefont{{Francke}}},
  \bibinfo{author}{\bibfnamefont{N.}~\bibnamefont{{Weber}}},
  \bibinfo{author}{\bibfnamefont{M.}~\bibnamefont{{Suffert}}},
  \bibinfo{author}{\bibfnamefont{M.}~\bibnamefont{{Hof}}},
  \bibinfo{author}{\bibfnamefont{W.}~\bibnamefont{{Menn}}},
  \bibinfo{author}{\bibfnamefont{M.}~\bibnamefont{{Simon}}},
  \bibinfo{author}{\bibfnamefont{S.~A.} \bibnamefont{{Stephens}}},
  \bibinfo{author}{\bibfnamefont{R.}~\bibnamefont{{Bellotti}}},
  \bibnamefont{et~al.}, \bibinfo{journal}{Astrophys. J.} \textbf{\bibinfo{volume}{532}},
  \bibinfo{pages}{653} (\bibinfo{year}{2000}).

\bibitem[{\citenamefont{{Casolino} et~al.}(2009)\citenamefont{{Casolino}, {de
  Simone}, {de Pascale}, {di Felice}, {Marcelli}, {Minori}, {Picozza},
  {Sparvoli}, {Castellini}, {Adriani} et~al.}}]{2009NuPhS.190..293C}
\bibinfo{author}{\bibfnamefont{M.}~\bibnamefont{{Casolino}}},
  \bibinfo{author}{\bibfnamefont{N.}~\bibnamefont{{de Simone}}},
  \bibinfo{author}{\bibfnamefont{M.~P.} \bibnamefont{{de Pascale}}},
  \bibinfo{author}{\bibfnamefont{V.}~\bibnamefont{{di Felice}}},
  \bibinfo{author}{\bibfnamefont{L.}~\bibnamefont{{Marcelli}}},
  \bibinfo{author}{\bibfnamefont{M.}~\bibnamefont{{Minori}}},
  \bibinfo{author}{\bibfnamefont{P.}~\bibnamefont{{Picozza}}},
  \bibinfo{author}{\bibfnamefont{R.}~\bibnamefont{{Sparvoli}}},
  \bibinfo{author}{\bibfnamefont{G.}~\bibnamefont{{Castellini}}},
  \bibinfo{author}{\bibfnamefont{O.}~\bibnamefont{{Adriani}}},
  \bibnamefont{et~al.}, \bibinfo{journal}{Nucl. Phys. B, Proc.
  Suppl.} \textbf{\bibinfo{volume}{190}}, \bibinfo{pages}{293}
  (\bibinfo{year}{2009}).

\bibitem[{\citenamefont{{Strong} and {Moskalenko}}(1998)}]{SM1998}
\bibinfo{author}{\bibfnamefont{A.~W.} \bibnamefont{{Strong}}} \bibnamefont{and}
  \bibinfo{author}{\bibfnamefont{I.~V.} \bibnamefont{{Moskalenko}}},
  \bibinfo{journal}{Astrophys. J.} \textbf{\bibinfo{volume}{509}}, \bibinfo{pages}{212}
  (\bibinfo{year}{1998}).

\bibitem[{\citenamefont{{Ptuskin} et~al.}(2006)\citenamefont{{Ptuskin},
  {Moskalenko}, {Jones}, {Strong}, and {Zirakashvili}}}]{ptuskin}
\bibinfo{author}{\bibfnamefont{V.~S.} \bibnamefont{{Ptuskin}}},
  \bibinfo{author}{\bibfnamefont{I.~V.} \bibnamefont{{Moskalenko}}},
  \bibinfo{author}{\bibfnamefont{F.~C.} \bibnamefont{{Jones}}},
  \bibinfo{author}{\bibfnamefont{A.~W.} \bibnamefont{{Strong}}},
  \bibnamefont{and} \bibinfo{author}{\bibfnamefont{V.~N.}
  \bibnamefont{{Zirakashvili}}}, \bibinfo{journal}{Astrophys. J.}
  \textbf{\bibinfo{volume}{642}}, \bibinfo{pages}{902} (\bibinfo{year}{2006}).

\bibitem[{\citenamefont{{Strong} et~al.}(2007)\citenamefont{{Strong},
  {Moskalenko}, and {Ptuskin}}}]{ARNPS}
\bibinfo{author}{\bibfnamefont{A.~W.} \bibnamefont{{Strong}}},
  \bibinfo{author}{\bibfnamefont{I.~V.} \bibnamefont{{Moskalenko}}},
  \bibnamefont{and} \bibinfo{author}{\bibfnamefont{V.~S.}
  \bibnamefont{{Ptuskin}}}, \bibinfo{journal}{Annual Review of Nuclear and
  Particle Science} \textbf{\bibinfo{volume}{57}}, \bibinfo{pages}{285}
  (\bibinfo{year}{2007}).

\bibitem[{\citenamefont{{Gleeson} and {Axford}}(1968)}]{Gleeson&Axford}
\bibinfo{author}{\bibfnamefont{L.~J.} \bibnamefont{{Gleeson}}}
  \bibnamefont{and} \bibinfo{author}{\bibfnamefont{W.~I.}
  \bibnamefont{{Axford}}}, \bibinfo{journal}{Astrophys. J.}
  \textbf{\bibinfo{volume}{154}}, \bibinfo{pages}{1011} (\bibinfo{year}{1968}).

\bibitem[{\citenamefont{{Clem} and {Evenson}}(2009)}]{clem}
\bibinfo{author}{\bibfnamefont{J.}~\bibnamefont{{Clem}}} \bibnamefont{and}
  \bibinfo{author}{\bibfnamefont{P.}~\bibnamefont{{Evenson}}},
  \bibinfo{journal}{Journal of Geophysical Research (Space Physics)}
  \textbf{\bibinfo{volume}{114}}, \bibinfo{pages}{10108}
  (\bibinfo{year}{2009}).

\bibitem[{\citenamefont{{Ferreira} and {Potgieter}}(2004)}]{potg}
\bibinfo{author}{\bibfnamefont{S.~E.~S.} \bibnamefont{{Ferreira}}}
  \bibnamefont{and} \bibinfo{author}{\bibfnamefont{M.~S.}
  \bibnamefont{{Potgieter}}}, \bibinfo{journal}{Astrophys. J.}
  \textbf{\bibinfo{volume}{603}}, \bibinfo{pages}{744} (\bibinfo{year}{2004}).

\bibitem[{\citenamefont{{Potgieter}}(1997)}]{Potgieter}
\bibinfo{author}{\bibfnamefont{M.~S.} \bibnamefont{{Potgieter}}},
  \bibinfo{journal}{Adv. Space Res.} \textbf{\bibinfo{volume}{19}},
  \bibinfo{pages}{883} (\bibinfo{year}{1997}).

\bibitem[{\citenamefont{{Abdo} et~al.}(2009{\natexlab{c}})\citenamefont{{Abdo},
  {Ackermann}, {Ajello}, {Anderson}, {Atwood}, {Axelsson}, {Baldini}, {Ballet},
  {Barbiellini}, {Bastieri} et~al.}}]{LatDiffuse}
\bibinfo{author}{\bibfnamefont{A.~A.} \bibnamefont{{Abdo}}},
  \bibinfo{author}{\bibfnamefont{M.}~\bibnamefont{{Ackermann}}},
  \bibinfo{author}{\bibfnamefont{M.}~\bibnamefont{{Ajello}}},
  \bibinfo{author}{\bibfnamefont{B.}~\bibnamefont{{Anderson}}},
  \bibinfo{author}{\bibfnamefont{W.~B.} \bibnamefont{{Atwood}}},
  \bibinfo{author}{\bibfnamefont{M.}~\bibnamefont{{Axelsson}}},
  \bibinfo{author}{\bibfnamefont{L.}~\bibnamefont{{Baldini}}},
  \bibinfo{author}{\bibfnamefont{J.}~\bibnamefont{{Ballet}}},
  \bibinfo{author}{\bibfnamefont{G.}~\bibnamefont{{Barbiellini}}},
  \bibinfo{author}{\bibfnamefont{D.}~\bibnamefont{{Bastieri}}},
  \bibnamefont{et~al.}, \bibinfo{journal}{Phys. Rev. Lett.}
  \textbf{\bibinfo{volume}{103}}, \bibinfo{pages}{251101}
  (\bibinfo{year}{2009}{\natexlab{c}}).

\bibitem[{\citenamefont{{Stawarz} et~al.}(2010)\citenamefont{{Stawarz},
  {Petrosian}, and {Blandford}}}]{2010ApJ...710..236S}
\bibinfo{author}{\bibfnamefont{{\L}.}~\bibnamefont{{Stawarz}}},
  \bibinfo{author}{\bibfnamefont{V.}~\bibnamefont{{Petrosian}}},
  \bibnamefont{and} \bibinfo{author}{\bibfnamefont{R.~D.}
  \bibnamefont{{Blandford}}}, \bibinfo{journal}{Astrophys. J.}
  \textbf{\bibinfo{volume}{710}}, \bibinfo{pages}{236} (\bibinfo{year}{2010}).

\bibitem[{\citenamefont{{Blasi}}(2009)}]{Blasi:2009hv}
\bibinfo{author}{\bibfnamefont{P.}~\bibnamefont{{Blasi}}},
  \bibinfo{journal}{Phys. Rev. Lett.} \textbf{\bibinfo{volume}{103}},
  \bibinfo{pages}{051104} (\bibinfo{year}{2009}).


\bibitem[{\citenamefont{{Kobayashi} et~al.}(1999)}]{Kobayashi1999}
\bibinfo{author}{\bibfnamefont{T.}~\bibnamefont{{Kobayashi}}}
  \bibnamefont{et~al.}, in \emph{\bibinfo{booktitle}{Proceedings of the 26th
  International Cosmic Ray Conference},
  \bibinfo{publisher}{{Salt Lake City,}}}, \bibinfo{year}{1999},
\urlprefix\url{http://adsabs.harvard.edu/abs/1999ICRC....3...61K}


\bibitem[{\citenamefont{{DuVernois} et~al.}(2001)\citenamefont{{DuVernois},
  {Barwick}, {Beatty}, {Bhattacharyya}, {Bower}, {Chaput}, {Coutu}, {de Nolfo},
  {Lowder}, {McKee} et~al.}}]{HEAT2001}
\bibinfo{author}{\bibfnamefont{M.~A.} \bibnamefont{{DuVernois}}},
  \bibinfo{author}{\bibfnamefont{S.~W.} \bibnamefont{{Barwick}}},
  \bibinfo{author}{\bibfnamefont{J.~J.} \bibnamefont{{Beatty}}},
  \bibinfo{author}{\bibfnamefont{A.}~\bibnamefont{{Bhattacharyya}}},
  \bibinfo{author}{\bibfnamefont{C.~R.} \bibnamefont{{Bower}}},
  \bibinfo{author}{\bibfnamefont{C.~J.} \bibnamefont{{Chaput}}},
  \bibinfo{author}{\bibfnamefont{S.}~\bibnamefont{{Coutu}}},
  \bibinfo{author}{\bibfnamefont{G.~A.} \bibnamefont{{de Nolfo}}},
  \bibinfo{author}{\bibfnamefont{D.~M.} \bibnamefont{{Lowder}}},
  \bibinfo{author}{\bibfnamefont{S.}~\bibnamefont{{McKee}}},
  \bibnamefont{et~al.}, \bibinfo{journal}{Astrophys. J.} \textbf{\bibinfo{volume}{559}},
  \bibinfo{pages}{296} (\bibinfo{year}{2001}).

\bibitem[{\citenamefont{{Torii} et~al.}(2001)\citenamefont{{Torii}, {Tamura},
  {Tateyama}, {Yoshida}, {Nishimura}, {Yamagami}, {Murakami}, {Kobayashi},
  {Komori}, {Kasahara} et~al.}}]{BETS}
\bibinfo{author}{\bibfnamefont{S.}~\bibnamefont{{Torii}}},
  \bibinfo{author}{\bibfnamefont{T.}~\bibnamefont{{Tamura}}},
  \bibinfo{author}{\bibfnamefont{N.}~\bibnamefont{{Tateyama}}},
  \bibinfo{author}{\bibfnamefont{K.}~\bibnamefont{{Yoshida}}},
  \bibinfo{author}{\bibfnamefont{J.}~\bibnamefont{{Nishimura}}},
  \bibinfo{author}{\bibfnamefont{T.}~\bibnamefont{{Yamagami}}},
  \bibinfo{author}{\bibfnamefont{H.}~\bibnamefont{{Murakami}}},
  \bibinfo{author}{\bibfnamefont{T.}~\bibnamefont{{Kobayashi}}},
  \bibinfo{author}{\bibfnamefont{Y.}~\bibnamefont{{Komori}}},
  \bibinfo{author}{\bibfnamefont{K.}~\bibnamefont{{Kasahara}}},
  \bibnamefont{et~al.}, \bibinfo{journal}{Astrophys. J.} \textbf{\bibinfo{volume}{559}},
  \bibinfo{pages}{973} (\bibinfo{year}{2001}).

\end{thebibliography}
\end{document}